%% file: main.tex
\documentclass[a4paper,onecolumn,accepted=2025-12-03]{quantumarticle}
\pdfoutput=1
\usepackage[numbers, sort]{natbib}
\usepackage{amsmath}
\usepackage{comment}
\usepackage{hyperref}
\usepackage{mathtools}
\input{macrosquantum}
\usepackage{comment}
\newenvironment{theo}
  {\begin{theorem}}
  {\end{theorem}}

\newcommand*\step{\eta}

\makeatletter
\newcommand\newtag[2]{#1\def\@currentlabel{#1}\label{#2}}
\makeatother
\newcommand{\edits}[1]{{\color{OliveGreen}{#1}}}
\title{Learning in Quantum Common-Interest Games and the Separability Problem
}

\author{Wayne Lin}
\email{wayne\_lin@sutd.edu.sg}
\affiliation{Singapore University of Technology and Design, Singapore}
\author{Georgios Piliouras}
\email{georgios@sutd.edu.sg}
\affiliation{Singapore University of Technology and Design, Singapore}

\author{Ryann Sim}
\email{ryann\_sim@sutd.edu.sg}
\affiliation{Singapore University of Technology and Design, Singapore}
\author{Antonios Varvitsiotis}
\email{antonios@sutd.edu.sg}
\affiliation{Singapore University of Technology and Design, Singapore}
\affiliation{Centre for Quantum Technologies, National University of Singapore, Singapore}
\affiliation{Archimedes/Athena RC, Greece}
\date{}

\begin{document}

\maketitle

\begin{abstract}

% Gamification is an emerging trend in the field of machine learning that presents a novel approach to solving optimization problems by transforming them into game-like scenarios. This paradigm shift allows for the development of robust, easily implementable, and parallelizable algorithms for hard optimization problems. In our work, we use  gamification  to tackle the Best Separable State (BSS) problem, a fundamental problem in quantum information theory that involves linear optimization over the set of separable quantum states.

% To achieve this we introduce and study quantum analogues of common-interest games~(CIGs) and potential games where players have density matrices  as strategies and their interests are perfectly aligned.

% We bridge the gap between optimization and game theory by establishing the equivalence between KKT (first-order stationary) points of a BSS instance  and the Nash equilibria of its corresponding quantum CIG. Taking the perspective of learning in games, we introduce non-commutative extensions of the continuous-time replicator dynamics and the discrete-time Baum-Eagon/linear multiplicative weights update for learning in quantum CIGs, which  also serve as decentralized algorithms for the BSS problem. We show that the common utility/objective value of a BSS instance is strictly increasing along trajectories of our algorithms, and finally corroborate our theoretical findings through extensive experiments. 

% [New abstract below:]

Learning in games has emerged as a powerful tool for machine learning with numerous applications. Quantum games model interactions between strategic players who have access to quantum resources, and several recent works have studied {learning in} the competitive regime of quantum zero-sum games.
% , {which are} extensions of classical zero-sum games.
% ,  from a learning perspective. 
Going beyond this setting, we introduce quantum common-interest games (CIGs) where players have density matrices as strategies and their interests are perfectly aligned. We bridge the gap between optimization and game theory by establishing the equivalence between KKT (first-order stationary) points of an instance of the Best Separable State (BSS) problem and the Nash equilibria of its corresponding quantum CIG. 
This allows learning dynamics for the quantum CIG to be seen as decentralized algorithms for the BSS problem.
Taking the perspective of learning in games, we then introduce non-commutative extensions of the continuous-time replicator dynamics and the discrete-time best response dynamics/linear multiplicative weights update for learning in quantum CIGs.
% , which also serve as decentralized algorithms for the BSS problem.
We prove analogues of classical convergence results of the dynamics and explore differences which arise in the quantum setting. Finally, we corroborate our theoretical findings through extensive experiments.

\end{abstract}

\section{Introduction}\label{sec:introduction}
\input{introquantumv6.tex}

\section{Quantum Common-Interest Games and the BSS problem}\label{sec:QCIG}

In this section, we introduce quantum common-interest games (CIGs) and establish their connection to the \ref{BSS} problem by showing that the Nash equilibria of a quantum CIG and the KKT points of the corresponding BSS problem instance coincide.

% A classical normal-form  game consists of $n$ players, where player $i$ has a finite set $\mca_i$ of $n_i$ pure actions and a payoff (aka utility) function $\payoff{i}: \prod_{i = 1}^n \mca_i \rightarrow \mathbb{R}$. More generally, players  are also allowed  to use mixed (randomized) strategies, corresponding to  elements of the simplex of dimension $n_i$, denoted by $\mcx_i$.  Moreover, the utility function $u_i$ extends to a multilinear mapping on the product of the corresponding simplices, i.e.,
% $$u_i(x)=\sum_{(a_1,\ldots,a_n)\in \mca_1\times \ldots \times \mca_n}x_{1,a_1}\ldots x_{n,a_n}u_i(a_1,\ldots, a_n).$$
% \paragraph{Preliminaries.} Density matrices are PSD matrices with trace 1. Hermitian matrices are matrices whi

%For the rest of the main paper, we will restrict our analysis to the setting of two-player quantum common-interest games to simplify notation.
% , though

\subsection{Quantum games}

\paragraph{Quantum games.} In this work we study non-interactive quantum games where each player $i$ controls a quantum register $\mathcal{H}_i$ and has as their strategy a density matrix $\rho_i \in {\mathrm{D}}(\mathcal{H}_i)$. The utility function of the $i$-th player is given by
the expected value of an observable $R_i$ on the strategy profile $\bigotimes_i \rho_i$, i.e.,
\begin{equation*}
%  \label{QG}
% \tag{QG}
	u_i\bigg(\bigotimes_i \rho_i \bigg)= \Tr(R_i\bigg(\bigotimes_i \rho_i \bigg)).
\end{equation*}

\paragraph{Quantum common-interest games.} A quantum CIG is a quantum game where every player has the same utility function. For simplicity, we shall in this paper consider two-player quantum CIGs, where two agents Alice and Bob 
 control  quantum registers $\A$ and $\B$ and play strategies given by density matrices in    $D(\A)$ and $ D(\B)$ respectively. Upon playing strategy profile $(\rho,\sigma)\in \ D(\A)\times  D(\B)$  both players receive a common utility $u(\rho, \sigma) = \la R, \rho\otimes\sigma\ra$,  where $R$ is a Hermitian matrix that we can assume without loss of generality to be positive definite.
 %\textcolor{red}{to transpose or not to transpose?}
% \subsection{Quantum common-interest games}
We refer to the matrix $R$ as the {\em game operator}. Equivalently,
using the Choi-Jamio\l{}kowski isomorphism defined in~\eqref{CJ}, it is useful to
also express the utility function as $u(\rho, \sigma) = \la \rho, \Phi(\sigma^\top) \ra$, since
\begin{align*}
\la \rho, \Phi(\sigma^\top) \ra &= \langle \rho,  \mathrm{Tr}_\mathcal{B} (R(\mathbb{1}_\mathcal{A}\otimes \sigma)\rangle =\langle \rho\otimes \id_\B,R(\mathbb{1}_\mathcal{A}\otimes \sigma)\rangle =\langle R, \rho\otimes \sigma\rangle,
\end{align*}
where $R$ is the Choi matrix of $\Phi$. 
Moreover, as $R$ is PSD it follows that $\Phi$ is positive. In order to simplify notation throughout the rest of the paper, we will drop the transpose from the utility and express it as $u(\rho,\sigma) = \la \rho, \Phi(\sigma) \ra$ where appropriate. This can be seen as Bob selecting $\sigma^\top$ as his strategy, instead of $\sigma$ as defined before.

A quantum CIG can also be defined  as the mixed extension of a %common-interest 
game  where the players' pure strategies are  complex unit vectors $x\in \mathbb{S}_\mathbb{C}^{n-1}, y\in\mathbb{S}_\mathbb{C}^{m-1}$  and the common utility  is biquadratic, i.e., $u(x,y)=(x\otimes y)^\dagger R(x\otimes y).$ In particular, if  the players randomize their play using   finitely supported distributions $\mathcal{D}_\A, \mathcal{D}_\B$
over their pure strategy spaces, i.e.,
$\mathcal{D}_\A$ has support
$\{x_i\}_{i=1}^k$
 and ${\rm Prob}(x_i)=\lam_i$
whereas  $\mathcal{D}_\B$ has support
$\{y_j\}_{j=1}^\ell$
 and ${\rm Prob}(y_j)=\mu_j$
, the expected payoff is bilinear in the density matrices $\rho=\sum_{i=1}^k\lam_ix_ix_i^\dagger$ and $\sigma=\sum_{j=1}^\ell \mu_jy_jy_j^\dagger$~as
$$\mathbb{E}[(x\otimes y)^\dagger R(x\otimes y)]=\Tr(R(\rho\otimes \sigma)),$$
where expectation is taken over $x\sim\mathcal{D}_\A$, $y\sim \mathcal{D}_\B$.

Lastly, a classical CIG with common utility $x^\top A y$ where $x\in \Delta_n, y\in \Delta_m$ is a special case of a quantum CIG. Indeed, consider  the quantum CIG with diagonal game operator $R\in \R^{nm\times nm}$ whose diagonal entries  are  $R_{ij,ij}=A_{ij}$. If we only consider  diagonal densities $\rho=\sum_{i=1}^n x_ie_ie_i^\dagger$ and $\sigma=\sum_{j=1}^my_je_je_j^\dagger$, it is straightforward to verify that  $x^\top A y={\rm Tr}(R(\rho \otimes \sigma))$.

\paragraph{Nash equilibria and exploitability.}  A strategy profile $(\rho^*, \sigma^*)$ is a Nash equilibrium of the quantum CIG with common interest $u(\rho, \sigma) = \rps$ if both strategies are best responses to the other, i.e.,
\begin{equation}
\label{NE} \tag{NE}
\innerprod{\rho^*}{\Phi(\sigma^*)} \geq \innerprod{\rho}{\Phi(\sigma^*)} \ \forall \; \rho \in D(\mathcal{A}) \quad \text{ and } \quad \innerprod{\rho^*}{\Phi(\sigma^*)} \geq \innerprod{\rho^*}{\Phi(\sigma)} \ \forall \; \sigma \in D(\mathcal{B}).
\end{equation}

% We experimentally test if a similar property holds for \ref{lin-QREP} 
To capture distance from the set of Nash equilibria we utilize the concept of \emph{exploitability} (see e.g. \cite{johanson2011accelerating}), defined as 
\begin{equation}\label{eqn:exploitability}
    \frac{1}{2}\left[\lambda_{\max} (\Phi(\sigma)) - \langle \rho, \Phi(\sigma)\rangle + \lambda_{\max} (\Phi^\dagger(\rho)) - \langle\Phi^\dagger(\rho), \sigma\rangle\right],
\end{equation}

where $\lambda_{\max} (\Phi(\sigma))$ and $\lambda_{\max} (\Phi^\dagger(\rho))$ are the maximum eigenvalues of $\Phi(\sigma)$ and $\Phi^\dagger(\rho)$ respectively. Using the variational characterization of eigenvalues,
$$\lambda_{\max}(\Phi(\sigma))=\max\{\la \rho', \Phi(\sigma)\ra:\ \rho'\in \da\},$$
the difference $\lambda_{\max} (\Phi(\sigma)) - \langle \rho, \Phi(\sigma)\rangle $ is exactly the maximum gain the $\rho$-player can attain by unilaterally deviating from $(\rho,\sigma)$.
Thus, if a profile $(\rho, \sigma)$ is $\epsilon-$exploitable, then it is a $2\epsilon-$\emph{approximate Nash equilibrium} (or simply a $2\epsilon-$Nash equilibrium), in the sense that no player can unilaterally improve their payoff by $\geq 2\epsilon$. 

\subsection{Relation between quantum CIGs and the BSS problem} 

In a quantum CIG, Alice and Bob  try to  jointly  maximize their common utility function $u(\rho, \sigma)=\la \rho, \Phi(\sigma) \ra$. Analogous to the classical case, there is a strong connection between the Nash equilibria of the game and the underlying \ref{BSS} optimization problem. Namely, the Karush-Kuhn Tucker (KKT) points of the \ref{BSS} problem are precisely the Nash equilibria of a corresponding two player quantum CIG.

% \textcolor{purple}{Have to address link between $R$ and $\Phi$ before this.}

%\begin{definition}[Optimization problem corresponding to a two-player common-interest game]\label{def:optiequiv}
%    Given a two player quantum common-interest game $G = [\{1, 2\}, \density{n} \times \density{m}, V]$, where $V: S \rightarrow \mathbb{R}, V(\rho, \sigma) = \innerprod{\rho}{\Phi(\sigma)}$, define the corresponding potential-maximizing problem, $\opt{G}$, to be the optimization problem:

%\end{definition}

\begin{theorem}\label{thm:_KKT_NE_equiv}
The Nash equilibria of a two-player quantum common-interest game with common utility function $u(\rho, \sigma)=\Tr(R(\rho \otimes \sigma))$
%\da\times \db\to \R, \; (\rho, \sigma) \mapsto \innerprod{\rho}{\Phi(\sigma)}$
correspond to the KKT points of~\ref{BSS}.
\end{theorem}

\begin{proof}
We shall prove instead that the Nash equilibria of the (transposed) two-player quantum common-interest game with common utility function $u(\rho, \sigma) = \innerprod{\rho}{\Phi(\sigma)}$ correspond to the KKT points of the transposed BSS problem
\begin{equation}
\label{BSS_trans}\tag{transposed-BSS}
    \max \{ \innerprod{\rho}{\Phi(\sigma)}:  \rho \in \da, \sigma \in \db\}.
\end{equation}
This correspondence is equivalent to the original correspondence we want to prove since $(\rho, \sigma)$ is a Nash equilibrium of the quantum CIG with common utility $\Tr(R(\rho \otimes \sigma))$ iff $(\rho, \sigma^\top)$ is a Nash equilibrium of the quantum CIG with common utility $\innerprod{\rho}{\Phi(\sigma)}$, and similarly $(\rho, \sigma)$ is a KKT point of the \ref{BSS} problem iff $(\rho, \sigma^\top)$ is a KKT point of the \ref{BSS_trans} problem.

Note that $\rho \in \da$ if and only if $\innerprod{\rho}{\ida} = \Tr \, \rho = 1$ and $\rho \succeq 0$, and similarly $\sigma \in \db$ if and only if $\innerprod{\sigma}{\idb} = \Tr \, \sigma = 1$ and $\sigma\succeq 0$.
The Lagrangian for the \ref{BSS_trans} problem 
is  given by
\begin{equation*}
     L = \innerprod{\rho}{\Phi(\sigma)}
    + \lambda (1 - \innerprod{\rho}{\ida})
    + \mu (1 - \innerprod{\sigma}{\idb})
    - \innerprod{\Lambda}{\rho}
    - \innerprod{M}{\sigma},
\end{equation*}
where the dual variables satisfy $\Lambda \preceq 0$, $M \preceq 0$.
Thus, the KKT conditions for the \ref{BSS_trans} problem are
\begin{subequations}
\label{eqn:_KKT_quantum_NablaL}
\begin{align}[left={\nabla L = 0: \empheqlbrace}]
    \pdv{L}{\rho} &= \Phi(\sigma) - \lambda \ida - \Lambda = 0, \label{eqn:_KKT_quantum_NablaL_a}\\
    \pdv{L}{\sigma} &= \Phi^\dagger (\rho) - \mu \idb - M = 0; \label{eqn:_KKT_quantum_NablaL_b}
\end{align}
\end{subequations}
\begin{subequations}
\label{eqn:_KKT_quantum_PrimalFeasibility}
\begin{align}[left={\text{primal feasibility:} \empheqlbrace}]
    \rho &\in \da, \label{eqn:_KKT_quantum_PrimalFeasibility_a}\\
    \sigma &\in \db; \label{eqn:_KKT_quantum_PrimalFeasibility_b}
\end{align}
\end{subequations}
\begin{subequations}
\label{eqn:_KKT_quantum_DualFeasibility}
\begin{align}[left={\text{dual feasibility:} \empheqlbrace}]
    \Lambda &\preceq 0, \label{eqn:_KKT_quantum_DualFeasibility_a}\\
    M &\preceq 0; \label{eqn:_KKT_quantum_DualFeasibility_b}
\end{align}
\end{subequations}
\begin{subequations}
\label{eqn:_KKT_quantum_CompSlack}
\begin{align}[left={\text{complementary slackness:} \empheqlbrace}]
    \innerprod{\Lambda}{\rho} = 0, \label{eqn:_KKT_quantum_CompSlack_a}\\
    \innerprod{M}{\sigma} = 0. \label{eqn:_KKT_quantum_CompSlack_b}
\end{align}
\end{subequations}
\linesmall
Suppose that $(\rho, \sigma, \lambda, \mu, \Lambda, M)$ is a KKT point of the \ref{BSS_trans} problem. We show that $(\rho, \sigma)$ is a Nash equilibrium.
Using  (\ref{eqn:_KKT_quantum_NablaL_a}), for  any density matrix $\rho'\in D(\A)$ we get that
\begin{equation*}
    \innerprod{\rho'}{\Phi(\sigma)} = \lambda \innerprod{\rho'}{\ida} + \innerprod{\rho'}{\Lambda} = \lambda +  \innerprod{\rho'}{\Lambda} \leq \lambda,
\end{equation*} where the inequality follows  since $\rho \succeq 0$ and  $\Lambda\preceq 0$ (recall \eqref{eqn:_KKT_quantum_DualFeasibility_a}). On the other hand, if we take the inner product of (\ref{eqn:_KKT_quantum_NablaL_a}) with $\rho$ instead we have that
\begin{equation*}
    \innerprod{\rho}{\Phi(\sigma)} = \lambda \innerprod{\rho}{\ida} + \innerprod{\rho}{\Lambda} = \lambda \innerprod{\rho}{I} = \lambda,
\end{equation*} 
where we used the complementary slackness condition   $\innerprod{\rho}{\Lambda} = 0$ (\ref{eqn:_KKT_quantum_CompSlack_a}) .
Summarizing, we have that $\innerprod{\rho'}{\Phi(\sigma)} \leq \lambda = \innerprod{\rho}{\Phi(\sigma)} \; \fa \rho' \in \da$, i.e., $\rho$ is a best response to $\sigma$. Similarly we get that $\sigma$ is a best response to $\rho$, and thus that $(\rho, \sigma)$ is a Nash equilibrium of the corresponding quantum CIG.

\linesmall
Next, suppose that $(\rho, \sigma) \in \da \times \db$ is a Nash equilibrium of the quantum CIG, and consider the point $(\rho, \sigma, \lambda, \mu, \Lambda, M)$ defined by
\begin{equation}
\begin{gathered}
    \lambda = \innerprod{\rho}{\Phi(\sigma)}, \quad
    \mu = \innerprod{\rho}{\Phi(\sigma)},
    \\
    \Lambda = \Phi(\sigma) - \lambda \ida,
    \quad
    M = \Phi^\dagger(\rho) - \mu \idb.
\end{gathered}
\end{equation}

The primal feasibility  constraints \eqref{eqn:_KKT_quantum_PrimalFeasibility_a} and \eqref{eqn:_KKT_quantum_PrimalFeasibility_b} are satisfied by construction. Furthermore, 
   (\ref{eqn:_KKT_quantum_NablaL_a}) is immediately satisfied by the definition of $\Lambda$ and $\lambda$, and similarly  \eqref{eqn:_KKT_quantum_NablaL_b} is  satisfied by the definition of $M$ and $\mu$.
The complementary slackness condition  (\ref{eqn:_KKT_quantum_CompSlack_a}) holds since \[\innerprod{\rho}{\Lambda} = \innerprod{\rho}{\Phi(\sigma)} - \lambda \innerprod{\rho}{\ida} = \innerprod{\rho}{\Phi(\sigma)} - \lambda = 0,\]
and similarly  (\ref{eqn:_KKT_quantum_CompSlack_b}) is also satisfied. Finally, since $\rho \in \BRa(\sigma)$ we have that $\fa v \in \mathbb{C}^m$ with $\norm{v}_2 = 1$ $$v^\dagger \Phi(\sigma) v = \innerprod{v v^\dagger}{\Phi(\sigma)} \leq \innerprod{\rho}{\Phi(\sigma)} = \lambda,$$ 
which in turn implies that $\Lambda \preceq 0$ as
$$v^\dagger \Lambda v = v^\dagger \Phi(\sigma) v - \lambda \norm{v}_2^2 = v^\dagger \Phi(\sigma) v - \lambda \leq 0.$$
Thus  (\ref{eqn:_KKT_quantum_DualFeasibility_a}) is satisfied  and a similar argument shows that \eqref{eqn:_KKT_quantum_DualFeasibility_b} is also satisfied.
\end{proof}

For a classical game, if $(x,y)$ is a Nash equilibrium, every pure strategy that is played by Alice with positive probability  is a best response to $y$, i.e., for each $i$ with  $x_i > 0$ we have $(Ay)_i=x^TAy$, and similarly for Bob. We shall prove the analogous statement for quantum CIGs using the notion of the minimal face. 

% We do not have an equivalent statement in the quantum case as the argument regarding support does not carry over quite so easily. However, we do have the quantum analogue of the weaker classical statement that if $(x,y)$ is a Nash equilibrium and $x$ has full support (i.e. $x_i > 0 \; \fa i$), then $(Ay)_i=x^TAy \; \fa i$, i.e. any strategy that Alice can play does equally well (and is a best response) to $y$: \\

%However, we do have the quantum analogue of the weaker classical statement that if $x$ has full support (i.e. $x_i > 0 \; \fa i$) and is a best response to $y$, then $(Ay)_i=x^TAy \; \fa i$, i.e. any strategy that Alice can play does equally well (and is a best response) to $y$. \\

%\begin{theorem}
%\label{thm:_BestResponse_payoffIdentity}
%    If $\rho \succ 0$ and $\rho \in \BRa(\sigma)$, then
%    \begin{equation}
%        \Phi (\sigma) = \innerprod{\rho}{\Phi(\sigma)}I,
%    \end{equation}
%    i.e. every possible density matrix that the Alice can play achieves the same payoff of $\innerprod{\rho}{\Phi(\sigma)}$.
%(The analogous statement holds for Bob: if $\sigma \succ 0$ and $\sigma \in \BRb(\rho)$, then $\Phi^\dagger (\sigma) = \innerprod{\rho}{\Phi(\sigma)}I$.)
%\end{theorem}
%
%The proof of Theorem \ref{thm:_interiorNE_equal_payoff} is deferred to Appendix \ref{proof:_BestResponse_payoffIdentity}. \\

% We can now state and prove the quantum analogue of the aforementioned implication of Nash equilibria:

\begin{theorem}
\label{thm:_interiorNE_equal_payoff}
    Let $(\rho, \sigma)$ be a Nash equilibrium of a two-player quantum CIG  with common utility  $u(\rho, \sigma) = \innerprod{\rho}{\Phi(\sigma)}$. 
    Then any $\rho' \in D(\A)$ satisfying $\range(\rho') \subseteq \range(\rho)$ is a best response to $\sigma$, and similarly any $\sigma' \in D(\B)$ satisfying $\range(\sigma') \subseteq \range(\sigma)$ is a best response to $\rho$.
    In particular, in the case that  $\rho \succ 0$ we have   
        $$\rho'\in \BRa(\sigma) \ \text{ for all } \rho'
        \in D(\A),$$
        and symmetrically  for the $\sigma $ player. 

\end{theorem}

\begin{proof}
    By Theorem~\ref{thm:_KKT_NE_equiv} there exist $\lambda, \mu \in \mathbb{R}$ and Hermitian matrices $\Lambda \in {\mathrm L}(\A)$, $M \in {\mathrm L}(\B)$ such that $(\rho, \sigma, \lambda, \mu, \Lambda, M)$  satisfy the KKT conditions (\ref{eqn:_KKT_quantum_NablaL_a})-(\ref{eqn:_KKT_quantum_CompSlack_b}).
     Since  $\rho \succeq 0$ and $\Lambda\preceq 0$, the complementary slackness condition \eqref{eqn:_KKT_quantum_CompSlack_a}  (i.e.,  $\innerprod{\Lambda}{\rho} = 0$)  implies that $\range(\Lambda) \subseteq \ker(\rho)$.
    On the other hand, we have that $\Phi(\sigma) = \lambda \ida - \Lambda$ from the KKT condition (\ref{eqn:_KKT_quantum_NablaL_a}), so for any $\rho' \in D(\A)$ satisfying $\range(\rho') \subseteq \range(\rho)\subseteq\ker(\Lambda)$ we have that $\innerprod{\rho'}{\Phi(\sigma)} = \lambda - \innerprod{\rho'}{\Lambda} = \lambda$, i.e., all such $\rho'$ do as well as $\rho$ against $\sigma$. Finally since $\rho$ is a best response to $\sigma$, all such $\rho'$ are best responses to $\sigma$.

    Finally, consider  the special case where $\rho \succ 0$ (the case $\sigma \succ 0$ is similar).
%Similarly,  if $\sigma \succ 0$ then $\Phi^\dagger(\rho) =~\innerprod{\rho}{\Phi(\sigma)}\id_\B$.
Since $\innerprod{\Lambda}{\rho} = 0 $  it follows that  $\Lambda = 0$. Thus, the KKT conditions imply that $   \Phi (\sigma) = \innerprod{\rho}{\Phi(\sigma)}\id_\A$ and  consequenbtly $\innerprod{\rho'}{\Phi(\sigma)} = \lambda$ for  all $ \rho' \in \range(\rho) = D(\A)$. Since $\rho$ is one such possible $\rho'$, we have that $\lambda = \innerprod{\rho}{\Phi(\sigma)}$ and hence $\Phi(\sigma) = \innerprod{\rho}{\Phi(\sigma)} \id_\A$. The arguments for the other player are completely symmetric.
\end{proof}

\section{Best Response Dynamics}\label{sec:brdynamics}

% \paragraph{Alternating best response dynamics.}

In classical potential games, the 
% naive 
alternating best response (BR) dynamics are known to converge to pure equilibria~(see e.g. \cite{roughgarden2010algorithmic,tardos2007network}). 
Generally, better response dynamics (having players alternately choose pure strategies that improves their utility) converge to pure Nash equilibria because there are a finite number of pure strategy profiles in a classical (finite) potential game, and at each timestep the potential increases, so this process must necessarily terminate. This means that, in finite time, the players reach a fixed point at which every player is already playing the best response to the others and does not need to make an update under the dynamics, which is exactly a Nash equilibrium. However, this termination condition does not occur in finite time in our setting, as there are an infinite number of pure actions (albeit in a finite-dimensional space).

Nevertheless, it turns out that in our setting alternating BR dynamics can also be defined, and indeed shown to converge to the set of Nash equilibria.
% to converge to pure equilibria. 
Specifically, in the best response dynamics, players compute and play the strategy which maximizes their utility, given what the other player has played prior, i.e.:
\begin{equation}
\tag{Quantum  BR} \label{BR}
\begin{split}
\new{\rho} &\in \argmax_{\rho' \in \da}\langle \rho', \Phi(\sigma)\rangle, \\
\new{\sigma} &\in \argmax_{\sigma' \in \db}\langle \new{\rho}, \Phi(\sigma')\rangle.
% \rho_{t+1} &\in \argmax_{\rho'}\langle \rho', \Phi(\sigma_t)\rangle, \\
% \sigma_{t+1} &\in \argmax_{\sigma'}\langle \rho_{t+1}, \Phi(\sigma')\rangle.
\end{split}
\end{equation} 
 
%  To establish convergence to the set of Nash equilibria,
% % We experimentally test if a similar property holds for \ref{lin-QREP} 
% we utilize the concept of \emph{exploitability} \cite{johanson2011accelerating}, defined as 
% \begin{equation}\label{eqn:exploitability}
%     \frac{1}{2}\left[\lambda_{\max} (\Phi(\sigma)) - \langle \rho, \Phi(\sigma)\rangle + \lambda_{\max} (\Phi^\dagger(\rho)) - \langle\Phi^\dagger(\rho), \sigma\rangle\right],
% \end{equation}

% where $\lambda_{\max} (\Phi(\sigma))$ and $\lambda_{\max} (\Phi^\dagger(\rho))$ are the maximum eigenvalues of $\Phi(\sigma)$ and $\Phi^\dagger(\rho)$ respectively. Using the variational characterization of eigenvalues,
% $$\lambda_{\max}(\Phi(\sigma))=\max\{\la \rho', \Phi(\sigma)\ra:\ \rho'\in \da\},$$
% the difference $\lambda_{\max} (\Phi(\sigma)) - \langle \rho, \Phi(\sigma)\rangle $ is exactly the maximum gain the $\rho$-player can attain by unilaterally deviating from $(\rho,\sigma)$.
% Thus, if a profile $(\rho, \sigma)$ is $\epsilon-$exploitable, then it is an $2\epsilon-$Nash equilibrium, in the sense that no player can unilaterally improve their payoff by $\geq 2\epsilon$. 
We have the following result for the alternating \ref{BR} dynamics in two-player quantum CIGs:

\medskip
\begin{theo}\label{thm:alternatingbrdynamics}
   Alternating \ref{BR} converges to the set of Nash equilibria in two-player quantum CIGs. Furthermore, for any $\epsilon > 0$, an $\epsilon$-approximate Nash equilibrium can be found in $O(\frac{1}{\epsilon})$ iterations.
\end{theo}

\begin{proof}
At timestep $t$, say after Bob's turn, Bob has just played his best response so the exploitability is only due to Alice:
\[
    \text{exploitability}_t
    =
    \frac{1}{2} \left(\max_{\rho'} \innerprod{\rho'}{\Phi(\sigma_t)} -
    \innerprod{\rho_t}{\Phi(\sigma_t)}\right).
\]
But at timestep $t+1$, Alice has just played her best response
$
    \rho_{t+1} \in \argmax_{\rho'} \innerprod{\rho'}{\Phi(\sigma_t)},
$
so
\begin{equation}
\label{u-exploit}
    u_{t+1} - u_t
    =
    \innerprod{\rho_{t+1}}{\Phi(\sigma_t)}
    -
    \innerprod{\rho_t}{\Phi(\sigma_t)}
    = \max_{\rho'} \innerprod{\rho'}{\Phi(\sigma_t)} -
    \innerprod{\rho_t}{\Phi(\sigma_t)}
    = 2 \times \text{exploitability}_t,
\end{equation}
i.e., from timestep $t$ to timestep $t+1$ the utility increases by twice the exploitability at time $t$.

Define $u_{\max} := \max_{\rho, \sigma} u(\rho, \sigma) < \infty$. The sequence $(u_t)_t$ is non-decreasing and upper bounded by $u_{\max},$ so the sequence converges, i.e., 
$u_t \to u_\infty$ for some $u_\infty \leq u_{\max} < \infty.$ Thus,
\[
    2 \times \text{exploitability}_t
    = u_{t+1} - u_t
    \leq
    u_\infty - u_t
    \to 0
    \qquad \text{as } t \to +\infty.
\]

Furthermore, to get to an $\epsilon$-approximate Nash equilibrium for a given $\epsilon > 0$, due to \eqref{u-exploit} we can terminate at the first timestep $T$ at which the utility improves by $< \epsilon$, i.e., $u_{T+1} - u_T < \epsilon$. By the pigeonhole principle, we will reach such a $T$ in at most $\frac{u_{\max} - u_0}{\epsilon}$ timesteps.
\end{proof}
% \begin{theorem}
%     In a quantum CIG, alternating best response dynamics converge to $\epsilon$-NE in $O(1/\epsilon)$ iterations. 
% \end{theorem}
% \begin{proof}
%     [TO ADD HERE]
% \end{proof}

We note that this proof of linear-time convergence, unlike the remaining proofs in this paper, is specific to the two-player setting that we focus on and not easily extendable to a higher number of players.
The fact that the alternating \ref{BR} dynamics converge in our setting (Theorem \ref{thm:alternatingbrdynamics}) recovers the well-known result that alternating \ref{BRc} dynamics converges to pure equilibria in classical potential games \cite{roughgarden2010algorithmic,tardos2007network}. Specifically, players using the BR algorithm in finite potential games are guaranteed to terminate at a pure strategy Nash equilibrium. Despite this, finding a pure Nash equilibrium in classical $N$-player finite CIGs (more broadly, $N$-player potential games) is known to be PLS-complete (see e.g \cite{tardos2007network,fabrikant2004complexity}), even when the best response can be computed in polynomial time. Comparatively, we are able to guarantee linear-time convergence to an $\epsilon$-approximate equilibrium in the two-player quantum CIG setting. Crucially, this setting does not possess a finite number of actions (in a finite-dimensional space). We leave the derivation of a similar PLS-complexity result in the vein of \cite{fabrikant2004complexity} for $N$-player quantum CIGs to future work.

% while Theorem \ref{thm:alternatingbrdynamics} is able to provide an algorithmic guarantee of reaching pure equilibria in our two-player setting, we 

% \subsection{Experiments}

\begin{figure}[!htb]
    \centering
    \begin{minipage}{.45\linewidth}
      \centering
      \includegraphics[width=.95\linewidth]{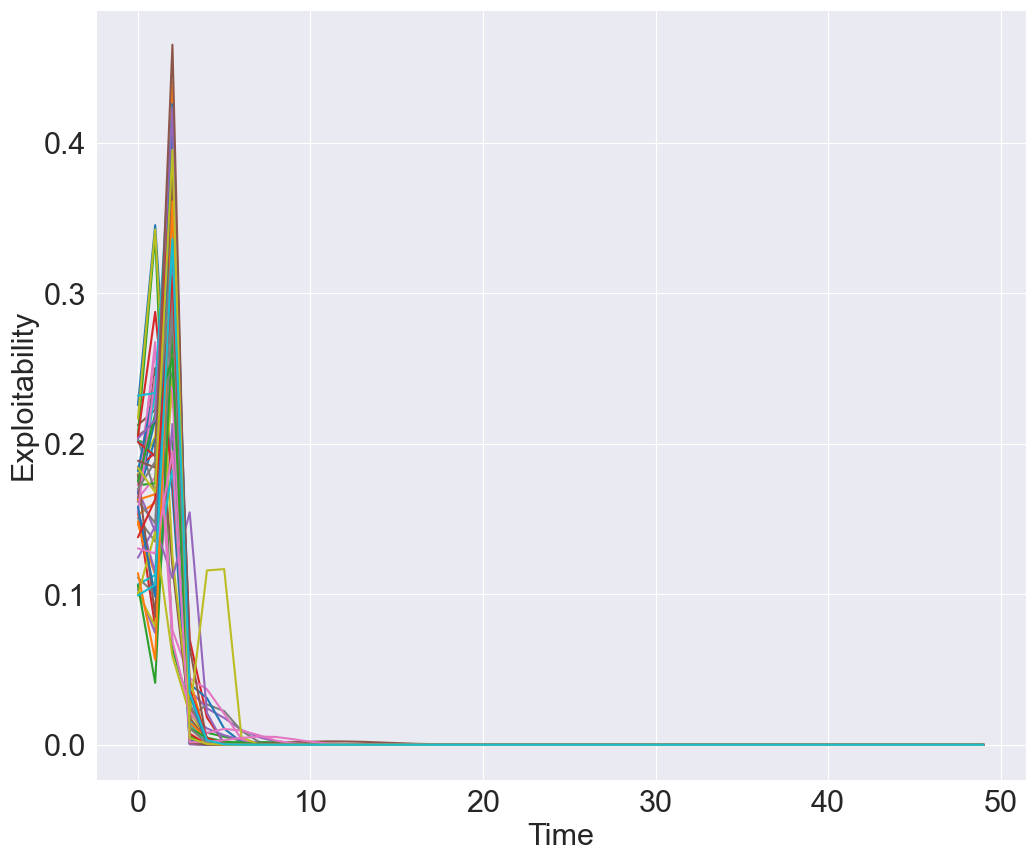}
    %   \label{fig:sub2}
    \end{minipage}
    \caption{Exploitability of trajectories under \ref{BR}. All exploitabilities go to zero quickly.}
    \label{fig:exploit_BR}
\end{figure}

\paragraph{Exploitability experiments.} 
% \paragraph{Exploitability experiments.}
In Figure \ref{fig:exploit_BR}  we plot the exploitability (as defined in Equation \ref{eqn:exploitability}) under the alternating \ref{BR} dynamics in a number of randomly-generated two-player quantum CIGs, and see that exploitability goes quickly to 0 (meaning that the limit points are Nash equilibria/KKT points), corroborating the convergence result of Theorem \ref{thm:alternatingbrdynamics}.
% }

\section{Linear Quantum Replicator Dynamics and Linear Matrix Multiplicative Weights Update}\label{sec:contdynamics}
In this section, we introduce and study dynamics which are closely related to the classical replicator and multiplicative weights update.
Consider a  quantum common-interest game  with  utility 
% $
% u(\rho, \sigma)= {\rm Tr}(R(\rho\otimes \sigma)),
% $
$
u(\rho, \sigma)= \innerprod{\rho}{\Phi(\sigma)},
$
where the Choi matrix $R$ corresponding to the superoperator $\Phi$ is  positive definite. We define the {\em Linear Quantum Replicator Dynamics}  as: 
\begin{equation} \label{lin-QREP}\tag{{\rm lin-QREP}}
        \dv{\rho}{t}
        = \rho^{\sfrac{1}{2}} \Big[\Phi(\sigma) - \innerprod{\rho}{\Phi(\sigma)} \id_\A \Big] \rho^{\sfrac{1}{2}}, \quad \text{ and } \quad 
        \dv{\sigma}{t}
        = \sigma^{\sfrac{1}{2}} \Big[\Phi^\dagger (\rho) -\innerprod{\rho}{\Phi(\sigma)} \id_\B \Big] \sigma^{\sfrac{1}{2}}.
\end{equation}
We derive the \ref{lin-QREP} dynamics---and indeed, the larger family of \emph{Linear Quantum $q$-Replicator Dynamics}---as a gradient flow of the common utility with respect to the quantum-Shahshahani metric. We defer this derivation to Appendix \ref{appsecs:gradient_flow}. We show that the set of density matrices is forward-invariant under \ref{lin-QREP} in Appendix \ref{sec:_QREP_props}.

% \textcolor{red}{Do we swap this around? i.e. define lin-MMWU first then matrix BE as a specialization as $\epsilon\rightarrow\infty$}

We also introduce a discretization of \ref{lin-QREP} called the {\em Linear Matrix Multiplicative Weights Update}, which includes in its definition a fixed step-size $\step \in (0, +\infty]$:
\begin{equation} \label{lin-MMWU} \tag{\rm{lin-MMWU}}
\new{\rho}  \leftarrow \frac{\powh{\rho}[\id_\A + \step \Phi(\sigma)] \powh{\rho}}{1+ \step \innerprod{\rho}{\Phi(\sigma)}},\quad   \text{ and } \quad \new{\sigma} \leftarrow \frac{\powh{\sigma} [\id_\B + \step \Phi^\dagger (\new{\rho})] \powh{\sigma}}{1+ \step \innerprod{\new{\rho}}{\Phi(\sigma)}}.
\end{equation}

% Finally, to  get a more adaptable  version of \ref{eqn:_DQREP}, we introduce   a fixed step-size  $\step \in (0, +\infty]$, and get the update 

% which we call the  {\em the Linear Matrix Multiplicative Weights Update}. 
Note that the updates described in \ref{lin-MMWU} are performed   in an alternating manner.
% , meaning $\rho$ and $\sigma$ are updated in turn.  First, $\rho$ is updated to $\new{\rho}$, and subsequently, $\sigma$ is updated using $\new{\rho}$. 
Moreover, both \ref{lin-QREP} and \ref{lin-MMWU} utilize only first-order information; to perform the update, each agent only needs to know the gradient of the utility with respect to their own state. 

We define \ref{lin-MMWU} with  infinite step-size  
to be the limit of the update as
% taking the limit of \ref{lin-MMWU} as the stepsize 
$\step \to +\infty$, i.e.,
\begin{equation}
\label{eqn:_DQREP} 
 \tag{\rm{Matrix BE}}
        \new{\rho}  \leftarrow 
        \frac{\powh{\rho} \Phi(\sigma) \powh{\rho}}{\innerprod{\rho}{\Phi(\sigma)}} \quad \text{ and }  \quad 
        \new{\sigma} \leftarrow
        \frac{\powh{\sigma} \adj{\Phi}(\new{\rho}) \powh{\sigma}}{\innerprod{\new{\rho}}{\Phi(\sigma)}},
\end{equation}
which we call the {\em Matrix Baum-Eagon} update. 
% \ref{eqn:_DQREP} is a non-commutative extension of the Baum Eagon update in the sense that  \ref{eqn:_DQREP} reduces to the Baum Eagon update (which is itself a special case of \ref{linear}) when the game operator is diagonal and $\rho, \sigma$ are diagonal~densities.  

% \ref{eqn:_DQREP} can be clearly viewed as a discretization of \ref{lin-QREP}, with \ref{lin-MMWU} serving as a more adaptable version that allows for a fixed step-size. 
The use of a fixed step size in \ref{lin-MMWU} enables finer control over the convergence process, allowing for adjustments to the rate at which updates are made. In the interest of notational brevity, we shall discuss and state results for \ref{lin-MMWU} explicitly, though these results hold for \ref{eqn:_DQREP} as well. We first make the remark that these updates are all non-commutative generalizations of their classical counterparts which were discussed in Section \ref{sec:classicaldynamics}.

% The use of a fixed step size in \ref{lin-MMWU} enables finer control over the convergence process, allowing for adjustments to the rate at which updates are made. Note that \ref{lin-MMWU} with infinite step-size should be understood as the limit of the update as $\step \to +\infty$, which is itself a non-commuatiexactly \ref{eqn:_DQREP}. 
% which we call the {\em Matrix Baum-Eagon} update:
%  \begin{equation}
% \label{eqn:_DQREP} 
%  \tag{\rm{Matrix BE}}
%         \new{\rho}  \leftarrow 
%         \frac{\powh{\rho} \Phi(\sigma) \powh{\rho}}{\innerprod{\rho}{\Phi(\sigma)}}, \quad \text{ and }  \quad 
%         \new{\sigma} \leftarrow
%         \frac{\powh{\sigma} \adj{\Phi}(\new{\rho}) \powh{\sigma}}{\innerprod{\new{\rho}}{\Phi(\sigma)}}.
% \end{equation}

\begin{remark}[Non-commutative extensions]
The \ref{lin-QREP} dynamics are a non-commutative generalization of the celebrated replicator dynamics \cite{weibull1997evolutionary,sandholm2010population}: specifically, the  \ref{lin-QREP} dynamics reduce to the usual replicator dynamics when applied to the quantum embedding of a classical game. Indeed, consider the following quantum embedding of a classical two-player CIG with common utility  $x^\top Ay$: let $\{e_i\}$ and $\{f_j\}$ be orthonormal bases for $\A$ and $\B$ respectively, define the diagonal game operator $R_{ij,ij}=A_{ij}$, and consider  diagonal density matrices $\rho=\sum_\ell x_\ell e_\ell e_\ell^\dagger$ and $\sigma=\sum_k y_k f_k f_k^\dagger $. Then, we have that
% \noindent
\begin{equation*}
%\label{diag}
\begin{split}
	\Phi(\sigma)=&
	\Tr_{\B}[R (\ida \otimes \sigma^\top)]\\
    =&\Tr_{\B}\left[
	\left( \sum_{ij} R_{ij,ij} (e_i \otimes f_j) (e_i \otimes f_j)^\dagger \right)\cdot\left(\ida \otimes \sum_k y_k f_k f_k^\dagger \right)\right]\\
 =&\sum_{ij}R_{ij,ij}y_je_ie_i^\dagger\\
=&\sum_{ij}A_{ij}y_ie_je_j^\dagger=\diag{Ay},
\end{split}
\end{equation*}
	and similarly $\Phi^\dagger (\rho) = \diag{A^\top x}$.
Consequently, if $\rho=\sum_\ell x_\ell e_ke_k^\dagger, \sigma=\sum_k y_k f_k f_k^\dagger $ are diagonal densities, we have that $ \dv{\rho}{t}$  and  $\dv{\sigma}{t}$ are diagonal. Specifically,
% densities, the \ref{lin-QREP} dynamics
% all the operators in the  dynamics
% are diagonal, and so $\rho$ and $\sigma$ remain as diagonal operators over the course of time, which is important so that we can extract the classical simplex vectors $x$ and $y$ from the density matrices $\rho$ and $\sigma$ via the relations $x_i = \rho_{ii}$, $y_j = \sigma_{jj}$ respectively.
 the time-evolution of $x$ is
\begin{equation}
\label{diag}
\begin{split}
	\dv{x_i}{t}
	=
	\dv{\rho_{ii}}{t}
        =&
        \rho^{\sfrac{1}{2}}_{ii} \Big[\Phi(\sigma)_{ii} - \innerprod{\rho}{\Phi(\sigma)} \Big] \rho^{\sfrac{1}{2}}_{ii}\\
        =&
        x_i \left[(Ay)_i - x^\top A y \right]
\end{split}
\end{equation}
and similarly
$
	\dv{y_j}{t}=
        y_j \left[(A^\top x)_j - x^\top A y \right],$
which correspond to replicator dynamics.
Similarly, \ref{lin-MMWU} can be understood as a non-commutative extension of \ref{linear} in the sense that \ref{lin-MMWU} reduces to \ref{linear} when the game operator is diagonal and $\rho, \sigma$ are diagonal densities. Indeed, define the diagonal game operator $R_{ij,ij}=A_{ij}$, and consider  diagonal density matrices $\rho=\sum_\ell x_\ell e_\ell e_\ell^\dagger$ and $\sigma=\sum_k y_k f_k f_k^\dagger $. By  \eqref{diag} we have  $\Phi(\sigma)={\rm diag}(Ay).$ 
Thus

$$ \new{\rho} 
        =\frac{\powh{\rho} [\id_\A + \step\Phi(\sigma)]\powh{\rho}}{1 + \step \innerprod{\rho}{\Phi(\sigma)}}
        = \frac{
            {\rm diag}(x \circ (1 + \step(Ay))
            }{
            1+ \step (x^\top A y)
            }
      $$
where $\circ$ denotes the componentwise  vector~product.

\end{remark}

We shall now provide some general properties of these dynamics in Section \ref{sec:_genprops}, prove our main theoretical results regarding their performance as learning dynamics in quantum CIGs in Section \ref{sec:_convprops}, and discuss some empirical simulations showcasing this performance in Section \ref{sec:_dynamics_empirical}.

%\ref{eqn:_DQREP} can be clearly viewed as a discretization of \ref{lin-QREP}, with \ref{lin-MMWU} serving as a more adaptable version that allows for a fixed step-size. T

% Note that the updates described in \ref{lin-MMWU} are performed   in an alternating manner, meaning that $\rho$ and $\sigma$ are updated in turn.  First, $\rho$ is updated to $\new{\rho}$, and subsequently, $\sigma$ is updated using $\new{\rho}$. Moreover, both \ref{lin-QREP} and \ref{lin-MMWU} utilize only first-order information: to perform the update, each agent only needs to know the gradient of the utility with respect to their own density matrix. 

\subsection{General properties of \ref{lin-QREP} and \ref{lin-MMWU}}
\label{sec:_genprops}
In this section we prove some general properties of \ref{lin-QREP} and \ref{lin-MMWU} as learning dynamics in an arbitrary quantum game. Namely, we prove that the minimal face of each player's strategy is invariant along trajectories (Theorem \ref{thm:minimalfaceinvariance}), that \ref{lin-QREP} and \ref{lin-MMWU} share the same set of fixed points (Theorem \ref{thm:_FixedPointsSame_ContTime_DiscreteTime}) and that these fixed points can be characterized as strategies for which all strategies in their support perform equally well (Theorem \ref{char:_fixedpts}). Finally, we show a partial analogue of the classical evolutionary property that the correlation with strategies that perform better than average increases under the update (Theorem \ref{thm:_q_compare}).

The first property we show is that the faces (see \eqref{eq:facerr}) of the set of density matrices are forward-invariant under either the continuous-time \ref{lin-QREP} dynamics or the discrete-time \ref{lin-MMWU}. This is an analogue of the classical property that the support of the strategy is invariant under \ref{replicator} and \ref{linear}, and has consequences on the geometry of the dynamics' fixed points in quantum games.

\begin{theorem}
\label{thm:minimalfaceinvariance}
    In a quantum game, under the continuous-time dynamic \ref{lin-QREP} and its discretization  \ref{lin-MMWU} (for any fixed step-size $\step \in (0, +\infty]$), the minimal face of each player's strategy is invariant along trajectories of the dynamics. As a result, all rank-one  density matrices  are fixed~points of the dynamics.
\end{theorem}
\begin{proof}
Consider the \ref{lin-MMWU} dynamics in a quantum game. 
% The definition of the minimal face of the set of density matrices that contains a given matrix $\rho$ is 
% $$\fc_{\da}(\rho)=\{Y\succeq 0: \range(Y)\subseteq \range(\rho), \ \tr(Y)=1\},$$
% % $${\rm face}(\rho, D)=\{Y\succeq 0: \range(Y)\subseteq \range(\rho), \ \tr(Y)=1\},$$
% and its relative interior is given by
% $${\rm relint} \, \fc_{\da}(\rho)=\{Y\succeq 0: \range(Y)= \range(\rho), \ \tr(Y)=1\}.$$
% % $${\rm relint} \, {\rm face}(\rho, D)=\{Y\succeq 0: \range(Y)= \range(\rho), \ \tr(Y)=1\}.$$
As $\rho$ is positive semidefinite we have that $\ker(\rho)=\ker(\rho^{1/2})$. Moreover, as $\Phi$ maps positive  semidefinite  matrices to positive semidefinite   matrices  we have that $\id+\step \Phi(\sigma)$ is positive definite for all $\sigma$ and $\step>0$. Thus,  it follows that 
$\range(\new{\rho}) = \range(\powh{\rho}( \id+\step \Phi(\sigma)) \powh{\rho}) = \range(\rho),$ which by \eqref{eq:facerr} implies that the minimal face of the strategy $\rho$ is invariant under the update. This holds for all players, and the proof is concluded.  

A similar argument holds in the case of \ref{lin-QREP}.
\end{proof}

We next demonstrate that \ref{lin-MMWU} is a faithful discretization of \ref{lin-QREP} in the sense that they share the same set of fixed points:

%between the fixed points of  %\ref{eqn:_DQREP}
%\ref{lin-MMWU} and~\ref{lin-QREP}.

%Finally, the fixed points of the discrete-time \ref{eqn:_DQREP} update and the continuous-time \ref{lin-QREP} dynamics coincide:

\begin{theorem}
\label{thm:_FixedPointsSame_ContTime_DiscreteTime}
    The discrete-time updates %\ref{eqn:_DQREP} and
    \ref{lin-MMWU} for any fixed step-size  $\step \in (0, +\infty] $ and the continuous-time gradient flow \ref{lin-QREP}  have the same set of fixed points.
\end{theorem}
% The proof is deferred to Appendix~\ref{proof:fixedpoint}.

\begin{proof}
    A density matrix  $\rho$ that is  stationary under the \ref{lin-QREP} flow satisfies:
    \[
        \powh{\rho} [ \Phi(\sigma) - \innerprod{\rho}{\Phi(\sigma)} \id_\A] \powh{\rho} = 0
        \Longleftrightarrow
        \powh{\rho} \Phi(\sigma) \powh{\rho} = \innerprod{\rho}{\Phi(\sigma)} \rho.  
    \]
  %  On the other hand, we have that $\rho$ is unchanged under a \ref{eqn:_DQREP} update, i.e.,
    % \[
    %     \rho = \frac{\powh{\rho} \Phi(\sigma) \powh{\rho}}{\innerprod{\rho}{\Phi(\sigma)}}
    %     \Longleftrightarrow
    %     \powh{\rho} \Phi(\sigma) \powh{\rho} = \innerprod{\rho}{\Phi(\sigma)} \rho,
    % \]
On the other hand, a density  matrix $\rho$   that is stationary  under a \ref{lin-MMWU} update satisfies:
    \[
        \rho = \frac{\powh{\rho}[\id_\A + \step \Phi(\sigma)] \powh{\rho}}{1 + \step \innerprod{\rho}{\Phi(\sigma)}}
        \Longleftrightarrow
        (1 + \step \innerprod{\rho}{\Phi(\sigma)})\rho =  \powh{\rho}[\id_\A + \step \Phi(\sigma)] \powh{\rho}
        \Longleftrightarrow
        \innerprod{\rho}{\Phi(\sigma)}\rho = \powh{\rho} \Phi(\sigma) \powh{\rho}.
    \]
Thus, a strategy is stationary under \ref{lin-QREP} if and only if it is stationary under a \ref{lin-MMWU} update, and hence the two dynamics have the same fixed points.
\end{proof}

Having shown that the fixed points of \ref{lin-QREP} and \ref{lin-MMWU} coincide, the following theorem then characterizes these fixed points, stating the analogue of the classical result for \ref{replicator} and \ref{linear} that a strategy is a fixed point if and only if all the pure actions in its support perform equally well:

% We have the following theorem characterizing fixed points. It is analogous to the classical result for rep and lin-MWU that a strategy is fixed under the update iff all pure strategies in the support do equally well.

% We can now prove the following theorem characterizing fixed points of lin-QREP:

\medskip
\begin{theorem}
\label{char:_fixedpts}
	Under \ref{lin-QREP} or \ref{lin-MMWU}, a strategy is a fixed point if and only if all strategies whose support is included in its support (i.e., all density matrices that lie in the same face of the set of density matrices as it) do equally well. Concretely,  we have for \ref{lin-QREP} that
    \begin{equation*}
        \dot{\rho} = 0 \Longleftrightarrow \innerprod{\rho'}{\Phi(\sigma)} = \innerprod{\rho}{\Phi(\sigma)} \; \forall \ \rho' \in \fc_{\da}(\rho)
    \end{equation*}
    and for \ref{lin-MMWU} that
    \begin{equation*}
        \new{\rho} = \rho \Longleftrightarrow \innerprod{\rho'}{\Phi(\sigma)} = \innerprod{\rho}{\Phi(\sigma)} \; \forall \ \rho' \in \fc_{\da}(\rho).
    \end{equation*}
 %    \begin{equation*}
 %    \begin{split}
 %        \dot{\rho} = 0 &\Longleftrightarrow \innerprod{\rho'}{\Phi(\sigma)} = \innerprod{\rho}{\Phi(\sigma)} \; \forall \ \rho' \in \fc_{\da}(\rho), \text{ and} \\
	% \dot{\sigma} = 0 &\Longleftrightarrow \innerprod{\rho}{\Phi(\sigma')} = \innerprod{\rho}{\Phi(\sigma)} \; \forall \ \sigma' \in \fc_{\db}(\sigma).
 %    \end{split}
 %    \end{equation*}
\end{theorem}

\begin{proof}
	As we have shown in Theorem \ref{thm:_FixedPointsSame_ContTime_DiscreteTime} that the fixed points of the two dynamics coincide, we need only prove this result for \ref{lin-QREP}.
 Let $\rho = \sum_{i=1}^n \lambda_i u_i u_i^\dagger$ where $\lambda_i > 0 \; \forall \ i \in [r]$ and $\lambda_i = 0 \; \forall \ i > r$, and $\Lambda := \Diag(\lambda_1, \ldots, \lambda_n)$. We first have the following characterization of a strategy $\rho$ being a fixed point (i.e., $\dot{\rho} = 0$), where $\vrr{A}$ denotes the $[r] \times [r]$ submatrix of a matrix $A$: %	and $S:= \{ i: \lambda_i > 0\} = [r]$. 
\begin{equation}
\label{char:rhofixed}
	\begin{split}
		\dot{\rho} = 0
		&\Longleftrightarrow  \rpr = \rps \rho \\
		&\Longleftrightarrow \uu{\rh} \uu{\ps} \uu{\rh} = \rps \uu{\rho} \\
		&\Longleftrightarrow \powh{\Lambda} \uu{\ps} \powh{\Lambda} = \rps \Lambda \\
		&\Longleftrightarrow \powh{\vrr{\Lambda}} \vrr{\uu{\ps}} \powh{\vrr{\Lambda}} = \rps
		\vrr{\Lambda} \\
		&\Longleftrightarrow \vrr{\uu{\ps}} = \rps \id_r.
	\end{split}
\end{equation}
Note that the second-to-last equivalence is due to the fact that $\powh{\Lambda}$ is 0 in all entries outside of the $[r] \times [r]$ submatrix, so only the $[r] \times [r]$ submatrix of $\uu{\ps}$ affects the matrix product $\powh{\Lambda} \uu{\ps} \powh{\Lambda}$.

We are now ready to prove the theorem. 
% We shall only show the proof for $\dot{\rho} = 0 \Longleftrightarrow \innerprod{\rho'}{\Phi(\sigma)} = \innerprod{\rho}{\Phi(\sigma)} \; \forall \ \rho' \in \fc(\rho)$, since the proof for $\sigma$ is the same.
	
	($\Rightarrow$) $\forall \ \rho'\in \fc(\rho)$ we have that
	\[
		\innerprod{\rho'}{\ps} 
		= \innerprod{\uu{\rho'}}{\uu{\ps}} 
		= \innerprod{\vrr{\uu{\rho'}}}{\vrr{\uu{\ps}}}
		= \rps,
	\]
	where the second equality is due to \eqref{eq:facerr} and the last equality is due to \eqref{char:rhofixed}.
		
	($\Leftarrow$) That $\innerprod{\rho'}{\Phi(\sigma)} = \innerprod{\rho}{\Phi(\sigma)} \; \forall \ \rho': \range(\rho') \leq \range(\rho)$ is equivalent to the statement
	\[
		\innerprod{U^\dagger \rho' U}{U^\dagger \Phi(\sigma) U} = \innerprod{\rho}{\Phi(\sigma)} \; \forall \ \rho': \range(\rho') \leq \range(\rho).
	\]
	
	Letting $\rho' = u_j u_j^\dagger$ for some $j \in [r]$, we have that
	\begin{equation}
	\label{eq:_Ujj}
		\rps = \innerprod{E_{jj}}{U^\dagger \ps U} = [U^\dagger \ps U]_{jj} \quad \forall \ j \in [r].
	\end{equation}
	
	Then, letting $\rho' = \frac{1}{2} (u_j + u_k) (u_j + u_k)^\dagger$ for some $j \neq k \in [r]$, we have that
	\begin{equation*}
	\begin{split}
		\rps
		&= \innerprod{\uu{\rho}}{\uu{\Phi}} \\
		&= \innerprod{\frac{1}{2}(e_j + e_k)(e_j + e_k)^\dagger}{U^\dagger \ps U} \\
		&= \frac{1}{2}
		 \left(
		 	\supu_{jj} + \supu_{jk} + \supu_{kj} +  \supu_{kk}
		 \right)
		 \quad \forall \ j \neq k \in [r].
	\end{split}
	\end{equation*}
	Since \eqref{eq:_Ujj} implies that $\rps = \h \left( \supu_{jj} + \supu_{kk} \right)$, we then have that
	\[
		\Re(\supu_{jk}) = \frac{1}{2} \left(\supu_{jk} + \supu_{kj}\right) = 0 \quad \forall \ j \neq k \in [r].
	\]
	On the other hand, letting $\rho' 
	= \h(u_j + i u_k) (u_j + i u_k)^\dagger 
	= \h \left(u_j u_j^\dagger - i u_j u_k^\dagger + i u_k u_j^\dagger + u_k u_k^\dagger \right)$ 
	gives
	\begin{equation*}
	\begin{split}
		\rps
		&= \innerprod{\uu{\rho}}{\uu{\Phi}} \\
		&= \innerprod{\frac{1}{2}\left(e_j e_j^\dagger - i e_j e_k^\dagger + i e_k e_j^\dagger + e_k e_k^\dagger \right)}{U^\dagger \ps U} \\
		&= \frac{1}{2}
		 \left(
		 	\supu_{jj} - i\supu_{jk} + i\supu_{kj} +  \supu_{kk}
		 \right)
		 \quad \forall \ j \neq k \in [r].
	\end{split}
	\end{equation*}
	Again, since \eqref{eq:_Ujj} implies that $\rps = \h \left( \supu_{jj} + \supu_{kk} \right)$, we have that
	\[
		\Im(\supu_{jk})
		=
		- \frac{i}{2} \left( \supu_{jk} - \supu_{kj} \right)
		=
		0
		\quad \forall \ j \neq k \in [r],
	\]
	and putting this together with the previous result that the real part is also 0 gives
	\[
		\supu_{jk}
		=
		0
		\quad \forall \ j \neq k \in [r].
	\]
	Combining this with \eqref{eq:_Ujj} gives us that $\vrr{\upu} = \rps \id$, which by \eqref{char:rhofixed} implies that $\dot{\rho} = 0$.
\end{proof}

% Now, let us prove some additional properties of the dynamics in quantum CIGs.

% \begin{theorem}\label{thm:fixedpoints}
% TODO: add lin-QREP.
%    The iterates of \ref{lin-MMWU} always remain within the relative interior of the minimal face of the set of density matrices  corresponding to the initial point. As a consequence, all rank-one  density  matrices  are fixed points of \ref{lin-MMWU}.
% \end{theorem}

% \begin{proof}
% The minimal face of the set of density matrices that contains a given matrix $\rho$ is 
% $${\rm face}(\rho, D)=\{Y\succeq 0: \range(Y)\subseteq \range(\rho), \ \tr(Y)=1\},$$
% and its relative interior is given by
% $${\rm relint} \, {\rm face}(\rho, D)=\{Y\succeq 0: \range(Y)= \range(\rho), \ \tr(Y)=1\}.$$

% As $\rho$ is positive semidefinite we have that $\ker(\rho)=\ker(\rho^{1/2})$. Moreover, as $\Phi$ maps positive  semidefinite  matrices to positive semidefinite   matrices  we have that $\id+\step \Phi(\sigma)$ is positive definite for all $\sigma$ and $\step>0$. Thus,  it follows that 
% $ \range(\rho)=\range(\new{\rho}),$
% and the proof is concluded.  
% \end{proof}

Finally, we recall a fundamental property of the classical \ref{replicator} and \ref{linear}: for any  strategy in the support, the probability of  being played  increases if and only if it performs better than average.  This principle is rooted in evolutionary game theory, where strategies outperforming the average are more likely to proliferate, mirroring natural selection.
More formally, and in discrete time for concreteness, if player \(i\) plays strategy \(k\) with positive probability \(p^{(t)}_{ik} > 0\) at time step \(t\), then we have that
\begin{equation} \label{classicalcondition}
u_i(k, p^{(t)}_{-i}) > u_i(p^{(t)}) \iff  p^{(t+1)}_{ik} > p^{(t)}_{ik}.
\end{equation}
The following theorem is a partial quantum analogue of this property for \ref{lin-QREP} and \ref{lin-MMWU} when looking at pure strategies in the state's eigenbasis:

\begin{theorem}
\label{thm:_q_compare}
% Under \ref{lin-MMWU}, the update $\new{\rho}$ is more correlated than $\rho$ with the positive eigendirections of $\rho$ that perform better than $\rho$ (and similarly for $\sigma$).  
Both \ref{lin-QREP} and \ref{lin-MMWU} increase the correlation of $\rho$ with positive eigendirections of $\rho$ that perform better than $\rho$ (and similarly for $\sigma$). 
Concretely, we have that:

    For all unit eigendirections \(u_i\) of the density matrix \(\rho\) with positive eigenvalue \(\lambda_i > 0\),
    \begin{equation}
    \label{correl:qrep}
        \la \Phi(\sigma), u_iu_i^\dagger\ra>\la \rho, \Phi(\sigma)\rangle \iff
        \innerprod{\dot{\rho}}{u_i u_i^\dagger} > 0 
    \end{equation}
    under \ref{lin-QREP} and
\begin{equation}\label{correl:linmmwu}
\la \Phi(\sigma), u_iu_i^\dagger\ra>\la \rho, \Phi(\sigma)\rangle \iff   \la \rho^{new}, u_iu_i^\dagger\ra> \la  \rho,  u_iu_i^\dagger\ra
\end{equation}
under \ref{lin-MMWU} with any stepsize $\step \in (0, +\infty]$.
    % If $u$ is a positive unit eigendirection of $\rho$ satisfying $\innerprod{u u^\dagger}{\Phi(\sigma)} > \innerprod{\rho}{\Phi(\sigma)}$, then $\innerprod{u u^\dagger}{\new{\rho}} > \innerprod{u u^\dagger}{\rho}$ under either \ref{lin-MMWU} with a fixed step-size $\step \in (0,+\infty].$
    \item \label{thm:_q_compare_gen}

\end{theorem}
\begin{proof}
We first note that $\lambda_i \innerprod{\ps}{u_i u_i^\dagger} = \innerprod{\ps}{\rh u_i u_i^\dagger \rh}$ and $\lambda_i \rps = \rps \innerprod{\rho}{u_i u_i^\dagger}$, so
\begin{equation}
\label{correl:char}
\begin{split}
    \innerprod{\ps}{u_i u_i^\dagger} > 
    \rps 
    &\iff
    \innerprod{\ps}{\rh u_i u_i^\dagger \rh} 
    > \rps \innerprod{\rho}{u_i u_i^\dagger} \\
    &\iff
    \innerprod{\rpr}{u_i u_i^\dagger} 
    > \rps \innerprod{\rho}{u_i u_i^\dagger}.
\end{split}
\end{equation}
For \ref{lin-QREP}, we recall that $\dot\rho = \powh{\rho} \left[\Phi(\sigma) - \innerprod{\rho}{\Phi(\sigma)} \id_\A \right] \powh{\rho}$ so
\[
    \innerprod{\dot{\rho}}{u_i u_i^\dagger}
    =
    \innerprod{\rpr}{u_i u_i^\dagger}
    - \rps \innerprod{\rho}{u_i u_i^\dagger},
\]
which together with \eqref{correl:char} gives \eqref{correl:qrep}. On the other hand, recall that \ref{lin-MMWU} with stepsize $\step > 0$ is given by
$$\rho^{new}={1\over 1+\step \la \rho, \Phi(\sigma)\ra} \rho^{1/2}(\id_\A +\step \Phi(\sigma))\rho^{1/2},$$
so\\
\begin{equation*}
\begin{split}
    \innerprod{\new{\rho}}{u_i u_i^\dagger}
    > \innerprod{\rho}{u_i u_i^\dagger}
    &\iff
    \innerprod{\rho^{1/2}(\id_\A +\step \Phi(\sigma))\rho^{1/2}}{u_i u_i^\dagger}
    > 
    (1+\step \la \rho, \Phi(\sigma)\ra)\innerprod{\rho}{u_i u_i^\dagger} \\
    &\iff
    \innerprod{\rho}{u_i u_i^\dagger}
    + \step \innerprod{\rpr}{u_i u_i^\dagger}
    >
    \innerprod{\rho}{u_i u_i^\dagger}
    +
    \step \rps \innerprod{\rho}{u_i u_i^\dagger} \\
    &\iff \innerprod{\rpr}{u_i u_i^\dagger}
    > \rps \innerprod{\rho}{u_i u_i^\dagger},
\end{split}
\end{equation*}
which together with \eqref{correl:char} gives \eqref{correl:linmmwu}.
% Consider an eigendirection $u_i$ of $\rho$ satisfying 
% $$\la \Phi(\sigma), u_iu_i^\dagger\ra>\la \rho, \Phi(\sigma)\rangle.$$ 
% Substituting   $\rho^{new}$  in $\la \rho^{new}, u_iu_i^\dagger\ra> \la  \rho,  u_iu_i^\dagger\ra$   this becomes   
% $$\la \rho^{1/2}(\id_\A+\step \Phi(\sigma))\rho^{1/2},  u_iu_i^\dagger\ra> \la  \rho,  u_iu_i^\dagger\ra(1+\step \la \rho, \Phi(\sigma)\ra).$$
% Using that $\rho^{1/2}u_i=\sqrt{\lambda_i}u_i$  this 
% is equivalent to 
% %$$\lambda_i+\step \lambda_i u_i^*\Phi(\sigma)u_i>\lambda_i(1+\step \la %\rho, \Phi(\sigma)\ra).$$
% %Equivalently, this is
% $$\lambda_i(1+\step  u_i^\dagger\Phi(\sigma)u_i)>\lambda_i(1+\step \la \rho, \Phi(\sigma)\ra).$$
% As $\lambda_i>0$, this inequality holds if  and only if  
% $\la \Phi(\sigma), u_iu_i^*\ra>\la \rho, \Phi(\sigma)\rangle.$
\end{proof}

   We note the following generalization to Theorem \ref{thm:_q_compare}:
   % (\ref{thm:_q_compare_eigen}): 
   suppose that $\rho$ has spectral decomposition $\rho = \sum_i \lambda_i u_i u_i^\dagger$, and let $S$ be the set of indices corresponding to eigendirections that perform better than average, i.e., 
   \[S = \left\{i : \lambda_i > 0, \, \innerprod{u_i u_i^\dagger}{\Phi(\sigma)} > \innerprod{\rho}{\Phi(\sigma)} \right\}.\] Then for any $W = \sum_{i \in S} \mu_i u_i u_i^\dagger$ with $\mu_i \geq 0 \, \forall \ i$ we have that $\innerprod{W}{\rho^{new}} > \innerprod{W}{\rho}$ under  \ref{lin-MMWU}. More generally though, if the direction of comparison is not an eigendirection of $\rho$, then it is possible for it to be in the support of $\rho$, perform better than $\rho$, and yet have the update $\new{\rho}$ under \ref{lin-MMWU} be less correlated with it than $\rho$ is. 
Concretely, there exist state $\rho$ and $\sigma$, game operator $\Phi$, and unit vector $v$ such that $\innerprod{vv^\dagger}{\Phi(\sigma)} > \innerprod{\rho}{\Phi(\sigma)}$ and $\innerprod{vv^\dagger}{\rho} > 0$ but $\innerprod{v v^\dagger}{\new{\rho}} < \innerprod{v v^\dagger}{\rho}$ under \ref{lin-MMWU} with any stepsize $\step \in (0, +\infty]$, and
one such example is the following:
\begin{equation}
\label{lin_evo_counter}
        v =
        \begin{pmatrix} \sqrt{\frac{2}{3}} \\  \sqrt{\frac{1}{3}} 
        \end{pmatrix},
        \qquad 
        \rho = 
        \begin{pmatrix} \frac{3}{4} & 0 \\  0 & \frac{1}{4} \end{pmatrix} ,
        \qquad 
        \Phi(\sigma) =
        \begin{pmatrix} 0 & 0 \\  0  & 1 \end{pmatrix}.
\end{equation}
    A simple calculation  shows that 
    $\innerprod{vv^\dagger}{\Phi(\sigma)} = \frac{1}{3} > \frac{1}{4} = \innerprod{\rho}{\Phi(\sigma)}$ 
   and  
    $\innerprod{vv^\dagger}{\rho} = \frac{7}{12} > 0$.  Consequently 
    \begin{equation*}
    % \label{ineq:_strat_comp_ave}
        \innerprod{vv^\dagger}{\powh{\rho}\Phi(\sigma)\powh{\rho}}
        = \frac{1}{12}
        < \frac{7}{48}
        = \innerprod{vv^\dagger}{\rho}\innerprod{\rho}{\Phi(\sigma)}. 
    \end{equation*}
    which means that $\innerprod{vv^\dagger}{\new{\rho}} < \innerprod{vv^\dagger}{\rho}$.

\subsection{Learning in quantum CIGs using \ref{lin-QREP} and \ref{lin-MMWU}}
\label{sec:_convprops}
In this section we provide theoretical results regarding the use of \ref{lin-QREP} and \ref{lin-MMWU} as learning dynamics for quantum common-interest games. Namely, we prove two main results: we first relate fixed points of the \ref{lin-QREP} and \ref{lin-MMWU} dynamics with Nash equilibria, showing that Nash equilibria are fixed points (Theorem \ref{thm:nashfixedpoints}); we then relate limit points of the dynamics' trajectories with the dynamics' fixed points, showing by a Lyapunov-type argument that limit points of dynamics are fixed points (Theorem \ref{thm:_a_quantShah}). These two theorems together imply that limit points of the dynamics form a superset of the Nash equilibria of the game. Finally, we round the section off with a discussion on why the discrete-time update \ref{lin-MMWU} fails to replicate the performance of its classical analogue \ref{linear} in converging to Nash equilibria under certain assumptions.

The first main result we show for \ref{lin-QREP} and \ref{lin-MMWU} in quantum CIGs is that Nash equilibria are fixed points. This is the first-order check for a learning dynamic that we want to go to Nash equilibria, since it means that if we are at a Nash equilibrium, we will stay there. 
% Together with Theorem \ref{thm:_interiorNE_equal_payoff} which we obtained from the equivalence of Nash equilibria and KKT points of the \ref{BSS} problem, this immediately gives us that Nash equilibria are fixed points of \ref{lin-QREP} in quantum CIGs. Moreover, by Theorem \ref{thm:_FixedPointsSame_ContTime_DiscreteTime}, Nash equilibria are also fixed points of \ref{lin-MMWU} in quantum CIGs. 
We can also prove a partial converse result that interior fixed points are Nash equilibria, and together these make up the first main theorem of this section, which relate Nash equilibria and fixed points of \ref{lin-QREP} and \ref{lin-MMWU}:

% \textcolor{red}{Main point of this subsection.}
\medskip
\begin{theo}
\label{thm:nashfixedpoints}
    In a quantum CIG, the continuous-time dynamic \ref{lin-QREP} and its discretization  \ref{lin-MMWU} (for any fixed step-size $\step \in (0, +\infty]$) exhibit the following properties:
    \begin{enumerate}
        \item Nash equilibria of the CIG are fixed points of the dynamics.
        \item Interior fixed points of the dynamics are Nash equilibria of the CIG.
    \end{enumerate}
\end{theo}

\begin{proof}
We utilize the results from Theorems \ref{thm:_interiorNE_equal_payoff}, which characterizes Nash equilibria in quantum CIGs, and Theorem \ref{char:_fixedpts}, which characterizes fixed points of \ref{lin-QREP} and \ref{lin-MMWU} in quantum games.
    \begin{enumerate}
        \item If $(\rho, \sigma)$ is a Nash equilibrium, we get from Theorem \ref{thm:_interiorNE_equal_payoff}  that 
	$\innerprod{\rho'}{\Phi(\sigma)} = \innerprod{\rho}{\Phi(\sigma)} \; \forall \ \rho' \in  \fc(\rho)$ and 
	$\innerprod{\rho}{\Phi(\sigma')} = \innerprod{\rho}{\Phi(\sigma')} \; \forall \ \sigma' \in \fc(\sigma)$
	. Then by Theorem \ref{char:_fixedpts}, $(\rho, \sigma)$ is a fixed point.
        \item Let  $(\rho, \sigma)$ be an interior fixed point of the \ref{lin-QREP} dynamics. As $\rho$ is invertible  and  $\dot{\rho}=0$ we immediately get that
$\Phi(\sigma) = \innerprod{\rho}{\Phi(\sigma)} \id_\A.$ In turn, this implies that
$\rho \in \BRa(\sigma)$ and similarly, $\sigma \in \BRb(\rho)$. Thus, $(\rho, \sigma)$ is an interior~NE. A similar argument holds for \ref{lin-MMWU}. \qedhere

    \end{enumerate}
\end{proof}

Next, we give the second main result of this section, which describes the convergence properties of the continuous and discrete-time dynamics introduced. Namely, we show that that they converge to the set of fixed points, which as we showed in Theorem \ref{thm:nashfixedpoints} are a superset of the set of Nash equilibria:

\medskip
 \begin{theo}
\label{thm:_a_quantShah}
The continuous-time dynamic  \ref{lin-QREP} and its discretization  \ref{lin-MMWU} (for any fixed step-size $\step \in (0, +\infty]$) have the following   properties:
\begin{enumerate}
    \item  The common utility  $u(\rho,\sigma)$ is strictly increasing  along  trajectories, except at fixed points.     

\item The set of $\omega$-limit points of a trajectory  is a compact, connected set of fixed points of the dynamics that all attain the same utility.
% \item    Nash equilibria are fixed points, and  interior fixed points are Nash equilibria.
% \item   The minimal face  is invariant along trajectories. As a result, all rank-one  density  matrices  are fixed~points.
	% %%%\item The interior $\omega$-limits of any trajectory  are Nash equilibria.

\end{enumerate}

\end{theo}
 
\begin{proof}  {\bf Property $(1)$:  \ref{lin-QREP}.} We first focus on the continuous-time dynamic  \ref{lin-QREP} and show that it is a gradient flow according to a specific geometry. (See Appendix \ref{appsecs:gradient_flow} for basic definitions of gradient flow dynamics.)
Specifically,  we will show that  \ref{lin-QREP} is a  gradient flow of   the scalar field    $u: \M \rightarrow \mathbb{R}, \; (\rho, \sigma) \mapsto \innerprod{\rho}{\Phi(\sigma)}$ corresponding to  the common utility function with  respect to the  \emph{quantum Shahshahani metric}  (also known as the intrinsic Riemannian metric, see e.g.  \cite{bhatia2009positive})
\begin{equation}
\label{cor:_a_quantShah}\tag{QShah}
    \innerprod{A}{B}_\rho := \Tr[\rho^{-\frac{1}{2}}A \rho^{-\frac{1}{2}}B], 
\end{equation}
defined on the product manifold $\M = \da \times \db$.  At any point  $(\rho, \sigma) \in \M$, we want to find ${\bf grad} u(\rho,\sigma)$, defined as  the unique vector $g = (g_\A, g_\B) \in T_{(\rho, \sigma)} \M = T_\rho D(\mathcal{A}) \times T_\sigma D(\mathcal{B})$ satisfying 
$$  D_{(\rho, \sigma)} u(\xi_\A, \xi_\B)
        = \langle(g_\A,g_\B), (\xi_\A, \xi_\B)\rangle_{(\rho, \sigma)}, \  \text{ for all } \xi = (\xi_\A, \xi_\B) \in T_{(\rho, \sigma)} \M.$$
Expanding the above we immediately get that 
\begin{equation}
\label{eqn:_gradflow_IntrinsicMetric1}
        D_{(\rho, \sigma)} u(\xi_\A, \xi_\B)        = \langle g_\A, \xi_\A\rangle_\rho + \langle g_\B, \xi_\B\rangle_\sigma 
        = \Tr[{\rho}^{-\frac{1}{2}}{g_\A} {\rho}^{-\frac{1}{2}} {\xi_\A}] +
        \Tr[{\sigma}^{-\frac{1}{2}}{g_\B}{\sigma}^{-\frac{1}{2}}{\xi_\B}],
\end{equation}
while on the other hand, as the Euclidean gradient $ \nabla u (\rho,\sigma)= \left(\Phi(\sigma), \Phi^\dagger(\rho) \right)$, we have that

\begin{equation}
\label{eqn:_gradflow_EuclideanMetric1}
  D_{(\rho, \sigma)} u(\xi_\A, \xi_\B) =\la \nabla u(\rho, \sigma),(\xi_\A, \xi_\B)\ra=
     \Tr[\Phi(\sigma)\xi_\A] + \Tr[\Phi^\dagger(\rho) \xi_\B].
\end{equation}

Equating (\ref{eqn:_gradflow_IntrinsicMetric1}) and (\ref{eqn:_gradflow_EuclideanMetric1}), we then have that 
${\bf grad} u(\rho,\sigma)$
is the unique element $(g_\A, g_\B) $ in the product of the tangent spaces  $T_\rho D(\mathcal{A}) \times T_\sigma D(\mathcal{B})$ with the following properties: 
\begin{itemize}
    \item $g_\A$ is the unique element in $T_\rho D(\mathcal{A})$ such that
$$
          \Tr[{\rho}^{-\frac{1}{2}}{g_\A} {\rho}^{-\frac{1}{2}} {\xi_\A}]= \Tr[\Phi(\sigma)\xi_\A], \quad \fa \xi_\A \in T_\rho D(\mathcal{A}).$$
    \item $g_\B$ is the unique element in $T_\sigma D(\mathcal{B})$ such that
    $$ \Tr[{\sigma}^{-\frac{1}{2}}{g_\B}{\sigma}^{-\frac{1}{2}}{\xi_\B}]=  \Tr[\Phi^\dagger(\rho)] \xi_\B,  \quad \fa \xi_\B \in T_\sigma D(\mathcal{B}).
              $$
\end{itemize}
A straightforward computation shows that for any constant $c$ we have
$$\begin{aligned}
\Tr[\Phi(\sigma)\xi_\A]&= \Tr[(\Phi(\sigma)-c\mathbb{1}_\A)\xi_\A]\\
&=\Tr[{\rho}^{-\frac{1}{2}}\underbrace{({\rho}^{\frac{1}{2}}(\Phi(\sigma)-c\mathbb{1}_\A){\rho}^{\frac{1}{2}})}_{g_\A}{\rho}^{-\frac{1}{2}}\xi_\A]
\end{aligned}$$
where for the first equality we used that all elements in the tangent space of $T_\rho D(\mathcal{A})$ have trace equal to zero  (see Appendix \ref{sec:_QREP_props}).  Lastly, to make $g_\A$ traceless we need to select the constant  $c$ so that 
$$\Tr({\rho}^{\frac{1}{2}}(\Phi(\sigma)-c\mathbb{1}_\A){\rho}^{\frac{1}{2}})=0 \iff c=  \frac{\Tr[\rho \Phi(\sigma)]}{\Tr[\rho]}.$$
Summarizing, we have established that 
$$g_\A=\rho^{\frac{1}{2}} \left[\Phi(\sigma) - \frac{\Tr[\rho \Phi(\sigma)]}{\Tr[\rho]}\mathbb{1}_\A \right] \rho^{\frac{1}{2}}$$
and symmetrically we also get that
$$g_\B = \sigma^{\frac{1}{2}} \left[\Phi^\dagger (\rho) -\frac{\Tr[\sigma \Phi^\dagger(\rho)]}{\Tr[\sigma]}\mathbb{1}_\B \right] \sigma^{\frac{1}{2}}.$$

Thus, 
the gradient flow on the product manifold $D(\A)\times D(\B)$ endowed  with the quantum Shahshahani metric
 is given by
\begin{equation*}
        \dv{\rho}{t}  = g_\A = \rho^{\frac{1}{2}} \left[\Phi(\sigma) - \frac{\Tr[\rho \Phi(\sigma)]}{\Tr[\rho]}\mathbb{1}_\A \right] \rho^{\frac{1}{2}}, \quad
        \dv{\sigma}{t}  = g_\B = \sigma^{\frac{1}{2}} \left[\Phi^\dagger (\rho) -\frac{\Tr[\sigma\Phi^\dagger(\rho)]}{\Tr[\sigma]}\mathbb{1}_\B \right] \sigma^{\frac{1}{2}}.
\end{equation*}

\paragraph{Property $(1)$:   \ref{lin-MMWU}.}   We show that the common utility is strictly increasing under a round of sequential updates, unless the strategy profile is a fixed point.   
     %We first prove this for the special case of \eqref{eqn:_DQREP}. Specifically, 
    Firstly, we   show that  under a $\rho$-update of \ref{lin-MMWU}, we have that
    \begin{align}\label{beineq}
        \innerprod{\new{\rho}}{\Phi(\sigma)}
        \geq
        \innerprod{\rho}{\Phi(\sigma)},
    \end{align}
    with equality iff %$\powh{\rho}\Phi(\sigma) \powh{\rho}$ is a scaling of $\rho$, 
 $\new{\rho} = \rho$ and similarly obtain that $\innerprod{\rho}{\Phi(\new{\sigma})} \geq \innerprod{\rho}{\Phi(\sigma)}$ iff $\sigma^{new}=\sigma$. Secondly, we show that 
       \begin{equation}\label{equalitycase}
       \innerprod{\new{\rho}}{\Phi(\sigma)}=         \innerprod{\rho}{\Phi(\sigma)} \iff \new{\rho}=\rho,
       \end{equation}
and similarly for the $\sigma$-update. 
Putting these two properties together we get that $\innerprod{\rho}{\Phi(\sigma)}$ is strictly increasing under \ref{lin-MMWU}  updates unless at a fixed point.

Substituting $\new{\rho}$ in \eqref{beineq} and clearing denominators, it is equivalent to   
    \begin{equation}  
    \label{incr_rho}
        \innerprod{\powh{\rho}(\id_\A+\step \Phi(\sigma))\powh{\rho}}{\Phi(\sigma)}
        	\geq
        	\innerprod{\rho}{\Phi(\sigma)}(1+\step \innerprod{\rho}{\Phi(\sigma)}).
    \end{equation}
    
    Expanding the left-hand side  we get
$$
\innerprod{\rho}{\Phi(\sigma)}+\step   \innerprod{\powh{\rho}\Phi(\sigma)\powh{\rho}}{\Phi(\sigma)}\geq
        	\innerprod{\rho}{\Phi(\sigma)}(1+\step \innerprod{\rho}{\Phi(\sigma)}).
$$
To prove this, it suffices to  show the inequality 
$$         
            \innerprod{\powh{\rho}\Phi(\sigma)\powh{\rho}}{\Phi(\sigma)}\geq \innerprod{\rho}{\Phi(\sigma)}^2. $$
   
      To see this, setting $\norm{A} = \sqrt{\innerprod{A}{A}}$, we have that
	\[	            
            \innerprod{\powh{\rho}\Phi(\sigma)\powh{\rho}}{\Phi(\sigma)}
		= \norm{\rho^{\sfrac{1}{4}} \Phi(\sigma) \rho^{\sfrac{1}{4}}}^2 
		= \norm{\rho^{\sfrac{1}{4}} \Phi(\sigma) \rho^{\sfrac{1}{4}}}^2 \norm{\powh{\rho}}^2 
		\geq \innerprod{\rho^{\sfrac{1}{4}} \Phi(\sigma) \rho^{\sfrac{1}{4}}}{\powh{\rho}}^2 
		= \innerprod{\rho}{\Phi(\sigma)}^2,
	\]
    where the inequality is due to Cauchy-Schwarz and the fact that $\norm{\powh{\rho}}^2 = \Tr(\rho) =~1$. 
    
    The second step is to  prove \eqref{equalitycase}. For this note that the  Cauchy-Schwarz equality used above holds  with equality  iff $\rho^{\sfrac{1}{4}} \Phi(\sigma) \rho^{\sfrac{1}{4}}$ is a scaling of $\powh{\rho}$, i.e.,
    $$  \innerprod{\new{\rho}}{\Phi(\sigma)}
        =
        \innerprod{\rho}{\Phi(\sigma)}\iff  \rho^{\sfrac{1}{4}} \Phi(\sigma) \rho^{\sfrac{1}{4}} = c\powh{\rho} \text{ for some }c \in \mathbb{R}.$$ 
        
        However, note that  $$ \rho^{\sfrac{1}{4}} \Phi(\sigma) \rho^{\sfrac{1}{4}} = c\powh{\rho}  \iff 
    \powh{\rho}\Phi(\sigma)\powh{\rho} 
    = c \rho. $$
    
    Indeed, if the LHS  is true we get that
    $$ \powh{\rho}\Phi(\sigma)\powh{\rho} =
    \rho^{\sfrac{1}{4}} (\rho^{\sfrac{1}{4}} \Phi(\sigma) \rho^{\sfrac{1}{4}}) \rho^{\sfrac{1}{4}} = \rho^{\sfrac{1}{4}}(c \powh{\rho})\rho^{\sfrac{1}{4}}
    = c \rho.$$
    Conversely if $\powh{\rho}\Phi(\sigma)\powh{\rho} = c\rho$, then letting $Q:= \sum_i f(\lambda_i) v_i v_i^\dagger$ where $\rho^{\sfrac{1}{4}}$ has spectral decomposition $\sum_i \lambda_i v_i v_i^\dagger$ and $f: \R_{\geq 0} \rightarrow  \R_{\geq 0}, \; f: x \mapsto \begin{cases} x^{-1} \quad \text{if }x > 0 \\ 0 \qquad \text{if } x = 0 \end{cases}$, we have that $$\rho^{\sfrac{1}{4}} \Phi(\sigma) \rho^{\sfrac{1}{4}} = Q\powh{\rho}\Phi(\sigma)\powh{\rho}Q = cQ\rho Q = c \powh{\rho}.$$

Finally, note that the condition $ \powh{\rho}\Phi(\sigma)\powh{\rho} 
    = c \rho$ is equivalent to $\new{\rho}=\rho.$ 
    
\paragraph{Property (2):} This is a
  direct  extension of a  fundamental convergence result  by Losert and Akin~\cite{losert1983dynamics} for classical games to the  the compact set of density matrices (see  Theorems~\ref{thm:_LimitSetCompactConnected_ContTime} and \ref{thm:_LimitSetCompactConnected_DiscreteTime} in the Appendix). It relies on the already established   fact that the utility function  is a Lyapunov function for both dynamics.  \qedhere
  
% \paragraph{Part (3):} 
% \paragraph{Part (4):}
  
\end{proof}

\textbf{Discussion on the non-convergence of \ref{lin-MMWU} to Nash equilibria.} The immediate consequence of Theorem \ref{thm:_a_quantShah} in tandem with Theorem \ref{thm:nashfixedpoints} is that the interior $\omega$-limits of any trajectory of the dynamics are Nash equilibria, but this is not sufficient to conclude that the dynamics in general converge to a Nash equilibrium. Indeed, as we shall see from our experiments in Section \ref{sec:_dynamics_empirical}, while the continuous-time \ref{lin-QREP} dynamics do appear to converge to Nash equilibria, there are cases where the discrete-time \ref{lin-MMWU} converges to states which are not Nash equilibria. In the following discussion, we shall show why the convergence result of \ref{linear} in classical common-interest games does \emph{not} carry over to the quantum setting.

% \subsection{Non-Convergence of \ref{lin-MMWU} to Nash equilibria}
In the classical case, it was shown in \cite{palaiopanos2017multiplicative} that, \emph{under the assumption that fixed points are isolated, any trajectory of \ref{linear} (with fixed stepsize) initialized in the interior converges to a Nash equilibrium of a classical common-interest game}. This result was proven in two steps: 1) if fixed points are isolated, then every trajectory converges, and 2) if a trajectory that begins in the interior converges, then it converges to a Nash equilibrium. The proof of Step 1 was a direct consequence of the analogous statement to our Theorem \ref{thm:_a_quantShah} that limit points are a compact, connected set of fixed points. However, while this assumption often holds in classical games, it is vacuous in quantum games as all rank-one density matrices are fixed points of \ref{lin-MMWU} (Theorem \ref{thm:minimalfaceinvariance}).

This means that the best we could have hoped to achieve in our setting is the statement that \emph{if a trajectory initialized in the interior converges, then it converges to a Nash equilibrium} (Step 2 of the proof of the classical theorem in \cite{palaiopanos2017multiplicative}). Classically the proof of this was due to:
\begin{enumerate}
    \item the property that \ref{linear} increases the weights of strategies that do better than average (Equation \eqref{classicalcondition}), and 
    \item the fact that \ref{linear} preserves the support of strategies.
\end{enumerate}
Together, these properties give us a result that any convergent trajectory initialized in the interior must converge to  Nash equilibrium via the following proof by contradiction: if $p^{(t)}$ converges as $t \to \infty$ to a point $p^*$ that is not a Nash equilibrium, then there exists player $i$ and pure strategy $k$ such that $u_i(k, p^*_{-i}) > u_i(p^*)$. By continuity of the utility function there then exists a neighborhood $U$ containing $p^*$ such that $u_i(k, p_{-i}) > u_i(p^*)$ for all $p \in U$, which means that after some time $T$ for which $p^{(t)} \in U$ for all $t \geq T$, we have by \eqref{classicalcondition} that $p^{(t)}_{ik} > p^{(T)}_{ik} > 0$ for all $t \geq T$, and hence $p^*_{ik} > 0$. (The fact that $p^{(T)}_{ik} > 0$ is due to the trajectory being initialized in the interior and the support-preserving property of \ref{linear}.) However, this is a contradiction since $p^*$ is a fixed point of \ref{linear} (since it is the limit point of a trajectory, so by Lyapunov analysis it is a fixed point by the analogous statement to our Theorem \ref{thm:_a_quantShah}) and hence it must be that all of the strategies in the support of $p^*_i$ perform equally well, i.e., $u_i(k', p^*_{-i}) = u_i(p^*)$ for all $k'$ for which $p^*_{ik} > 0$, which means, from the assumption that $u_i(k, p^*_{-i}) > u_i(p^*)$, that $p^*_{ik} = 0$. 

In our setting of quantum common-interest games, we similarly get that if an interior trajectory $(\rho(t), \sigma(t)) \to (\rho^*, \sigma^*)$ but $(\rho^*, \sigma^*)$ is not a Nash equilibrium, i.e., without loss of generality there exists a unit vector $v$ satisfying $\innerprod{vv^\dagger}{\Phi(\sigma^*)} > \innerprod{\rho^*}{\Phi(\sigma^*)}$, then there exists a neighborhood $U$ containing $(\rho^*\sigma^*)$ for which $\innerprod{vv^\dagger}{\Phi(\sigma)} > \innerprod{\rho}{\Phi(\sigma)}$ for all $(\rho, \sigma) \in U$ and a time $T$ after which all future iterates $(\rho(t),\sigma(t)) \in U$. It also similarly holds from the fact that $(\rho^*, \sigma^*)$ is a fixed point of \ref{lin-MMWU} by Theorem \ref{thm:_a_quantShah} that $\innerprod{vv^\dagger}{\rho^*} = 0$ due to the characterization that all strategies in the support of a fixed point perform equally well (Theorem \ref{char:_fixedpts}). 

However, the desired contradiction cannot be reached because, unlike in the classical case, the fact that $\innerprod{vv^\dagger}{\Phi(\sigma(t))} > \innerprod{\rho}{\Phi(\sigma(t))}$ for all $t \geq T$ is not enough to show that the quantity $\innerprod{vv^\dagger}{\Phi(\sigma(t))}$ increases at each $t$. This is because the direction of comparison $v$ is not necessarily an eigendirection of $\rho(t)$, and we only have the theorem that ``correlation with directions that do better than average increases under \ref{lin-MMWU}'' for eigendirections (Theorem \ref{thm:_q_compare}), but not for general directions (see the negative example in \eqref{lin_evo_counter}). As the direction of comparison $v$ is not guaranteed to be an eigenvector of the iterates $\rho(t)$ and, moreover, the eigenbasis of the iterates can, in general, change over time (Theorem \ref{thm:minimalfaceinvariance} only says that \ref{lin-MMWU} preserves the \emph{support}, i.e., the span of the positive eigenvectors, of the strategies), the classical argument no longer holds. Indeed, as we shall see in the following experiments, \ref{lin-MMWU} can converge to points that are not Nash equilibria.

\subsection{Empirical simulations}
\label{sec:_dynamics_empirical}
\paragraph{Convergence of \ref{lin-QREP} to the set of Nash equilibria.}
% \paragraph{Experiments.} In order to corroborate our theoretical results, we performed various simulations. }

\begin{figure}[H]
    \centering
    \begin{minipage}{.50\linewidth}
      \centering
      \includegraphics[width=.95\linewidth]{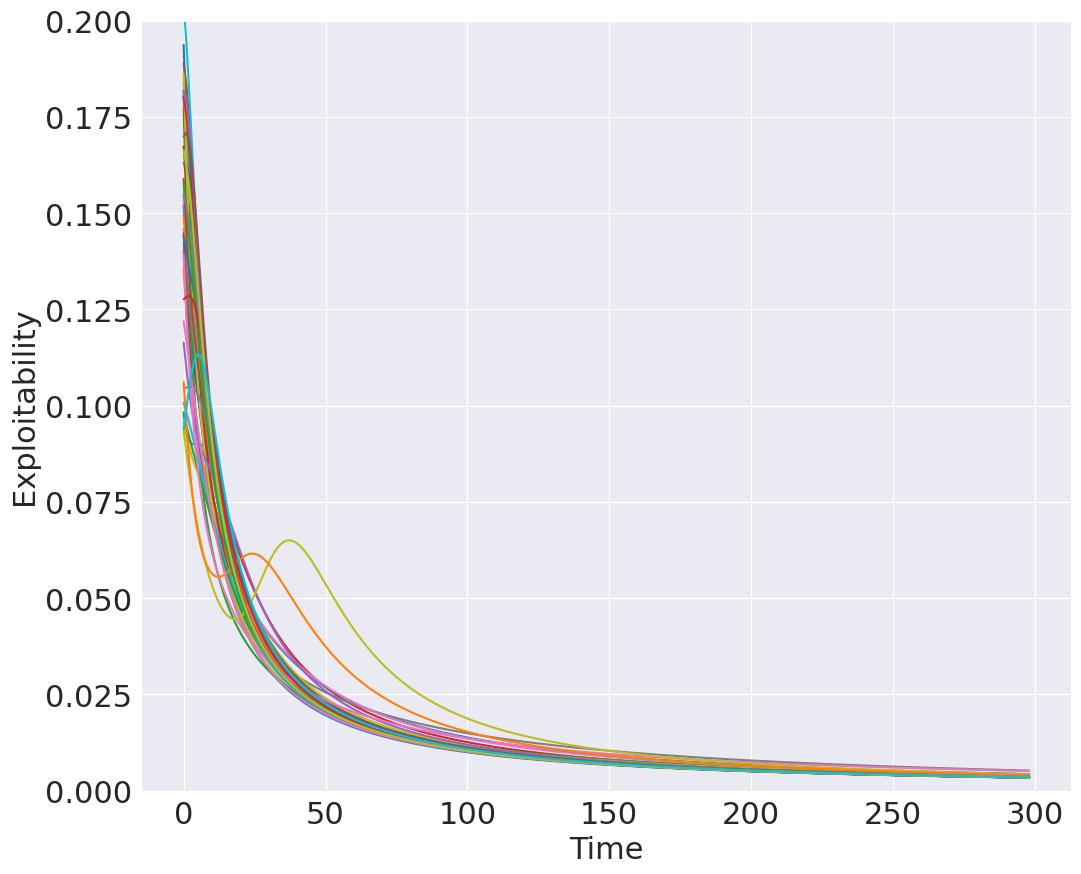}
     % \ref{lin-QREP}
    %   \label{fig:sub2}
    \end{minipage}
    % \begin{minipage}{.45\linewidth}
    %   \centering
    %   \includegraphics[width=.95\linewidth]{Images/exploit_MBE300new.png}
    %   \ref{eqn:_DQREP}
    % %   \label{fig:sub1}
    % \end{minipage}
    \caption{Exploitability of trajectories under \ref{lin-QREP}. The exploitability of all trajectories goes to zero.}
    \label{fig:exploit}
\end{figure}

% It is well known that natural learning dynamics converge to pure Nash equilibria in classical CIGs \cite{kleinberg2009multiplicative}.
In order to corroborate our theoretical results, we experimentally test the convergence of \ref{lin-QREP} to the set of Nash equilibria. In Figure \ref{fig:exploit} we plot  the exploitability (as defined in Equation \ref{eqn:exploitability})  of \ref{lin-QREP} in 100 randomly generated $\mathcal{H}_2 \otimes \mathcal{H}_2$ quantum CIG instances with uniform initialization, where $\mathcal{H}_n$ denotes an $n$-level quantum system. In all runs, the exploitabilities of \ref{lin-QREP} go to zero, meaning 
% they arrive at a 
that their limit points are
KKT points/Nash equilibria of the problem.

\begin{figure}[!h]
    \centering
    \begin{minipage}{.45\linewidth}
      \centering
      \includegraphics[width=.95\linewidth]{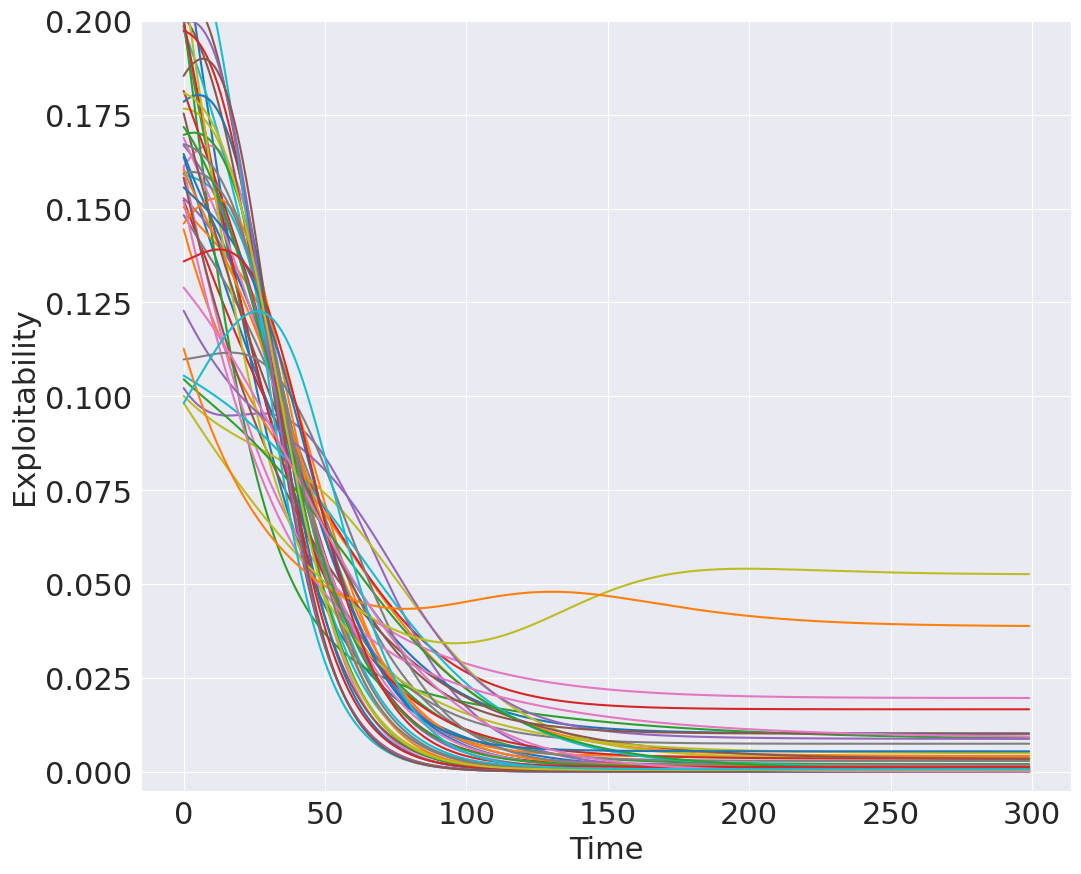}
     \ref{lin-MMWU} ($\step = 0.1$)
    %   \label{fig:sub2}
    \end{minipage}
    \begin{minipage}{.45\linewidth}
      \centering
      \includegraphics[width=.95\linewidth]{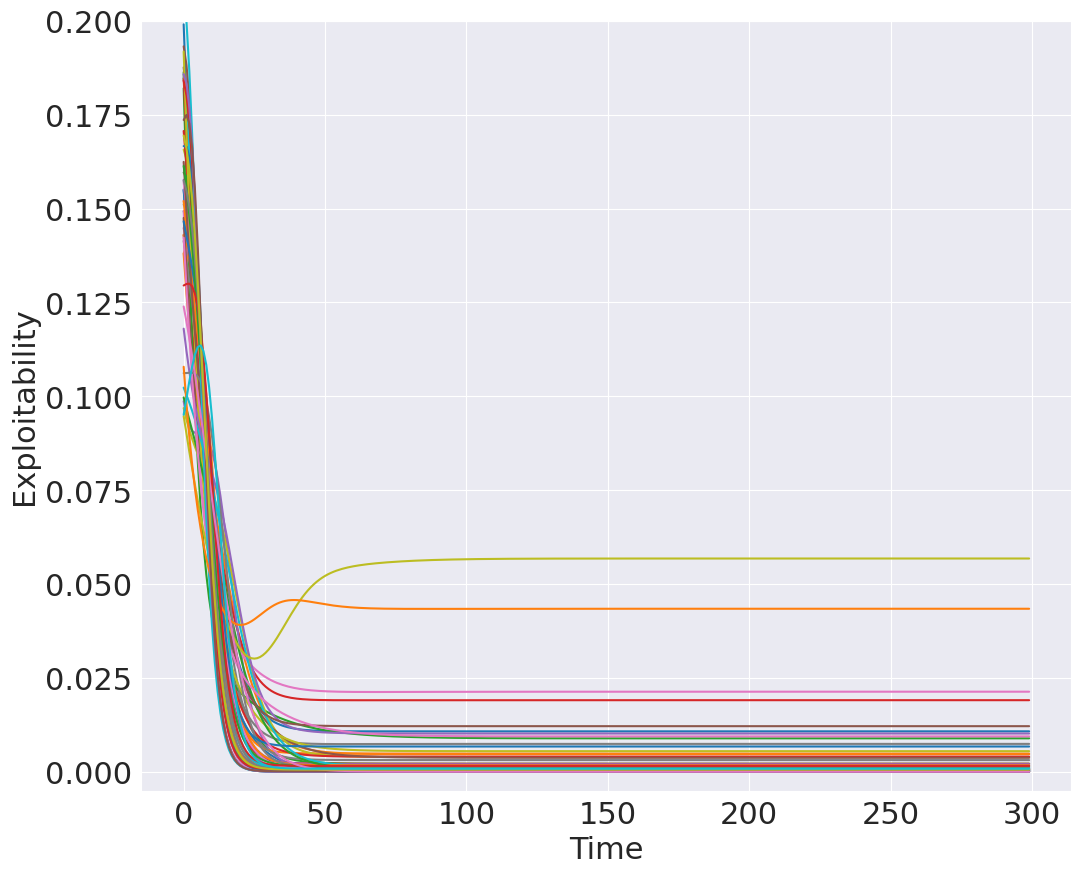}
     \ref{lin-MMWU} ($\step = 0.5$)
    %   \label{fig:sub2}
    \end{minipage}
    \begin{minipage}{.45\linewidth}
      \centering
      \includegraphics[width=.95\linewidth]{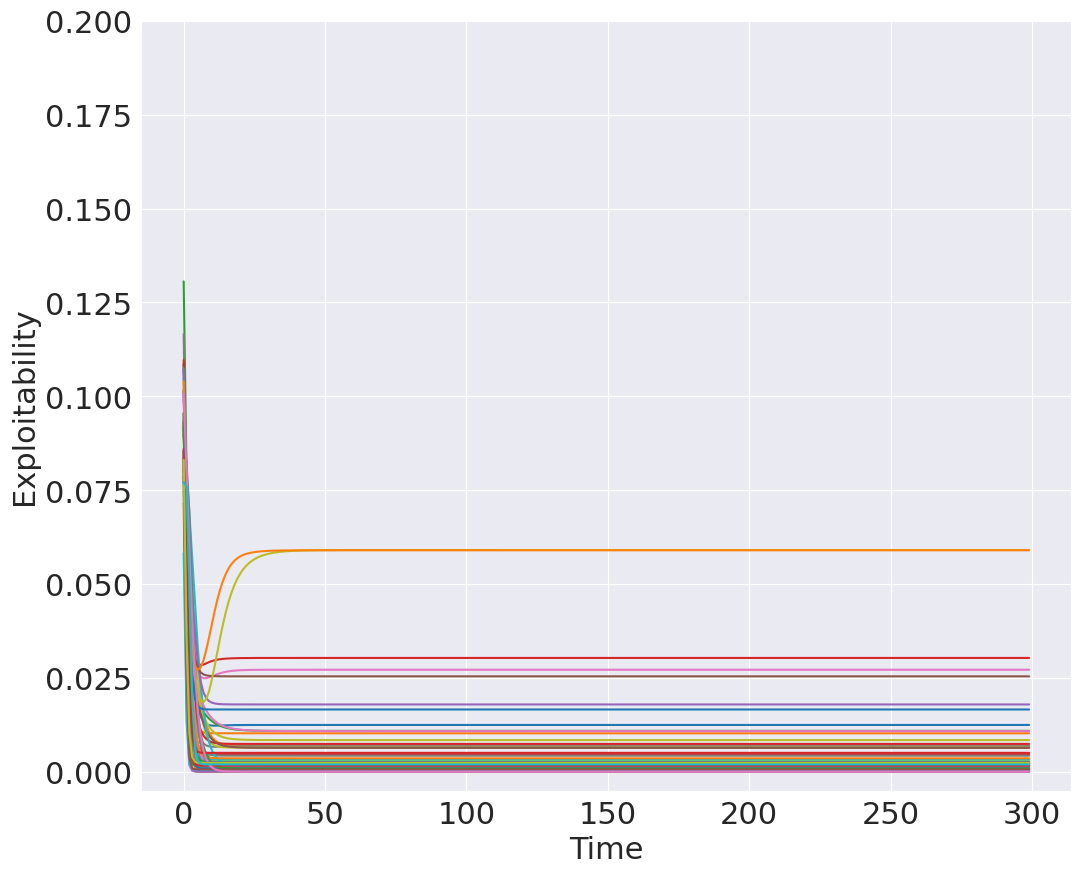}
      \ref{eqn:_DQREP} (\ref{lin-MMWU} with $\step = +\infty$)
    %   \label{fig:sub1}
    \end{minipage}
    \caption{
        Comparing exploitability of \ref{lin-MMWU} with different stepsizes $\step$. Trajectories with the same color were for the same game, with uniform initialization. 
        % Note that with larger stepsize the exploitability tends to converge faster, but with smaller stepsize the trajectories tend to end up with lower exploitability.
    }
    \label{fig:exploitlinmmwu}
\end{figure}

\paragraph{Exploitability experiments for \ref{lin-MMWU}.}
% Unlike the continuous \ref{lin-QREP} dynamics, we do not have theoretical convergence results to Nash equilibria for \ref{eqn:_DQREP} and \ref{lin-MMWU}. Nevertheless,
Similar to the continuous \ref{lin-QREP} dynamics, we empirically compute the exploitability of both \ref{eqn:_DQREP} and \ref{lin-MMWU}. 
% Unlike \ref{lin-QREP}, Figure \ref{fig:exploitlinmmwu} shows that \ref{eqn:_DQREP} and \ref{lin-MMWU} converge in some cases to states with positive exploitability, meaning they do not converge to pure Nash equilibria.
% \paragraph{Comparing \ref{lin-MMWU} to \ref{eqn:_DQREP}.} 
% \ref{lin-MMWU} is a more tunable version of \ref{eqn:_DQREP}, and we are able to empirically show a difference in performance between these two algorithms. 
Figure \ref{fig:exploitlinmmwu} compares the exploitability of \ref{lin-MMWU} with different stepsizes and \ref{eqn:_DQREP}, showing that some runs clearly attain a lower exploitability when using \ref{lin-MMWU}. In other words, while \ref{eqn:_DQREP} is \ref{lin-MMWU} with infinite stepsize $\step$, there is a qualitative difference in the states converged to from an exploitability standpoint when using a slower/smoother update. Notice that in both \ref{lin-MMWU} and \ref{eqn:_DQREP}, there exist examples of non-convergence to a pure Nash equilibrium (positive exploitability). 
% We perform a more theoretical analysis of this non-convergence phenomenon in Appendix \ref{app:_linMMWU_noncomm}, which explores why the classical Nash convergence result of \ref{linear} does not hold in our setting.

\begin{figure}[!h]
    \centering
    \begin{minipage}{.47\linewidth}
      \centering
      \includegraphics[width=.95\linewidth]{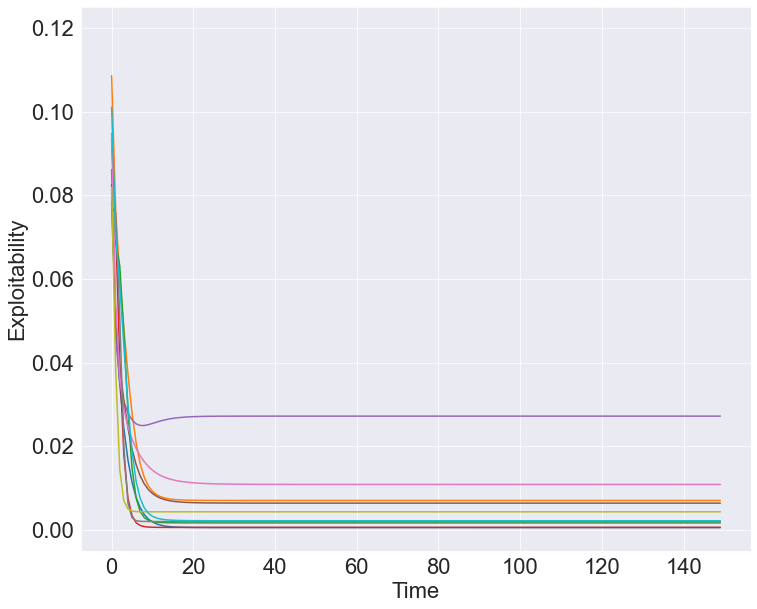}
      Exploitability.
    %   \label{fig:sub2}
    \end{minipage}
    \begin{minipage}{.47\linewidth}
      \centering
      \includegraphics[width=.95\linewidth]{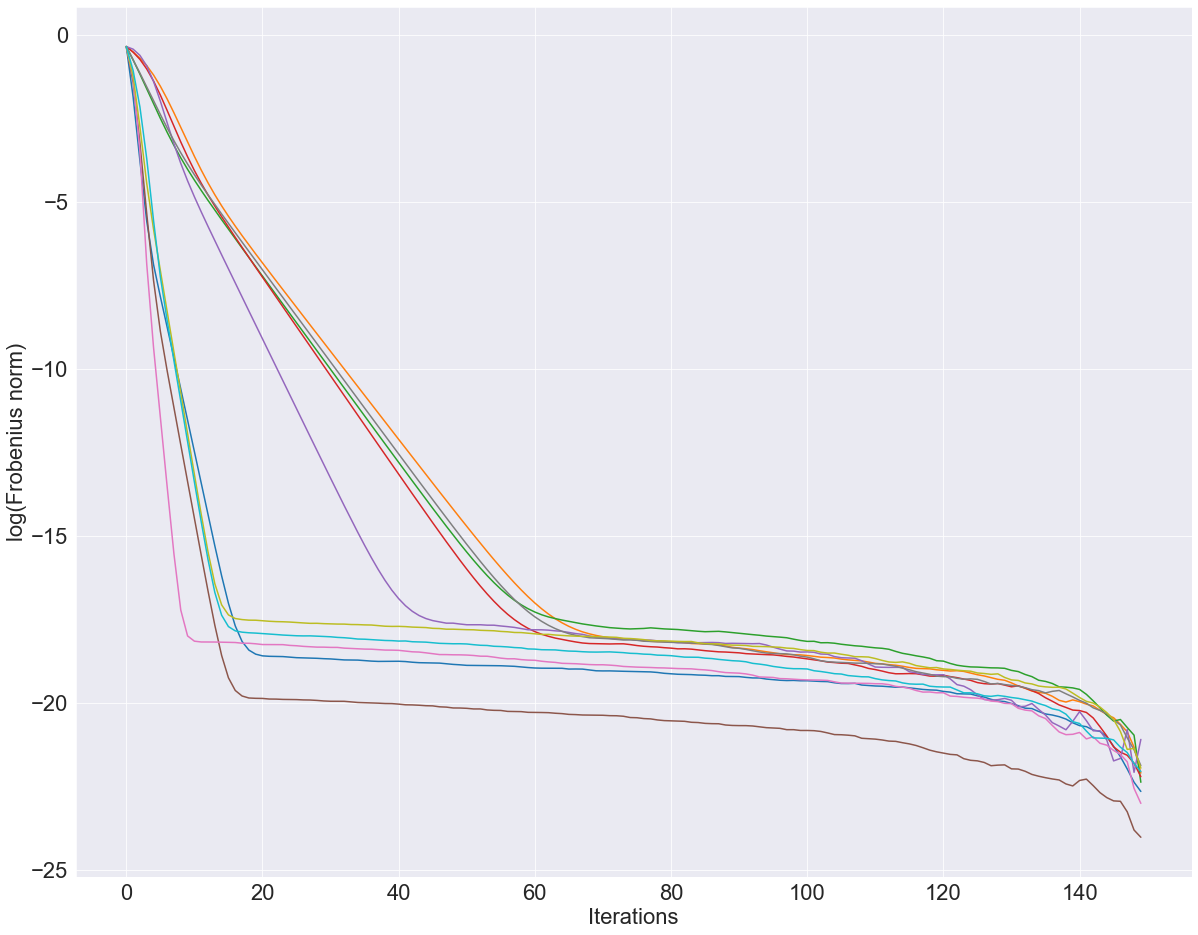}
      $log$ (Frobenius norm).
    %   \label{fig:sub1}
    \end{minipage}
    \caption{Counterexamples showing that convergence to a fixed point (low Frobenius norm) does not imply zero exploitability. All of the runs of the experiment converge to a fixed point, but several remain bounded away from zero exploitability.}
    \label{fig:counterexample1}
\end{figure}

From the discussion 
at the end of Section \ref{sec:_convprops},
% in Section \ref{sec:nonconvergenceLinMMWU}, 
it stands to reason that in our setting, the trajectories which do not converge to Nash equilibria could still be converging to fixed points. Indeed, we show in Figure \ref{fig:counterexample1} that it is possible to find such a trajectory. We plot the Frobenius norm between the dynamics at each time step and the next iterate. Intuitively, if the log of the Frobenius norm decreases over time, the dynamics stabilize and do not exhibit any oscillating behavior. We see that these trajectories, though convergent to a point, do not have zero exploitability, meaning that the fixed point is not a Nash equilibrium.
% , and indeed we have empirical evidence in Section \ref{sec:experiments} that this statement is not true.

\section{Experiments for the BSS Problem}\label{sec:bss}
With the connection between Nash equilibria and KKT points of the \ref{BSS} problem established in Section \ref{sec:QCIG}, we are able to empirically test our discrete-time dynamics as decentralized methods to approximately solve the \ref{BSS} problem. 

Recall that classically, both best response dynamics and linear multiplicative updates with constant step-size converge to Nash equilibria. However, we have shown in Section \ref{sec:_convprops} that the classical proof technique does not carry over to \ref{lin-MMWU}. Moreover, simulations in Section \ref{sec:_dynamics_empirical} experimentally confirm that \ref{lin-MMWU} might fail to converge to Nash equilibria. In this section, we will compare the performance of \ref{BR} to \ref{lin-MMWU} variants for solving the \ref{BSS} problem.
% and motivated by the well-known  classical result that `natural' learning dynamics converge to Nash equilibria in classical CIGs, in the next section
% % (see e.g. \cite{hofbauer1998evolutionary, hofbauer2002global, kleinberg2009multiplicative}).
% we propose a non-commutative extension of one such family of gradient flow dynamics and study their theoretical convergence properties.

% \paragraph{\ref{lin-MMWU} as an algorithm for the \ref{BSS} problem.}
%As described in Section \ref{sec:introduction}, the dynamic we propose for quantum CIGs can also be applied to the \ref{BSS} problem. In particular, 
% In this section we evaluate the performance of  \ref{lin-MMWU} applied to 
%we wish to test if \ref{eqn:_DQREP} converges to an approximate solution of 
% the \ref{BSS} problem. 
%problem given a positive definite matrix $R$ which is unknown to agents. 
% The global optimum \texttt{OPT} of the \ref{BSS} problem for $\mathcal{H}_2 \otimes \mathcal{H}_2$ and $\mathcal{H}_2 \otimes \mathcal{H}_3$ systems can be obtained exactly by solving a semidefinite program.
% (see Appendix \ref{appsecs:sdp} for a detailed explanation). 
Moreover, recall that the  \ref{BSS} problem corresponds to linear optimization over the set of separable states, which is in general hard to compute.
In order to benchmark the performance of \ref{BR} and \ref{lin-MMWU}, we first identify instances of the  \ref{BSS} problem that can be solved to optimality.
%several standard results from the quantum foundations literature. %Specifically, \cite{doherty2004complete} have described a hierarchy of semidefinite programs (SDPs), 
%that can solve the \ref{BSS} problem utilizing the well-known 
For this, we rely  on the  {\em Positive Partial Transpose} (PPT)  criterion for separability~\cite{peres1996separability, horodecki1996necessary}. Specifically,   any density matrix  $\rho$ describing the joint system $\A\otimes \B$  where $n=\dim\A $ and $m=\dim \B$, can be written as a block matrix 
$$\rho=
\begin{pmatrix}A_{11} & \ldots & A_{1n}\\
\vdots & & \vdots\\
A_{n1}& \ldots & A_{nn}
\end{pmatrix},$$
where each block is an $m\times m$   matrix. The partial transpose of $\rho$ with respect to $\B$ is the matrix obtained from $\rho$  transposing each block $A_{ij}$, namely 
$$\rho^{T_\B}=
\begin{pmatrix}A_{11}^T & \ldots & A_{1n}^T\\
\vdots & & \vdots\\
A_{n1}^T& \ldots & A_{nn}^T
\end{pmatrix},$$
and analogously we can also define the partial transpose of $\rho$ with respect to $\A$.
The PPT criterion states that a {\em necessary condition} for $\rho$ to be separable is that the partial transpose $\rho^{T_\B}$ is positive semidefinite. %As transposition preserves eigenvalues, it can be easily seen that the PPT
Moreover, \cite{horodecki1996necessary}, based on previous work from \cite{woronowicz1976positive} also show that the PPT criterion is  {\em necessary and sufficient} when both $\A$ and $\B$ are qubit systems (i.e., $\mathcal{H}_2 \otimes \mathcal{H}_2$) or when one of them is a  qubit system and the other a qutrit (i.e.,  $\mathcal{H}_2 \otimes \mathcal{H}_3)$. 
Consequently, in these two regimes, the \ref{BSS} problem corresponds to  the following  Semidefinite Program:
\begin{equation}\label{problem:sdp1}
    \max\{  \langle R, \rho\rangle:   \rho^{T_B} \succeq 0, \rho \succeq 0,\  \Tr(\rho) = 1\},
\end{equation}
and consequently, it can be efficiently solved to optimality. 
Hence, we can benchmark the performance of \ref{BR} and \ref{lin-MMWU} in the qubit vs. qubit or qubit vs. qutrit regimes  
by first computing the ground truth by solving the SDP \eqref{problem:sdp1}  which we then compare to the last iterate of the respective dynamic used.

% We benchmark \ref{eqn:_DQREP} against this ground truth  optimal value.
In each run of the experiments, we randomly generate a Hermitian positive definite  matrix $R$ and standardize a uniform diagonal initialization (i.e. $\mathbb{1}_\mathcal{H}/n$) for the dynamic.
% \ref{lin-MMWU}.  
Subsequently, we run the dynamic until convergence, which can be detected by checking the moving average (window size $= 5$) of the players' utility and we terminate the update if the moving average stabilizes  for several iterations.  As a benchmarking metric, we report the mean relative accuracy of the output of the dynamic compared to the optimal solution for the problem instance \eqref{problem:sdp1} (denoted \texttt{OPT} and computed using CVXPY \cite{diamond2016cvxpy,agrawal2018rewriting}) across 100 runs. 
%This is obtained by taking the average ratio over 100 runs of the utility value from running.
We also report the average number of iterations needed to find a fixed point/solution, along with the  standard deviation of the accuracy across the 100 runs. 

\begin{figure}[!htb]
    \centering
    \begin{minipage}{.45\linewidth}
      \centering
      \includegraphics[width=.95\linewidth]{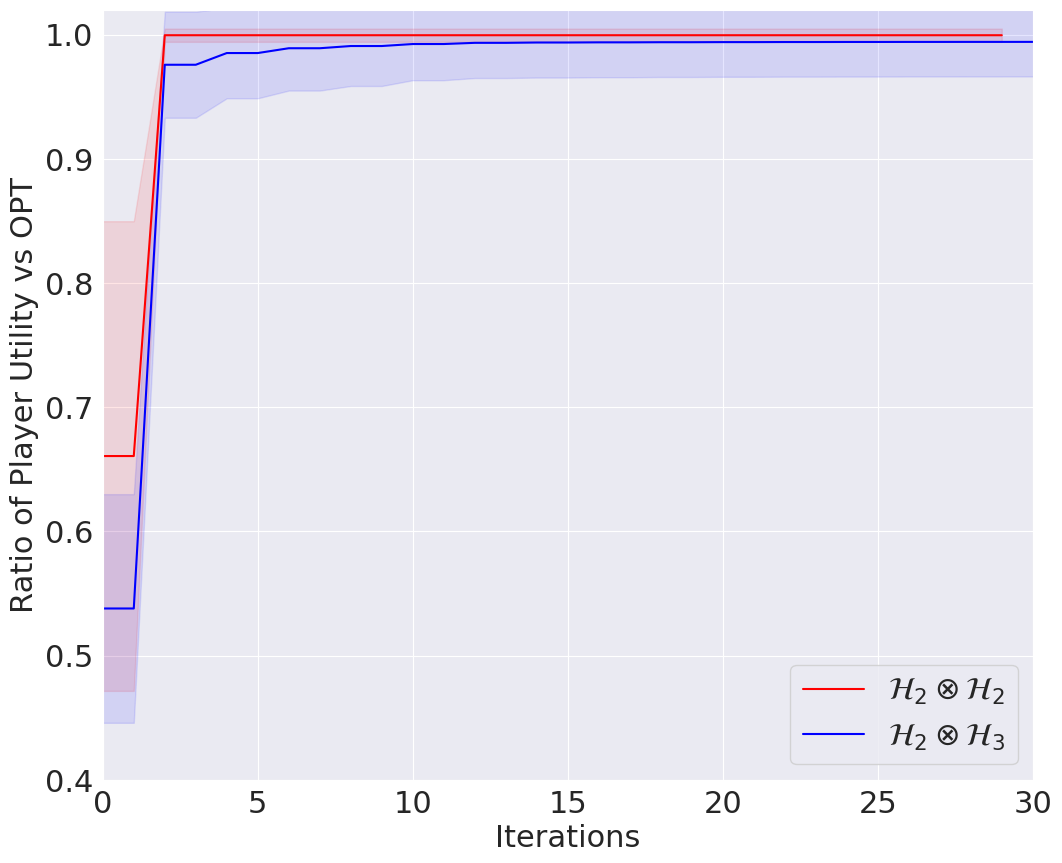}
     Random games, uniform initialization
    %   \label{fig:sub2}
    \end{minipage}
    \begin{minipage}{.45\linewidth}
      \centering
      \includegraphics[width=.95\linewidth]{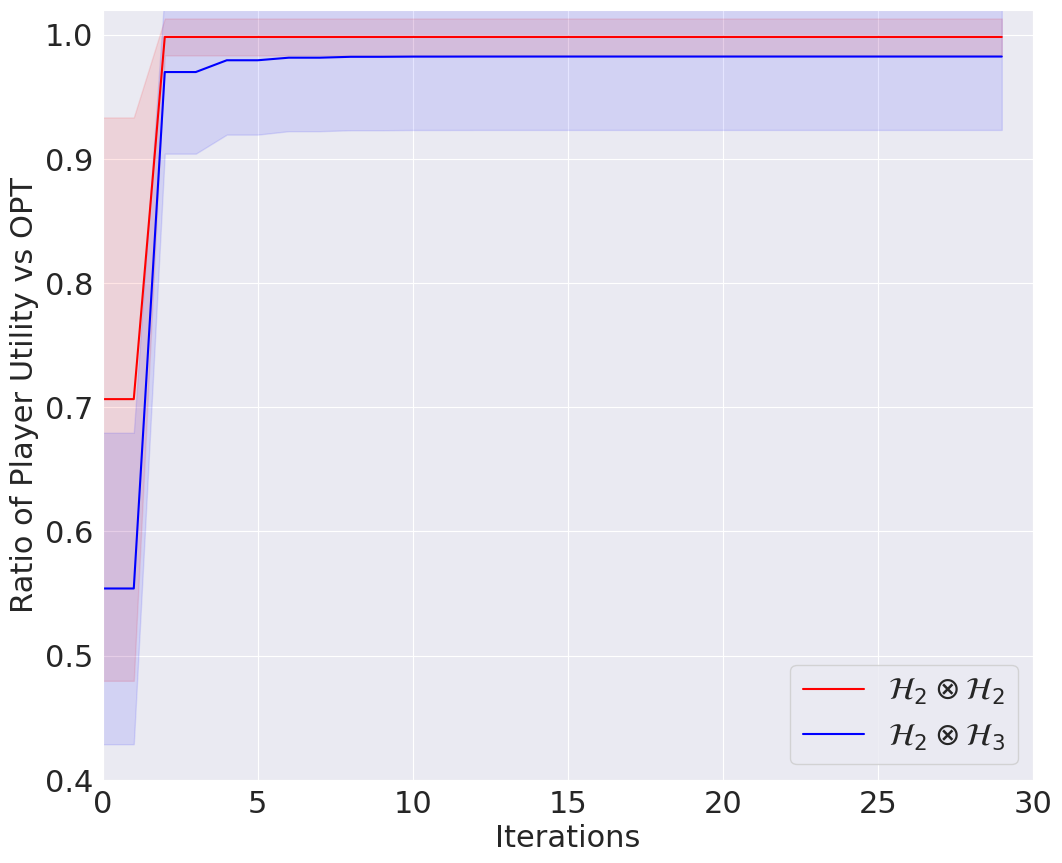}
      Random games, random initializations
    %   \label{fig:sub1}
    \end{minipage}
    \caption{Ratio of the utility attained using \ref{BR} vs \texttt{OPT}, averaged over 100 random BSS problem instances. Shaded region represents $\pm 1$ standard deviation from the mean, and each iteration represents alternating updates for $\rho$ and $\sigma$.}
    \label{fig:BRperformance}
\end{figure}

In the first set of experiments, we focus on alternating \ref{BR} dynamics and show that the dynamic converges very quickly to nearly optimal solutions to the \ref{BSS} problem. All these results are summarized  in Table \ref{table:sdpresultsbr}. Figure \ref{fig:BRperformance} visualizes our results, and we also include a version of the experiment where the initializations for each player are random density matrices instead of uniform diagonal matrices.
% terminating the algorithm after the average utility value for both players has remained invariant for 5 iterations.
% As a benchmark, the optimal solution to \eqref{problem:sdp1} is computed using CVXPY \cite{diamond2016cvxpy,agrawal2018rewriting}.

\begin{table*}[!tb]
\centering
\caption{Empirical performance of \ref{BR} (uniform initialization) for the BSS problem.}
\label{table:sdpresultsbr}
\begin{tabular}{@{}ccccc@{}}
\toprule
\multirow{2}{*}{\textbf{Problem Dimensions}}        & \multirow{2}{*}{\textbf{Runs}} & \multicolumn{2}{c}{\textbf{Accuracy}} & \multirow{2}{*}{\textbf{Average Iterations to Convergence}} \\ \cmidrule(lr){3-4}
                                      &                                & \textbf{Mean}   & \textbf{Std. Dev.}  &                                                             \\ \midrule
$\mathcal{H}_2 \otimes \mathcal{H}_2$ & 100                            & 0.998           & 0.007              & 8.81                                                       \\
$\mathcal{H}_2 \otimes \mathcal{H}_3$ & 100                            & 0.995           & 0.020              & 9.61                                                        \\ \bottomrule
\end{tabular}
\end{table*}

\begin{figure}[!htb]
    \centering
    \begin{minipage}{.45\linewidth}
      \centering
      \includegraphics[width=.95\linewidth]{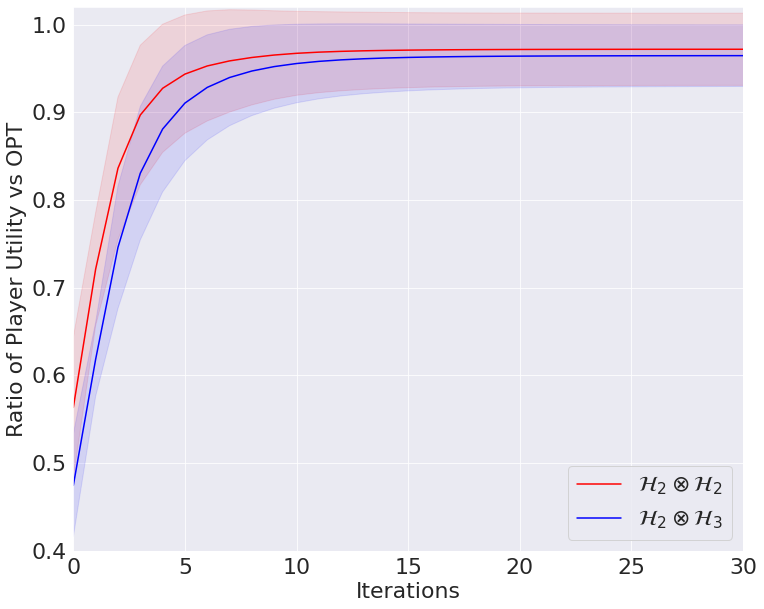}
     Random games, uniform initialization
    %   \label{fig:sub2}
    \end{minipage}
    \begin{minipage}{.45\linewidth}
      \centering
      \includegraphics[width=.95\linewidth]{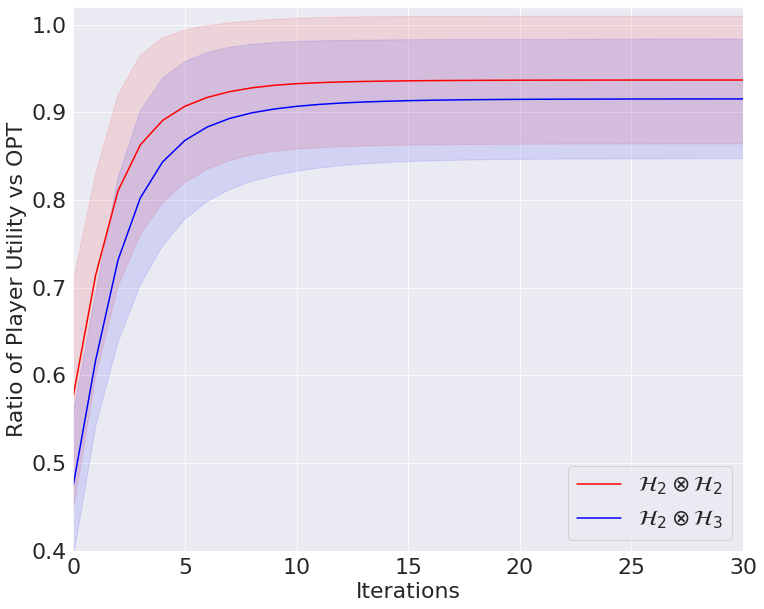}
      Random games, random initializations
    %   \label{fig:sub1}
    \end{minipage}
    \caption{Ratio of the utility attained using \ref{eqn:_DQREP} (\ref{lin-MMWU} with $\step= +\infty$) vs \texttt{OPT}, averaged over 100 random BSS problem instances. Shaded region represents $\pm 1$ standard deviation from the mean, and each iteration represents alternating updates for $\rho$ and $\sigma$. }
    % For uniform initialization, we see that \ref{eqn:_DQREP} attains a $0.97$ of  $\texttt{OPT}$ on average, while it does comparatively worse with random initialization.}
    \label{fig:dqrepperformance}
\end{figure}

In the subsequent experiment, we focus on \ref{eqn:_DQREP}, which is \ref{lin-MMWU} with stepsize $\step= +\infty$.
All these results are summarized  in Table \ref{table:sdpresults} and visualized in Figure \ref{fig:dqrepperformance}.
% visualizes our results, and we also include a version of the experiment where the initializations for each player are random density matrices instead of uniform diagonal matrices.
% terminating the algorithm after the average utility value for both players has remained invariant for 5 iterations.
% The solution to \eqref{problem:sdp1} is computed using CVXPY \cite{diamond2016cvxpy,agrawal2018rewriting}.
% and our findings are described in Section \ref{sec:experiments} and~{Table \ref{table:sdpresults}}.
% Moreover, we also keep track of the number of iterations needed for the discrete algorithm to converge. The results derived are summarized in Table \ref{table:sdpresults}.
% Beyond experiments of these dimensions, we are only able to obtain approximate optimal values by utilizing a hierarchy of semidefinite relaxations. As such, we omit numerics in higher dimensions from the main text.
% , but remark that empirically, our algorithm performs well even for larger scale problems.

\begin{table*}[!tb]
\centering
\caption{Empirical performance of \ref{eqn:_DQREP} (\ref{lin-MMWU} with $\step= +\infty$, uniform initialization) for the BSS problem.}
\label{table:sdpresults}
\begin{tabular}{@{}ccccc@{}}
\toprule
\multirow{2}{*}{\textbf{Problem Dimensions}}        & \multirow{2}{*}{\textbf{Runs}} & \multicolumn{2}{c}{\textbf{Accuracy}} & \multirow{2}{*}{\textbf{Average Iterations to Convergence}} \\ \cmidrule(lr){3-4}
                                      &                                & \textbf{Mean}   & \textbf{Std. Dev.}  &                                                             \\ \midrule
$\mathcal{H}_2 \otimes \mathcal{H}_2$ & 100                            & 0.972           & 0.0410              & 14.12                                                       \\
$\mathcal{H}_2 \otimes \mathcal{H}_3$ & 100                            & 0.965           & 0.0351              & 16.97                                                        \\ \bottomrule
\end{tabular}
\end{table*}

% \begin{figure*}[!htb]
%     \centering
%     \includegraphics[width=0.9\linewidth]{Images/bloch/Slide1.PNG}
%     \caption{Trajectories going to the boundary of the Bloch sphere in a fixed game instance with 100 randomized density initializations. 
%     Points are color-coded based on distance to boundary, with yellow denoting points that are close the the boundary.
%     % The values of the distance from the boundary are normalized such that yellow only denotes points which are a small $\step$-value from the boundary.
%     }
%     \label{fig:blochfixedgame}
% \end{figure*}

\begin{figure}[!htb]
    \centering
    \begin{minipage}{.45\linewidth}
      \centering
      \includegraphics[width=.95\linewidth]{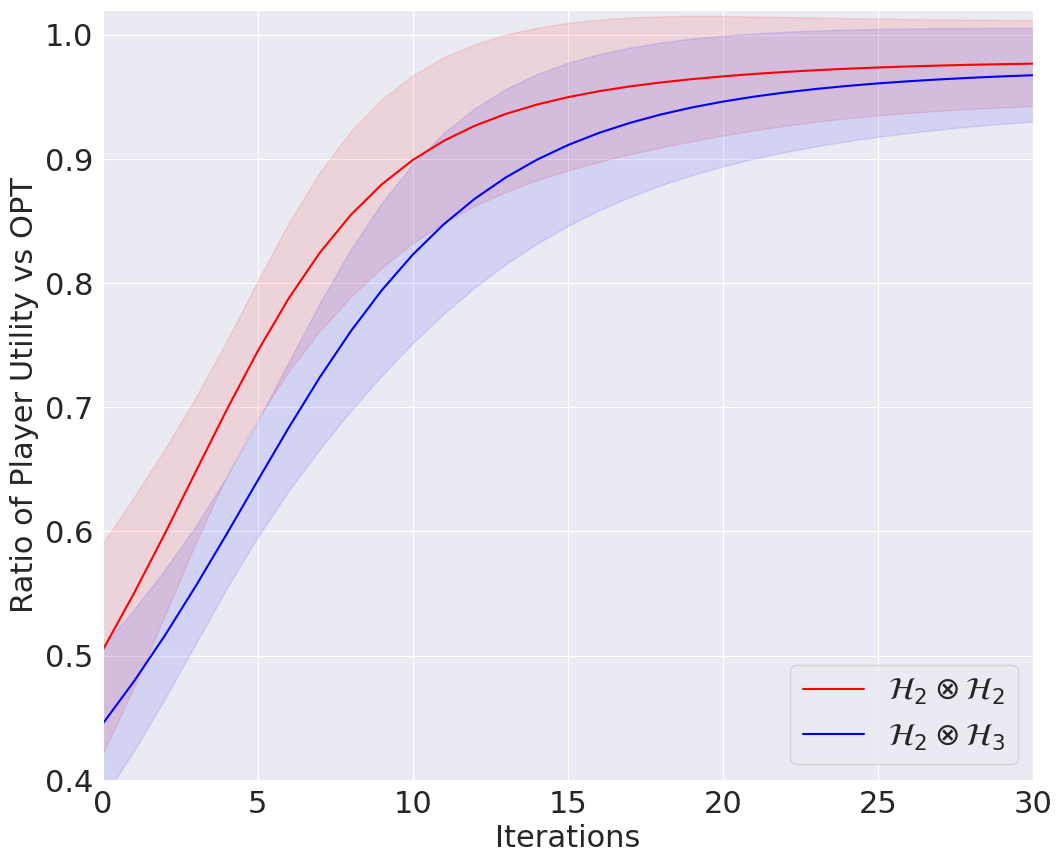}
     Random games, uniform initialization
    %   \label{fig:sub2}
    \end{minipage}
    \begin{minipage}{.45\linewidth}
      \centering
      \includegraphics[width=.95\linewidth]{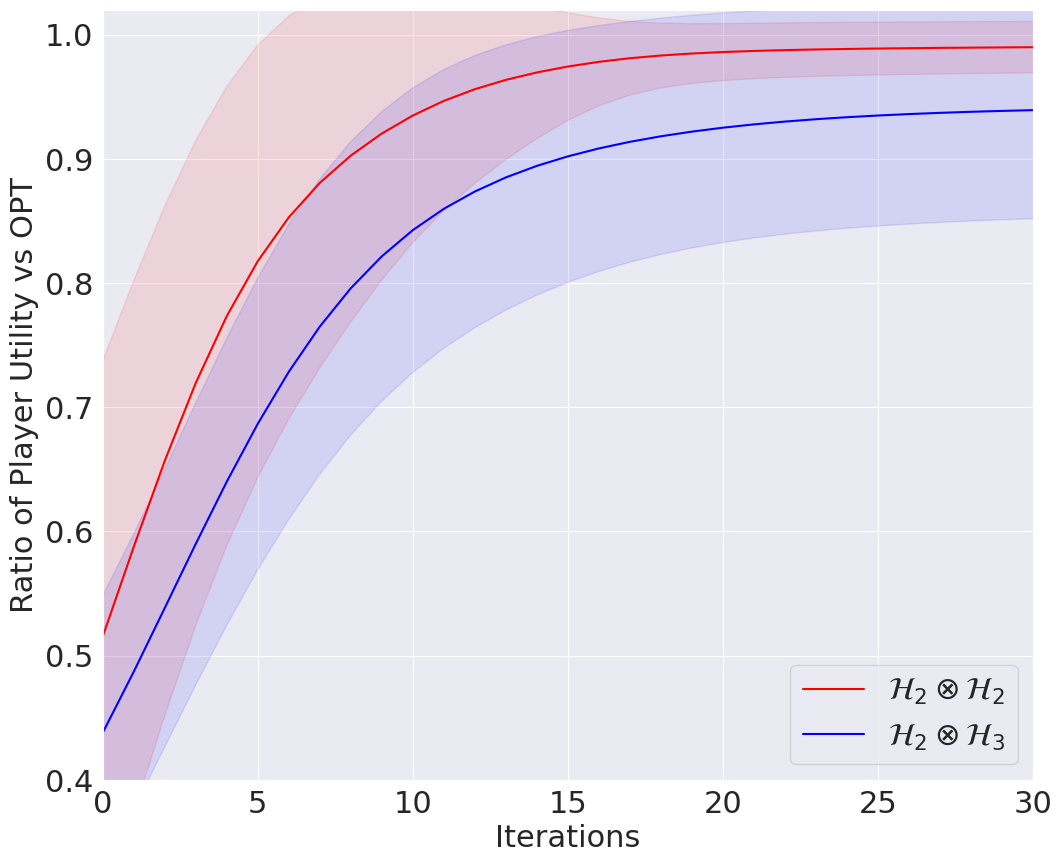}
      Random games, random initializations
    %   \label{fig:sub1}
    \end{minipage}
    \caption{Ratio of the utility attained using \ref{lin-MMWU} vs \texttt{OPT}, averaged over 100 random BSS problem instances. Shaded region represents $\pm 1$ standard deviation from the mean.}
    \label{fig:linmmwuperformance}
\end{figure}

% \paragraph{Comparing \ref{lin-MMWU} to \ref{eqn:_DQREP}.} 
% \ref{lin-MMWU} is a more tunable version of \ref{eqn:_DQREP}, and we are able to empirically show a difference in performance between these two algorithms for the \ref{BSS} problem. We compare the average case performance of each algorithm using the SDP benchmarking technique. The experimental setup is the same as before, and we set stepsize $\step = 0.9$. 
To explore the performance of \ref{lin-MMWU} with different stepsizes, we perform the same series of experiments on \ref{lin-MMWU} with a smaller stepsize of $0.9$. The results of these experiments are shown in Table \ref{table:sdpresultslinmmwu} and Figure \ref{fig:linmmwuperformance}. 
% Notice that in both \ref{lin-MMWU} and \ref{eqn:_DQREP}, there exist examples of non-convergence to a pure Nash equilibrium (positive exploitability). 
% We perform a more theoretical analysis of this non-convergence phenomenon in Appendix \ref{app:_linMMWU_noncomm}, which explores why the classical Nash convergence result of \ref{linear} does not hold in our setting.} 

\begin{table*}[!tb]
\centering
\caption{Empirical performance of \ref{lin-MMWU} (stepsize $=0.9$, uniform initialization) for the BSS problem.}
\label{table:sdpresultslinmmwu}
\begin{tabular}{@{}ccccc@{}}
\toprule
\multirow{2}{*}{\textbf{Problem Dimensions}}        & \multirow{2}{*}{\textbf{Runs}} & \multicolumn{2}{c}{\textbf{Accuracy}} & \multirow{2}{*}{\textbf{Average Iterations to Convergence}} \\ \cmidrule(lr){3-4}
                                      &                                & \textbf{Mean}   & \textbf{Std. Dev.}  &                                                             \\ \midrule
$\mathcal{H}_2 \otimes \mathcal{H}_2$ & 100                            & 0.977           & 0.0343              & 29.76                                                       \\
$\mathcal{H}_2 \otimes \mathcal{H}_3$ & 100                            & 0.972           & 0.0324              & 35.69                                                        \\ \bottomrule
\end{tabular}
\end{table*}

In our experiments, it is clear that best response dynamics perform well for the small, verifiable BSS problem instances. However, we observe that on average, the \ref{lin-MMWU} dynamics also perform competitively on average. 
Beyond benchmarking average case performance via SDPs, we are also interested in comparing the per-iteration performance of \ref{lin-MMWU} to \ref{eqn:_DQREP}. For this, we consider the metrics of exploitability and utility attained. We performed 1000 runs of \ref{lin-MMWU} and \ref{eqn:_DQREP} in small BSS instances, and found that in $93.1\%$ of runs, \ref{lin-MMWU} with stepsize $=0.9$ obtained lower exploitability than \ref{eqn:_DQREP}, while in $99.6\%$ of runs \ref{lin-MMWU} obtained higher utility than \ref{eqn:_DQREP}. The average percentage decrease in exploitability is $36.3\%$, while the percentage improvement in utility is merely $0.635\%$. Thus, we observe empirically that \ref{lin-MMWU} consistently finds less exploitable fixed points than \ref{eqn:_DQREP} for solving the \ref{BSS} problem. We note that the stepsize of $0.9$ is selected in order to balance performance and convergence at a reasonable rate.

\section{Additional Experiments}\label{sec:experiments}
% % \textcolor{red}{In this section... what does each experiment imply, what is the goal of each experiment?}
In this section, we present a suite of additional experimental results that 
% either corroborate our theoretical results from prior sections, or 
provide new insights into the empirical behavior of our dynamics, primarily focusing on \ref{lin-MMWU}.
% and \ref{eqn:_DQREP}, which is \ref{lin-MMWU} with $\step = + \infty$.  
% First, \edits{for the continuous-time dynamics,} we empirically show that \ref{lin-QREP} converges to \edits{the set of} Nash equilibria by utilizing the exploitability metric. \edits{Moving on to discrete-time dynamics, we then similarly show that \ref{BR} dynamics converges to the set of Nash equilibria, corroborating our theoretical results.} Next, we test \ref{eqn:_DQREP} as an algorithm for the \ref{BSS} problem, showing that it achieves $\approx 97\%$ of the optimal value for small problem instances. Moreover, 
First, we explore the potential for \ref{eqn:_DQREP} (\ref{lin-MMWU} with $\step = + \infty$) as an algorithm for biquadratic optimization, showing that it converges to rank-1 densities. To complete our experiments for \ref{eqn:_DQREP}, we provide some larger scale examples that show our convergence results in higher dimensions. Finally, we compare the \ref{lin-MMWU} update with smaller stepsizes to \ref{eqn:_DQREP}, showing that they achieve comparable performance, though perturbation can be used to improve the performance of \ref{lin-MMWU}.

\begin{figure*}[!htb]
    \centering
    \includegraphics[width=0.91\linewidth]{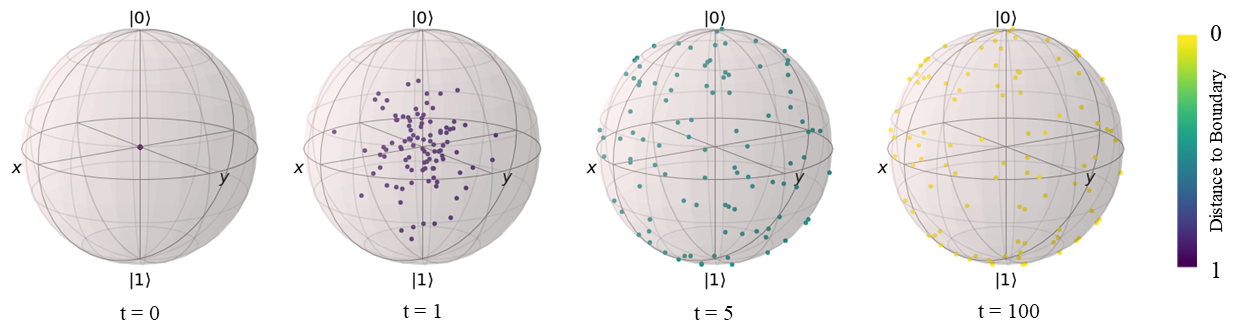}
    \caption{\ref{eqn:_DQREP} 
    % (\ref{lin-MMWU} with $\step= +\infty$) 
    trajectories going to the boundary of the Bloch-sphere in 100 random game instances with uniform density initializations.
    Points are color-coded based on distance to boundary, with yellow denoting points that are close the the boundary.
    }
    \label{fig:blochrandgame}
\end{figure*}

\paragraph{\ref{eqn:_DQREP} as an algorithm for biquadratic optimization.} 
Classically, replicator dynamics and their discretizations converge to pure equilibria in almost all generic common-interest games~\cite{kleinberg2009multiplicative, mertikopoulos2016learning, panageas2019multiplicative, mehta2014natural}. We experimentally verify that this behavior carries over to the quantum setting. In the quantum CIG setting, the players' strategies are density matrices, which (via the SVD) correspond to distributions  over rank-1 densities. Consequently,  
a quantum CIG can be viewed as the mixed extension of a common-interest game where players choose unit vectors and share a biquadratic utility. 
 Our experiments suggest that  when the players in a  quantum CIG with game operator $R$ use \ref{eqn:_DQREP},  their states converge to rank-1 density matrices, an intriguing  analogue of the result that classical CIGs converge to pure equilibria, and an empirical confirmation of our result in Theorem \ref{thm:minimalfaceinvariance} that rank-one density matrices are fixed points of the dynamics.
 Specifically, for a fixed, randomly generated game instance (i.e. a $4\times4$ Hermitian $R\succ 0$), we run \ref{eqn:_DQREP}
 % \ref{lin-MMWU} with $\step= +\infty$ 
 on 100 randomly generated $\mathcal{H}_2 \otimes \mathcal{H}_2$ games with uniform initialization for both players and visualize them on the Bloch sphere (Figure \ref{fig:blochrandgame}), which is a standard technique for visualizing $2\times2$ density matrices and achieved using the QuTiP package \cite{johansson2012qutip}. 

\begin{figure*}[!htb]
    \centering
    \includegraphics[width=0.9\linewidth]{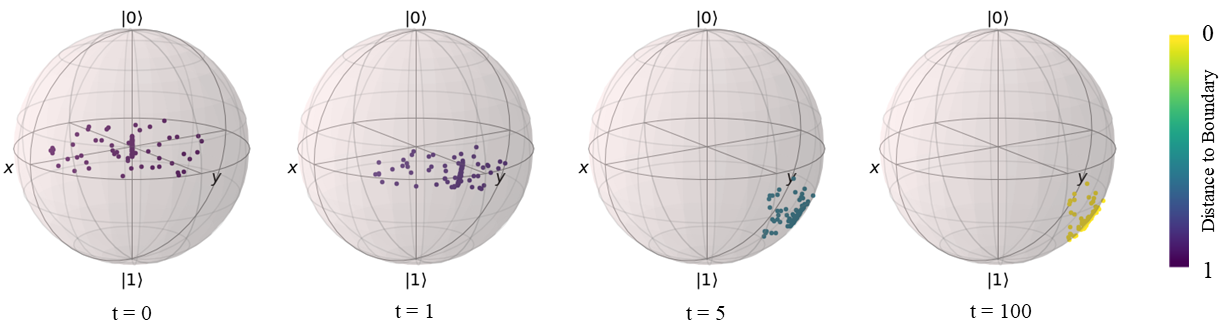}
    \caption{\ref{eqn:_DQREP} (\ref{lin-MMWU} with $\step= +\infty$) trajectories going to the boundary of the Bloch sphere in a fixed game instance with 100 randomized density initializations. 
    Points are color-coded based on distance to boundary, with yellow denoting points that are close the the boundary.
    % The values of the distance from the boundary are normalized such that yellow only denotes points which are a small $\step$-value from the boundary.
    }
    \label{fig:blochfixedgame}
\end{figure*}

In addition to the specific game instance with random initializations in Figure \ref{fig:blochrandgame}, we also explore \ref{lin-MMWU} for a fixed game instance. In particular, we run \ref{eqn:_DQREP} (\ref{lin-MMWU} with $\step = +\infty$) on a fixed, randomly generated game with 100 randomly generated density initialization. The game generated is in $\mathcal{H}_2 \otimes \mathcal{H}_2$. We then observe a similar phenomenon as before, with \ref{lin-MMWU} generating trajectories that converge to the boundary of the Bloch sphere (Figure \ref{fig:blochfixedgame}) for each initialization.

 % (For a detailed explanation of the Bloch sphere visualization see Appendix~\ref{appsecs:bloch}.)  
 % We observe that the trajectories of \ref{eqn:_DQREP} converge to the boundary of the Bloch sphere. 
 Since rank-1 densities in the quantum CIG correspond to unit vectors in the biquadratic optimization problem over the product of unit spheres 
$\max\{(x\otimes y)^\dagger R(x\otimes y):\|x\|_2=1, {\|y\|_2=1}\}$, this means that \ref{eqn:_DQREP} can be interpreted as a learning algorithm for solving the biquadratic problem. 

\paragraph{Large-scale experiments.}
Next, we present larger-scale experiments on \ref{eqn:_DQREP} and \ref{lin-MMWU} which show that our results for convergence to fixed points still holds in systems of larger dimensions.

\begin{figure}[!htb]
    \centering
    \begin{minipage}{.45\linewidth}
      \centering
      \includegraphics[width=.95\linewidth]{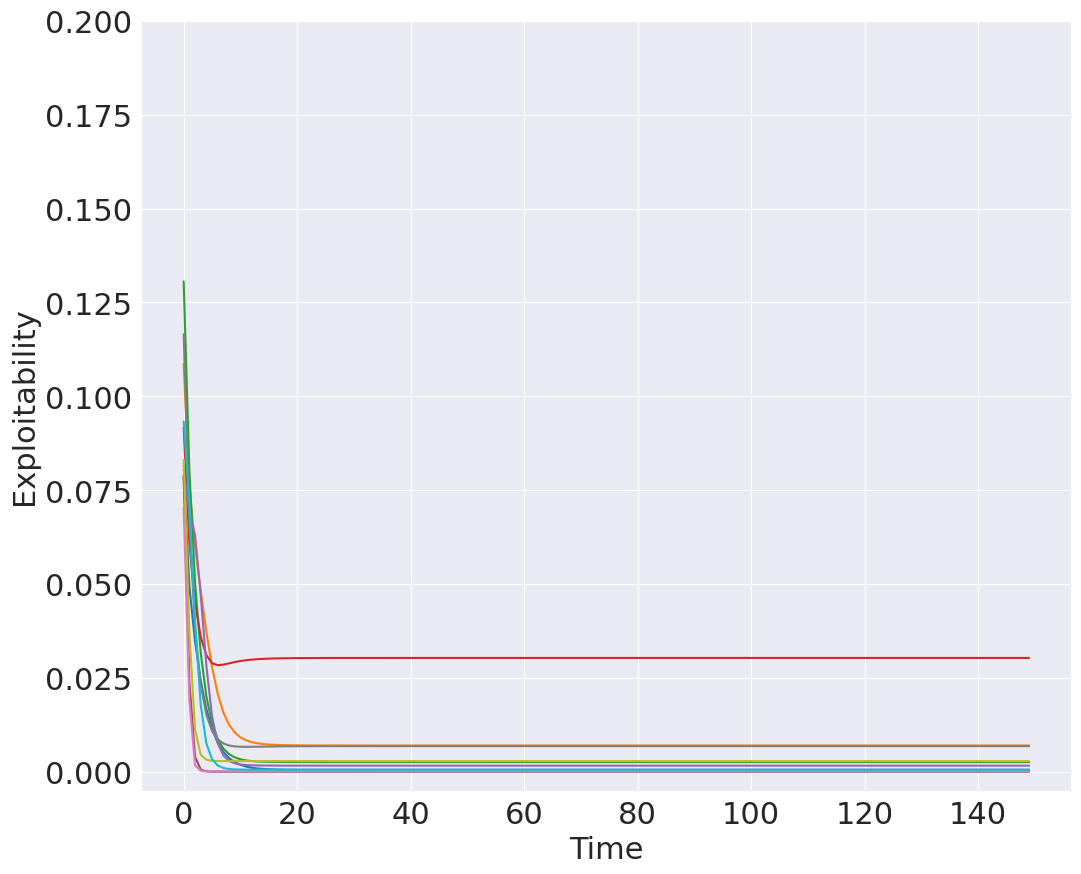}
      $\mathcal{H}_{10}\otimes\mathcal{H}_{10}$
    %   \label{fig:sub2}
    \end{minipage}
    \begin{minipage}{.45\linewidth}
      \centering
      \includegraphics[width=.95\linewidth]{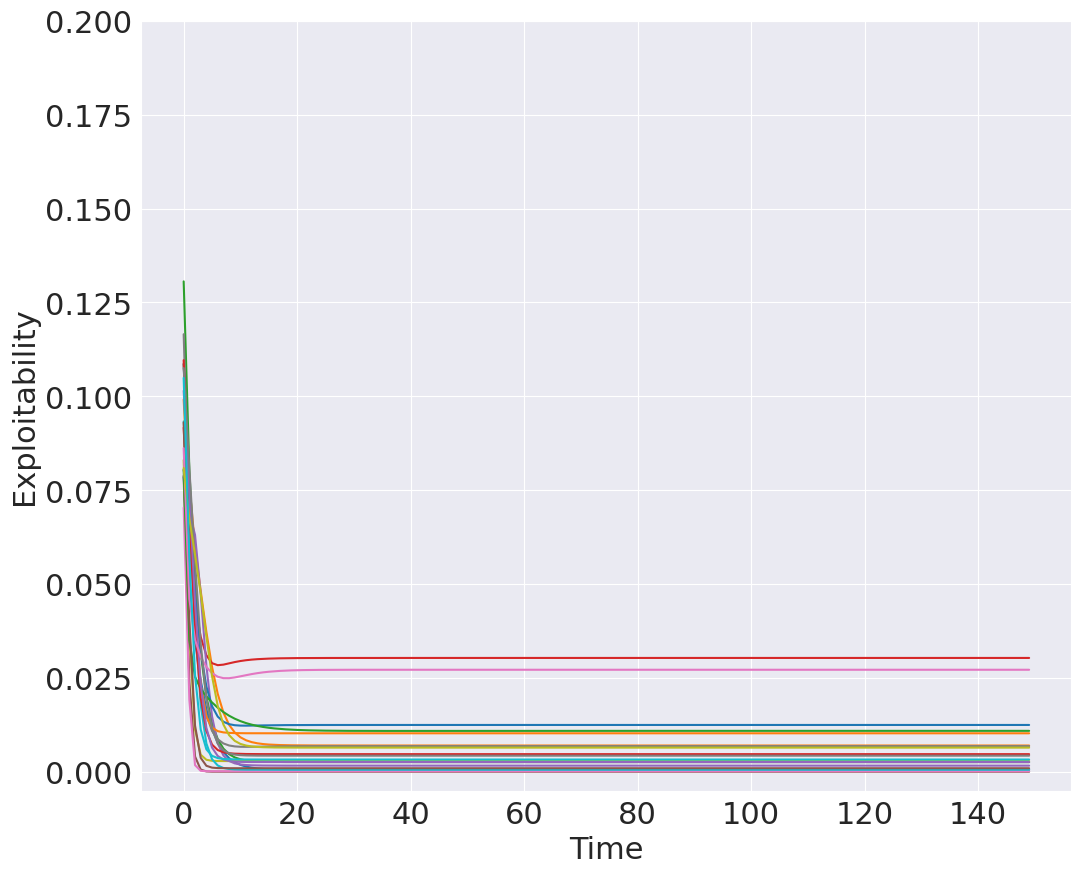}
      $\mathcal{H}_{20}\otimes\mathcal{H}_{20}$
    %   \label{fig:sub1}
    \end{minipage}
    \caption{Exploitability of \ref{eqn:_DQREP} (\ref{lin-MMWU} with $\step = +\infty$) when applied to 10 randomly generated common-interest games.}
    \label{fig:largescale}
\end{figure}

Thus far we have focused on problems of dimension $\mathcal{H}_2\otimes\mathcal{H}_2$ and $\mathcal{H}_2\otimes\mathcal{H}_3$ since we can efficiently compute the optimal value. In order to test the efficacy of \ref{eqn:_DQREP} for larger-scale problems, we run \ref{eqn:_DQREP} in randomly generated problems of size $\mathcal{H}_{10}\otimes\mathcal{H}_{10}$ and $\mathcal{H}_{20}\otimes\mathcal{H}_{20}$. In Figure \ref{fig:largescale} we see that the dynamics converge to fixed points, like in the smaller scale experiments.

\begin{figure}[!htb]
    \centering
    \begin{minipage}{.45\linewidth}
      \centering
      \includegraphics[width=.95\linewidth]{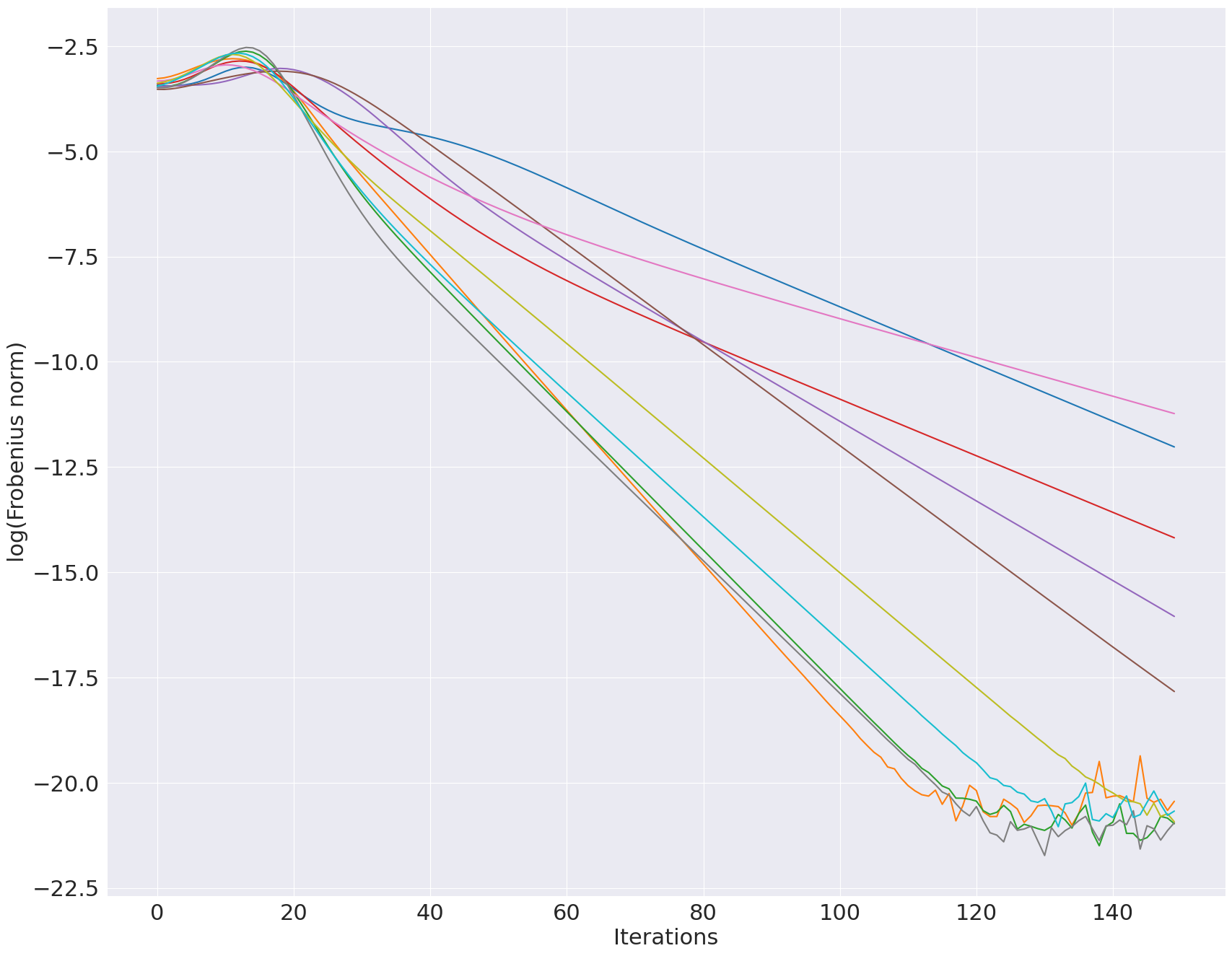}
      $\mathcal{H}_{10}\otimes\mathcal{H}_{10}$
    %   \label{fig:sub2}
    \end{minipage}
    \begin{minipage}{.45\linewidth}
      \centering
      \includegraphics[width=.95\linewidth]{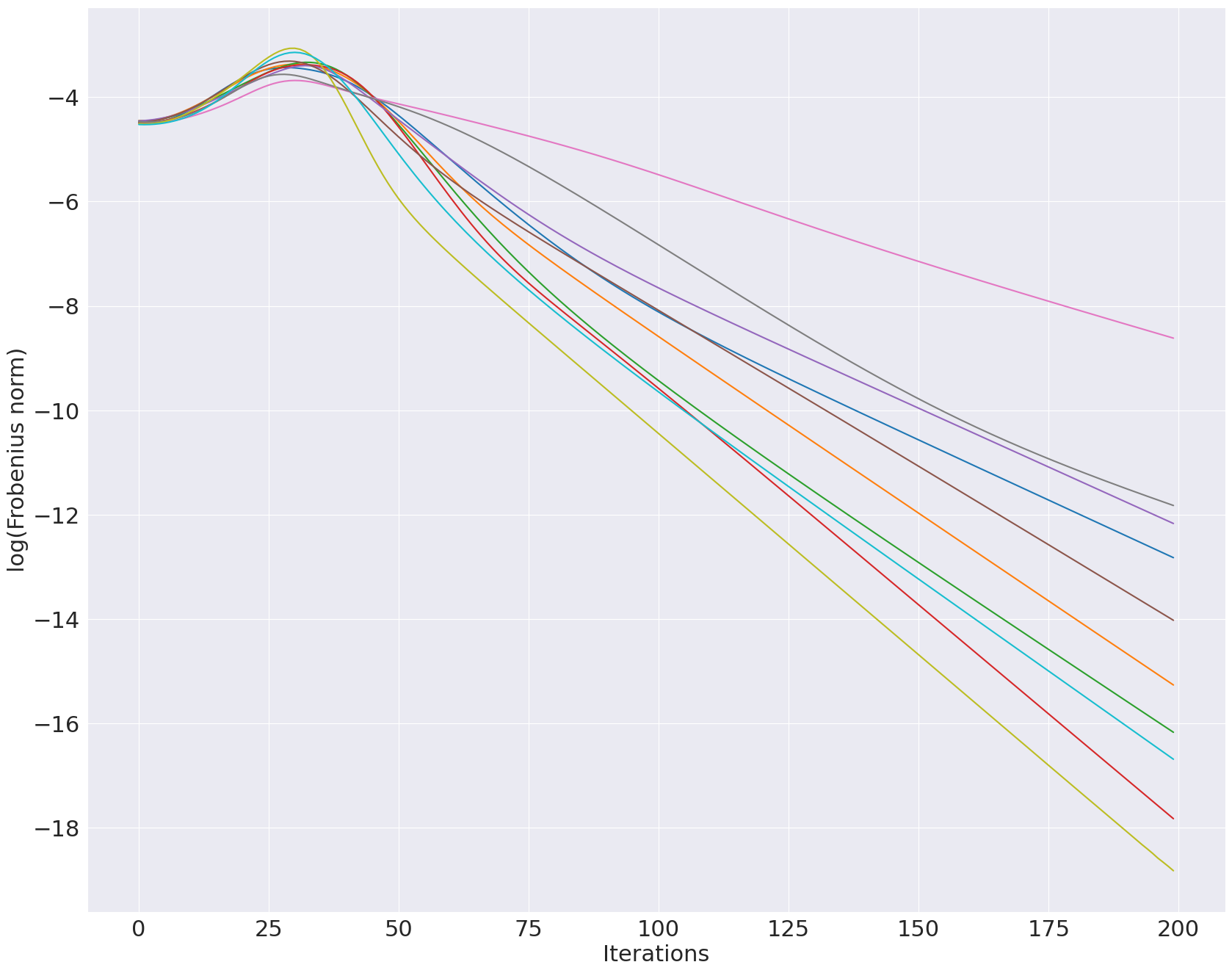}
      $\mathcal{H}_{20}\otimes\mathcal{H}_{20}$
    %   \label{fig:sub1}
    \end{minipage}
    \caption{Frobenius norm between \ref{eqn:_DQREP} (\ref{lin-MMWU} with $\step = +\infty$) at each time step and the next iterate.}
    \label{fig:largescalefrob}
\end{figure}

Despite the low exploitability, it is not a guarantee that the dynamics have fully stabilized. Hence, for each run of the simulation, we additionally visualize the Frobenius norm between the dynamics at each timestep and the next iterate of the dynamics. Notice that in Figure \ref{fig:largescalefrob}, the logarithm of the Frobenius norm generally decreases steadily over time, implying the dynamics stabilize and do not exhibit any oscillating behaviour.

\begin{figure}[!htb]
    \centering
    \begin{minipage}{.45\linewidth}
      \centering
      \includegraphics[width=.95\linewidth]{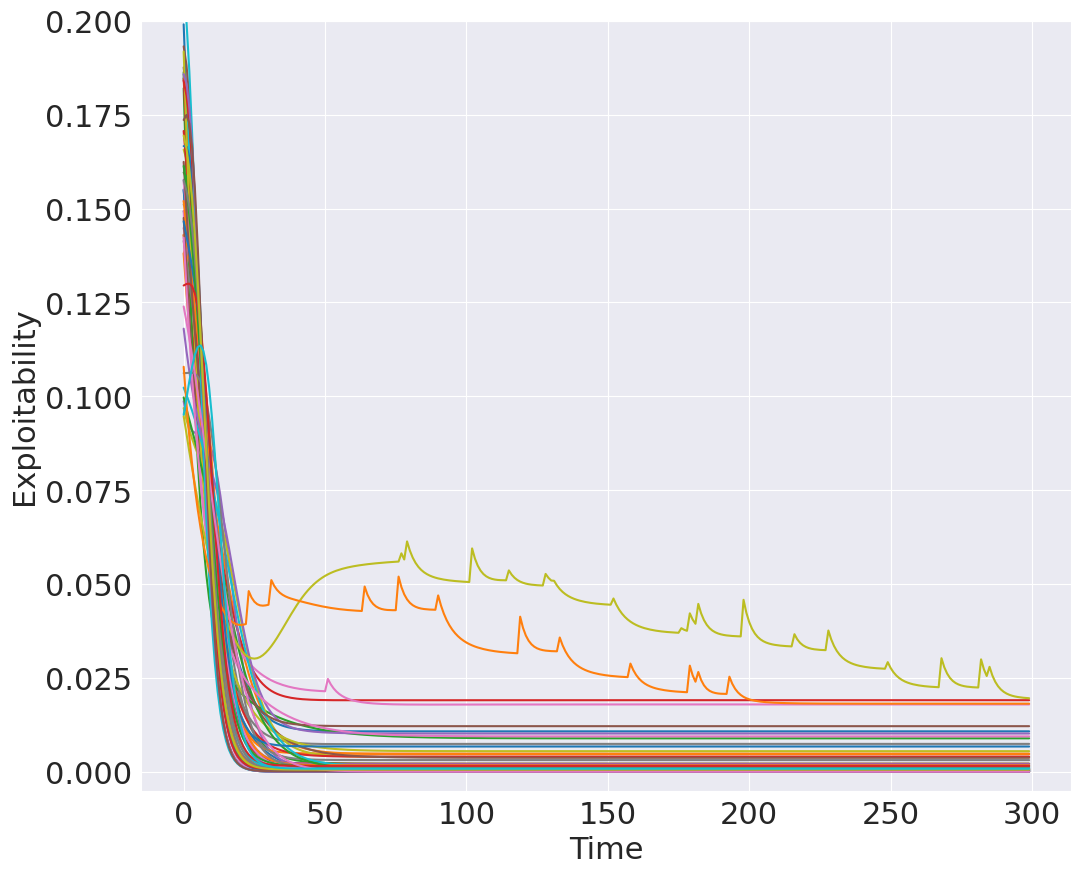}
      Random perturbation
    %   \label{fig:sub2}
    \end{minipage}
    \begin{minipage}{.45\linewidth}
      \centering
      \includegraphics[width=.95\linewidth]{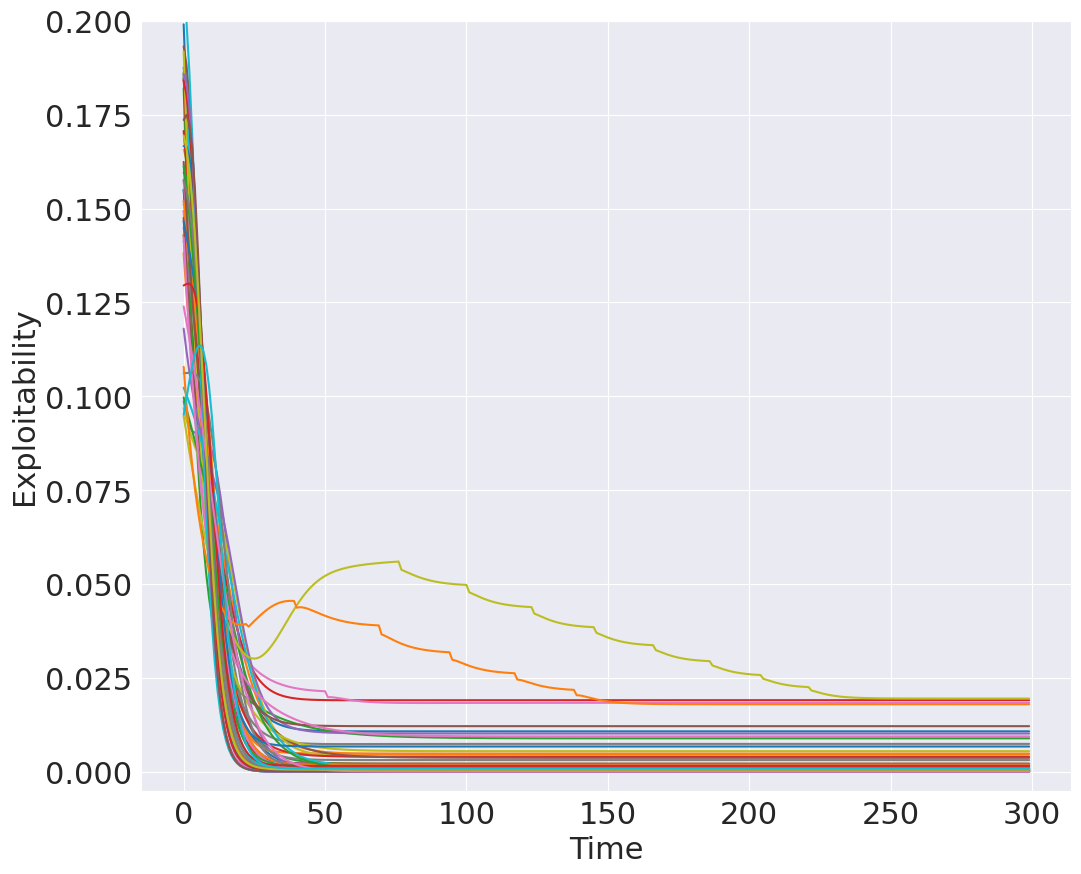}
      Best-response perturbation
    %   \label{fig:sub1}
    \end{minipage}
    \caption{Exploitability of \ref{lin-MMWU} ($\step = 0.9$) after perturbation. Both random perturbations and best-response perturbations can help the dynamics escape points of high exploitability. However, neither kind of perturbation is sufficient to bring the exploitability to zero.}
    \label{fig:perturb}
\end{figure}

\paragraph{Improving empirical performance using perturbation.} Notice that in Figure \ref{fig:exploitlinmmwu}, there are two game instances that perform poorly (i.e., high exploitability) for both \ref{eqn:_DQREP} and \ref{lin-MMWU}. This indicates that the dynamics converge to a sub-optimal fixed point. In order to improve performance, we introduce the following modification of both algorithms:
whenever a sub-optimal stationary point is reached, the players apply a random perturbation to their strategy and perform the subsequent update using this perturbed strategy. In principle, players may choose to perturb in the direction of their best response, but if the game matrix is unknown to players they can also perturb randomly, which suffices to drastically reduce exploitability (see Figure \ref{fig:perturb}). We also note that while the fixed points converged to in these two game examples are entirely different after perturbation, they are still rank-1 density matrices. We provide further examples and explanation of the perturbation in Appendix \ref{appsecs:additionalexperiments}.

\section{Conclusion}
% \textcolor{red}{Add something about BSS, unify table 5 with table 2.}
% extends the existing framework for learning in quantum games beyond the zero-sum setting and studies learning in quantum potential games. 

% This paper uses gamification to study the Best Separable State problem, and introduces several new algorithms that solve the problem approximately.
% In order to accomplish this, we introduce and study quantum common-interest games (CIGs) and establish and equivalence between the first-order stationary points of a BSS instance and the Nash equilibria of a corresponding quantum CIG. For learning in quantum CIGs we introduce non-commutative extensions of continuous and discrete dynamics used for learning in classical potential games, completing the picture of analogues of the classical replicator and multiplicative weight update learning dynamics (see Table \ref{table:quantumdy_incomplete}), and study their convergence properties. Our work establishes deep connections between (online) optimization theory (i.e. multiplicative weights update), traditional as well as evolutionary game theory (i.e. replicator dynamics), and quantum randomness/entanglement (i.e. density matrices).

In this paper, we introduce the class of quantum common-interest games (CIGs), which serve as cooperative counterparts to the quantum zero-sum game formulation studied in previous works. Toward studying the behavior of learning dynamics in quantum CIGs, we introduce non-commutative extensions of continuous and discrete dynamics used for learning in classical common-interest games, completing the picture of analogues of the classical replicator, best-response and multiplicative weight update learning dynamics (see Table \ref{table:quantumdy_incomplete}), and study their convergence properties. Along the way, we bridge game theory and optimization by establishing an equivalence between the first-order stationary points of an instance of the Best Separable State problem and the Nash equilibria of a corresponding quantum CIG, and experimentally show that our dynamics are able to converge close to the optimal solution of randomly generated BSS instances.
% In this work we introduce several new concepts which meaningfully extend our understanding of learning in quantum games. 
% We begin by formulating the class of quantum potential games and show that from the perspective of first-order learning dynamics, they are equivalent to quantum common-interest games. We then introduce a non-commutative generalization of $q$-replicator dynamics called \ref{eqn:_aQREP_family}. By setting $q=1$ we obtain the \ref{lin-QREP} dynamics, which is the quantum analogue of classical replicator dynamics. Discretizing this gradient flow, we recover the matrix analogue of the linear multiplicative weights update, which we call \ref{eqn:_DQREP}. The primary application which we explore in this paper is solving the \ref{BSS} problem. In particular, this discrete update rule gives a decentralized first-order iterative method for solving the \ref{BSS} problem, and empirically performs well, closely approximating the optimal value for several problem instances.
%

This work opens up several exciting new research directions, the first of which is to theoretically corroborate the experimental findings for convergence of \ref{lin-QREP} to Nash equilibria.
Another intriguing open question is to what extent and under what conditions can the \ref{lin-MMWU} dynamics be provably shown to converge to a rank-1 matrix for both players in quantum CIGs in analogy to the aforementioned classical results~\cite{kleinberg2009multiplicative, heliou2017learning,panageas2019multiplicative}. 
% More broadly, the ultimate goal of this line of research is to develop a general theory for learning in arbitrary quantum games. 
More broadly, our research contributes to the general theory for learning in quantum games by studying analogues of well-known classical dynamics in the important class of quantum common-interest games, exploring which results carry over, and understanding why some results can break down in the quantum setting.
Furthering the development of learning in general quantum games would require new notions of equilibration and convergence in quantum games, 
% echoing 
and the careful exploration of analogues of other
results from learning in classical games~\cite{BaileyEC18,cesa2006prediction,fudenberg1998theory, roughgarden2010algorithmic}.

% the study of learning dynamics such as \ref{eqn:_DQREP} in other quantum games such as quantum zero-sum games (beyond time-average convergence) is also a fascinating question with well-explored classical analogues~\cite{BaileyEC18}.

% Other directions for future research include exploring theoretically \ref{eqn:_DQREP} in the context of biquadratic optimization and applications thereof, as well as further studying of the properties of quantum potential games.

% \begin{table*}[ht]
% \centering
% \caption{Learning dynamics for quantum games, after this work. Each dynamic is a noncommutative generalization of its counterpart in Table \ref{table:classicaldy}.}
% \label{table:quantumdy_complete}
% \begin{tabular}{|c|c|c|}
% % \toprule
% \hline
% & Continuous-time & Discrete-time \\
% \hline
% Linear variant & \ref{lin-QREP} & 
% \begin{tabular}{c} \ref{eqn:_DQREP} \\ \ref{lin-MMWU} \end{tabular}
% % ref{BE}, \ref{linear} 
% \\
% \hline
% Exponential variant & \ref{exp-QREP} & \ref{MMWU} \\
% \hline
% % \multirow{2}{*}{\textbf{Problem Dimensions}}        
% % \bottomrule
% \end{tabular}
% \end{table*}

\section*{Acknowledgments}
This research is supported by the National Research Foundation, Singapore and DSO National Laboratories under its AI Singapore Program (AISG Award No: AISG2-RP-2020-016), grant PIESGP-AI-2020-01, AME Programmatic Fund (Grant No.A20H6b0151) from A*STAR and Provost’s Chair Professorship grant RGEPPV2101, the MOE Tier 2 Grant (MOE-T2EP20223-0018), the National Research Foundation, Singapore, under its QEP2.0 programme (NRF2021-QEP2-02-P05), the CQT++ Core Research Funding Grant (SUTD) (RS-NRCQT-00002) and partially by Project MIS 5154714 of the National Recovery and Resilience Plan, Greece 2.0, funded by the European Union under the NextGenerationEU Program. Wayne Lin and Ryann Sim acknowledge support from the SUTD President's Graduate Fellowship (SUTD-PGF). The authors also thank anonymous referees for insightful comments that greatly improved the presentation of the paper.
\bibliographystyle{quantum}
\bibliography{qrep_doi}

\newpage
\appendix
\onecolumn

\section{Linear Quantum q-Replicator Dynamics}\label{appsecs:gradient_flow}
\paragraph{Gradient flow dynamics.}
% \textcolor{purple}{To fill in details and motivate gradient flow better.}
While the implementation of learning in game theory often requires algorithms in discrete time, past work has shown that continuous-time dynamics can give rise to families of discrete dynamics. The most relevant such examples to our work are in the context of gradient-based optimization algorithms \cite{wibisono2016variational, nemirovskij1983problem} and evolutionary game dynamics \cite{mertikopoulos2018riemannian, shahshahani1979new}. Our definition of quantum replicator dynamics (\ref{lin-QREP}) can be generalized to a broader family of gradient flow dynamics which arise from the generalized family of Riemannian metrics on the PSD manifold, parametrized by $q\in\mathbb{R}$.
% In this section, we will derive a family of (continuous) gradient flow dynamics which arise from the generalized family of Riemannian metrics on the PSD manifold, effectively extending the prior work to the non-commutative setting and providing a family of continuous algorithms which can be discretized in various ways to obtain non-commutative versions of well known classical algorithms, such as the Baum-Eagon iterative process and Multiplicative Weights Update.

Consider a differentiable manifold $\M$ equipped with a differentiable scalar field $u : \M \rightarrow \mathbb{R}$ and a symmetric, positive-definite inner product $\innerprod{\cdot}{\cdot}_p : T_p \M \times T_p \M \rightarrow~\mathbb{R}_{\geq 0}$ defined at all $p \in \M$. (Here $T_p \M$ is the tangent space of $\M$ at $p$.) 
% At every point $p \in \M$ and for every $\xi \in T_p \M$, there exists a directional derivative $D_p u (\xi) = \dv{(u \circ \alpha)}{t}\vert_{t=0}$ for all differentiable curves $\alpha: (-\step, \step) \rightarrow \M$ satisfying $\alpha(0) = p$ and $\alpha'(0) = \xi$.
By the Riesz Representation Theorem (see, e.g., \cite{rudin1987real}), at each  $p \in \M$ there exists a \emph{unique} vector $\grads{} u(p) \in T_p \M$~with
\begin{equation}\label{gengrad}
    D_pu(\xi) = \innerprod{\grads{} u(p)}{\xi}_p \quad \fa \xi \in T_p \M,
\end{equation}
where $D_pu(\xi) : T_p \M \rightarrow \mathbb{R}$ is the directional derivative of $u$ at the point $p$ in direction $\xi$, i.e., %differential one-form of the scalar $V$
$D_pu(\xi)~=~\la \nabla u(p), \xi\ra$ where $ \nabla u(p)$ is the usual Euclidean gradient of $u$ at $p$ and $\la \cdot, \cdot \ra$ is the Euclidean inner product. Equation  \eqref{gengrad} allows us to associate to each point $p\in \M$ a vector $\grads{}u(p)\in~T_p~\M$, or in other words, to define a gradient flow on the manifold $\M$ given explicitly by  $  \dot{p} = \grads{} u(p)$.
Moreover, simply by construction, it follows that the function $u(p)$ is nondecreasing along the trajectories of the gradient flow, i.e., $\dv{u(p)}{t}\ge 0$ since $\dv{u(p)}{t}
% =\sum_i {\partial u\over \partial p_i}{d p_i\over dt} 
=\la \nabla u(p), \dot{p}\ra
=D_pu(\dot{p})=\innerprod{\grads{} u(p)}{\dot{p}}_p = \innerprod{\dot{p}}{\dot{p}}_p \ge0,$
% \begin{align*}
% \dv{u(p)}{t}=\sum_i {\partial u\over \partial p_i}{d p_i\over dt} =\la \nabla u(p), \dot{p}\ra
% &=D_pu(\dot{p})\\
% &=\innerprod{\grads{} u(p)}{\dot{p}}_p \\
% &= \innerprod{\dot{p}}{\dot{p}}_p \ge0,
% \end{align*}
and moreover $\dv{u(p)}{t}=0$ if and only if $\dot{p}=0$ (as the inner product $\innerprod{\cdot}{\cdot}_p $ is positive definite), so $u$ is in fact strictly increasing along gradient flow trajectories unless at a fixed~point.

\paragraph{Quantum Shahshahani gradient flow.}
% >>>>>>> main
\label{subsec:_GradFlow_IntrinsicMetric}
Consider a two-player quantum CIG  with common utility 
$
u(\rho, \sigma)= \la\rho, \Phi(\sigma) \ra.
$
% <<<<<<< overleaf-2022-05-06-0726
% Our goal is to provide a dynamic that improves the expected utility $V(\rho,\sigma)$. The state space we are operating in is the manifold $M \coloneqq \mathcal{D}^n \times \mathcal{D}^m$, so all that remains is to select a metric on the manifold of density matrices which would imbue the product manifold $M$ with the product metric, giving a gradient flow.
% % by Theorem \ref{thm:potential_increasing_grad_flow}.
% =======
Our goal is to provide continuous-time dynamics that improve the utility $u(\rho,\sigma)$. The state space we are operating in is the manifold
$\M~=~\da~\times~\db$,
%$\M = \mathcal{D}^n \times \mathcal{D}^m$,
so all that remains is to select a metric on the manifold of density matrices which would imbue the product manifold $\M$ with the product metric, giving a gradient flow.
% by Theorem \ref{thm:potential_increasing_grad_flow}.
% >>>>>>> main
To accomplish this, we consider the generalized family of Riemannian metrics on the manifold of  PSD matrices parametrized by $q \in \mathbb{R}$, which we call the \emph{quantum $q$-Shahshahani metric}:
\begin{equation}
\label{cor:_a_quantShahq}\tag{QShah}
    \innerprod{A}{B}_\rho^{(q)} := \Tr[\rho^{-\frac{q}{2}}A \rho^{-\frac{q}{2}}B].
\end{equation}
Indeed, in the case of diagonal matrices this family of metrics reduces to the $q$-Shahshahani family of metrics on the simplex (see e.g. \cite{mertikopoulos2018riemannian}).  On the PSD manifold, $q = 0$ gives the Euclidean inner product $\Tr[AB]$, while $q = 2$ gives the intrinsic Riemannian metric (see e.g. \cite{bhatia2009positive}). In addition, $q=1$ reduces to the Shahshahani metric on the simplex in the case of diagonal~matrices.

\begin{theorem}[{Linear quantum $q$-replicator dynamics}]
\label{thm:_a_quantShahGeneral}
Consider a quantum CIG with utility function    $u(\rho,\sigma)=\la \rho, \Phi(\sigma) \ra$ where $\rho \in D(\A), \sigma \in D(\B)$. The dynamics
\begin{equation} %
\begin{gathered} \label{eqn:_aQREP_family} \tag{\rm{lin-QREP$_q$}}
    \dv{\rho}{t}  = \rho^\frac{q}{2} \left[\Phi(\sigma) - \frac{\Tr[\rho^q \Phi(\sigma)]}{\Tr[\rho^q]}\id_\A \right] \rho^\frac{q}{2}, \\
    \dv{\sigma}{t}  =  \sigma^\frac{q}{2} \left[\Phi^\dagger (\rho) -\frac{\Tr[\sigma^q \Phi^\dagger(\rho)]}{\Tr[\sigma^q]}\id_\B \right] \sigma^\frac{q}{2}
\end{gathered}
\end{equation}
% which we call the \emph{$q$-quantum replicator dynamics},
define a gradient flow of the utility function $u(\rho, \sigma)$ on the product manifold $D(\A)\times D(\B)$ imbued  with the quantum $q$-Shahshahani metric. Moreover, the utility  $u(\rho,\sigma)$ is strictly increasing  along the trajectories of the \ref{eqn:_aQREP_family} dynamics, except at fixed points.
\end{theorem}

\begin{proof}
To derive the \ref{eqn:_aQREP_family} dynamics as a gradient flow with respect to the quantum $q$-Shahshahani metric $\innerprod{A}{B}_\rho^{(q)} := \Tr[\rho^{-\frac{q}{2}}A \rho^{-\frac{q}{2}}B]$ given in \eqref{cor:_a_quantShahq}, 
we operate on the product manifold $\M = \da \times \db$ endowed with the product metric and use the scalar field    $u: \M \rightarrow \mathbb{R}, \; (\rho, \sigma) \mapsto \innerprod{\rho}{\Phi(\sigma)}$ corresponding to  the common utility function. At any point  $(\rho, \sigma) \in \M$, we want to find ${\bf grad}^{(q)} u(\rho,\sigma)$, defined as  the unique vector $g = (g_\A, g_\B) \in T_{(\rho, \sigma)} \M = T_\rho D(\mathcal{A}) \times T_\sigma D(\mathcal{B})$ satisfying 
$$  D_{(\rho, \sigma)} u(\xi_\A, \xi_\B)
        = \langle(g_\A,g_\B), (\xi_\A, \xi_\B)\rangle^{(q)}_{(\rho, \sigma)}, \  \text{ for all } \xi = (\xi_\A, \xi_\B) \in T_{(\rho, \sigma)} \M.$$
Expanding the above we immediately get that 
\begin{equation}
\label{eqn:_gradflow_IntrinsicMetric}
        D_{(\rho, \sigma)} u(\xi_\A, \xi_\B)        = \langle g_\A, \xi_\A\rangle^{(q)}_\rho + \langle g_\B, \xi_\B\rangle^{(q)}_\sigma 
        = \Tr[{\rho}^{-\frac{q}{2}}{g_\A} {\rho}^{-\frac{q}{2}} {\xi_\A}] +
        \Tr[{\sigma}^{-\frac{q}{2}}{g_\B}{\sigma}^{-\frac{q}{2}}{\xi_\B}],
\end{equation}

while on the other hand, as the Euclidean gradient $ \nabla u (\rho,\sigma)= \left(\Phi(\sigma), \Phi^\dagger(\rho) \right)$, we have that

\begin{equation}
\label{eqn:_gradflow_EuclideanMetric}
  D_{(\rho, \sigma)} u(\xi_\A, \xi_\B) =\la \nabla u(\rho, \sigma),(\xi_\A, \xi_\B)\ra=
     \Tr[\Phi(\sigma)\xi_\A] + \Tr[\Phi^\dagger(\rho) \xi_\B].
\end{equation}

Equating (\ref{eqn:_gradflow_IntrinsicMetric}) and (\ref{eqn:_gradflow_EuclideanMetric}), we then have that 
${\bf grad}^{(q)} u(\rho,\sigma)$
is the unique element $(g_\A, g_\B) $ in the product of the tangent spaces  $T_\rho D(\mathcal{A}) \times T_\sigma D(\mathcal{B})$ with the following properties: 
\begin{itemize}
    \item $g_\A$ is the unique element in $T_\rho D(\mathcal{A})$ such that
$$
          \Tr[{\rho}^{-\frac{q}{2}}{g_\A} {\rho}^{-\frac{q}{2}} {\xi_\A}]= \Tr[\Phi(\sigma)\xi_\A] \quad \fa \xi_\A \in T_\rho D(\mathcal{A}).$$
    \item $g_\B$ is the unique element in $T_\sigma D(\mathcal{B})$ such that
    $$ \Tr[{\sigma}^{-\frac{q}{2}}{g_\B}{\sigma}^{-\frac{q}{2}}{\xi_\B}]=  \Tr[\Phi^\dagger(\rho)] \xi_\B  \quad \fa \xi_\B \in T_\sigma D(\mathcal{B}).
              $$
\end{itemize}
A straightforward computation shows that for any constant $c$ we have
$$\begin{aligned}
\Tr[\Phi(\sigma)\xi_\A]&= \Tr[(\Phi(\sigma)-c\mathbb{1}_\A)\xi_\A]\\
&=\Tr[{\rho}^{-\frac{q}{2}}\underbrace{({\rho}^{\frac{q}{2}}(\Phi(\sigma)-c\mathbb{1}_\A){\rho}^{\frac{q}{2}})}_{g_\A}{\rho}^{-\frac{q}{2}}\xi_\A]
\end{aligned}$$
where for the first equality we used that tangent space of $T_\rho D(\mathcal{A})$ consists of traceless matrices. Lastly, to make $g_\A$ traceless we need to select the constant  $c$ so that 
$$\Tr({\rho}^{\frac{q}{2}}(\Phi(\sigma)-c\mathbb{1}_\A){\rho}^{\frac{q}{2}})=0 \iff c=  \frac{\Tr[\rho^{q} \Phi(\sigma)]}{\Tr[\rho^{q}]}.$$
Summarizing, we have established that 
$$g_\A=\rho^{\frac{q}{2}} \left[\Phi(\sigma) - \frac{\Tr[\rho^{q} \Phi(\sigma)]}{\Tr[\rho^{q}]}I \right] \rho^{\frac{q}{2}}$$
and symmetrically we also get that
$$g_\B = \sigma^{\frac{q}{2}} \left[\Phi^\dagger (\rho) -\frac{\Tr[\sigma^{q} \Phi^\dagger(\rho)]}{\Tr[\sigma^{q}]}I \right] \sigma^{\frac{q}{2}}.$$

Thus, 
the gradient flow on the product manifold $D(\A)\times D(\B)$ endowed  with the quantum $q$-Shahshahani metric
 is given by
\begin{equation*}
        \dv{\rho}{t}  = g_\A = \rho^{\frac{q}{2}} \left[\Phi(\sigma) - \frac{\Tr[\rho^{q} \Phi(\sigma)]}{\Tr[\rho^{q}]}I \right] \rho^{\frac{q}{2}}, \quad
        \dv{\sigma}{t}  = g_\B = \sigma^{\frac{q}{2}} \left[\Phi^\dagger (\rho) -\frac{\Tr[\sigma^{q} \Phi^\dagger(\rho)]}{\Tr[\sigma^{q}]}I \right] \sigma^{\frac{q}{2}}.
\end{equation*}
That the utility $u = \innerprod{\rho}{\Phi(\sigma)}$ is strictly increasing unless at fixed points follows directly from the fact that this dynamic is a gradient flow. 
\end{proof}

% The proof that \ref{eqn:_aQREP_family}  is a gradient flow is given in Appendix \ref{sec:_GradFlowDerivation}.

In terms of the convergence properties of \ref{eqn:_aQREP_family} we have the following result:

%for which $u = \innerprod{\rho}{\Phi(\sigma)}$ is non-decreasing along trajectories. $u=\la\rho, \Phi(\sigma)\ra$ can be viewed as a Lyapunov function for dynamics (\ref{eqn:_aQREP_family}), an explicit computation of which is given in Appendix \ref{appsecs:explicit}.

\begin{corollary}\label{cor:omegalimits}
    The set of $\omega$-limit points of a trajectory $\{\rho(t), \sigma(t)\}_{t\ge 0}$ of the \ref{eqn:_aQREP_family} dynamics is a compact, connected set of fixed points of the dynamics that all attain the same utility.
\end{corollary}

%\begin{lemma}\label{omegalimitsarerest}
%    If a dynamic admits a Lyapunov function, then the set of $\omega$-limit points of the dynamic is a compact, connected set of rest points of the dynamic, on which the Lyapunov function has the same value.
%\end{lemma}
The proof of this result follows directly from  an extension of the fundamental convergence theorem by \cite{losert1983dynamics} to general compact sets,
%relies on the fact that the utility $u(\rho,\sigma)= \innerprod{\rho}{\Phi(\sigma})$ is a Lyapunov function of the dynamics and $\density{n} \times \density{m}$ is compact, this is a direct application
which we prove in   Theorem \ref{thm:_LimitSetCompactConnected_ContTime}.

\section{Forward invariance of \ref{lin-QREP}}
\label{sec:_QREP_props}

% \color{purple}
%     TODO: add references, name facts appropriately.
% \color{black}

In this section we 
show that time derivative of the state $(\rho_0, \sigma_0)$ under the quantum replicator dynamics (\ref{lin-QREP}) is always in the product of the tangent cones of $\rho_0 \in \da$ and $\sigma_0 \in \db$, i.e., it does not point outside of the state space. To do this we first recall the notion of the cone of feasible directions and the tangent cone:

Given a closed convex set $C$ in a Hilbert space and a point $x \in C$, the cone of feasible directions at $x$ is given by
\[
    \feasdir_C(x) = 
    \{ 
        d : x + td \in C \text{ for some } t > 0
    \}
\]
and the tangent cone $\tcone_C(x)$ at $x$ (see, e.g, \cite{rockafellar1979clarke}) is given by the closure of the cone of feasible directions, i.e.,
\[
    \tcone_C(x) = \overline{\feasdir_C(x)}.
\]

We shall characterize the tangent cones of points in the set of density matrices using the fact that the tangent cone is the polar to the normal cone. We first make the following observation characterizing the normal cones to the PSD cone:

\begin{lemma}
\label{lem:_normalcone_PSD}
The normal cone to a matrix $X$ in the PSD cone, $N_{\psdcone} (X) \equiv \{Z: \innerprod{Z}{Y-X} \leq 0 \; \forall \, Y \succeq 0\}$, is equal to the set $\{Z \preceq 0: \innerprod{Z}{X} = 0\}$.
\end{lemma}

\begin{proof}
    If $Z \preceq 0$ and $\innerprod{Z}{X} = 0$, then $ \forall \, Y \succeq 0$ we have that $\innerprod{Z}{Y-X} = \innerprod{Z}{Y} - \innerprod{Z}{Y} = \innerprod{Z}{X} \leq 0$, so $Z \in N_{\psdcone} (X)$.
    
    On the other hand, if $Z \in N_{\psdcone} (X)$ so that $\innerprod{Z}{Y - X} \leq 0 \; \forall \, Y \succeq 0$, then we can in particular consider the PSD matrix $Y := X + v v^\dagger$ for any vector $v$. Then $v^\dagger Z v = \innerprod{Z}{v v^\dagger} = \innerprod{Z}{Y-X} \leq 0 \; \forall \, v$, so we have that $Z \preceq 0$. It then follows that $\innerprod{Z}{X} \leq 0$. However, taking $Y := \frac{1}{2} X$ in the definition of the normal cone, we have that $\frac{1}{2} \innerprod{Z}{X} = - \innerprod{Z}{- \frac{1}{2} X} = - \innerprod{Z}{Y-X} \geq 0$. Thus $\innerprod{Z}{X} = 0$.
\end{proof}

This gives us the following characterization of the tangent cones to the PSD cone:

\begin{lemma}
\label{lem:_TCone}
    $\tcone_{\da} (\rho) = \{W: \tr(W) = 0, \; u^\dagger W u \geq 0 \; \forall u \in \ker \rho \}$.
\end{lemma}

\begin{proof}
    Given a closed convex set $C$ and a point $x \in C$, the tangent cone at $x$ is the polar of the normal cone at $x$, i.e.  $
        \tcone_C(x) = N_C(x)^\circ
    $
    where the normal cone at $x$ is given by $
        N_C(x) \equiv \{z: \innerprod{z}{y - x} \leq 0 \; \forall y \in C \}
    $, and the polar of a set $S$ is given by $
    S^\circ =
        \{
            w: \innerprod{w}{z} \leq 0 \; \forall z \in S
        \}
    $ 
    (see, e.g., \cite{rockafellar1979clarke})
    . 
    
    For a Hermitian matrix $X$ in the PSD cone, we have from Lemma \ref{lem:_normalcone_PSD} that the normal cone at $X$ is given by
    \begin{align*}
        N_{\psdcone} (X) =&
        \{
            Z : \innerprod{Z}{Y - X} \leq 0 \; \forall \, Y \succeq 0
        \}\\
        =&
        \{
            Z \preceq 0: \innerprod{Z}{X} = 0
        \},
    \end{align*}

    and so the tangent cone at $X$ is given by
    \[
        \tcone_{\psdcone} (X) = N_{\psdcone} (X) ^\circ = \{ W : u^\dagger W u \geq 0 \; \forall u \in \ker X \} .
    \]

    Finally, we have that
    \begin{equation*}
    \begin{split}
            &\tcone_{\da} (X)\\ 
            =& \tcone_{\psdcone} (X) \cap \tcone_{\tr= 1} (X) \\
            =& \{ W : u^\dagger W u \geq 0 \; \forall u \in \ker X \} \cap (\tr=0) \\
            =& \{ W : \tr W = 0, \; u^\dagger W u \geq 0 \; \forall u \in \ker X \} .
    \end{split}
    \end{equation*}
\end{proof}

With this characterization of the tangent cone in hand, we are now ready to prove the theorem:

\begin{theorem}
    At any point $(\rho, \sigma) \in \da \times \db$, the time derivatives $\dot{\rho}
        = \rho^{\sfrac{1}{2}} \Big[\Phi(\sigma) - \innerprod{\rho}{\Phi(\sigma)} \id_\A \Big] \rho^{\sfrac{1}{2}}$, $\dot{\sigma}
        = \sigma^{\sfrac{1}{2}} \Big[\Phi^\dagger (\rho) -\innerprod{\rho}{\Phi(\sigma)} \id_\B \Big] \sigma^{\sfrac{1}{2}}$ given by \ref{lin-QREP} lie in the tangent cones $\tcone_{\da}(\rho)$, $\tcone_{\db}(\sigma)$ respectively.
\end{theorem}

\begin{proof}
    Firstly, the time derivative $\dot{\rho}$ is traceless since 
    \begin{align*}
        &\Tr( \rho^{\sfrac{1}{2}} \Big[\Phi(\sigma) - \innerprod{\rho}{\Phi(\sigma)} \id_\A \Big] \rho^{\sfrac{1}{2}}) \\
        =& \langle\rho, \Phi(\sigma)\rangle - \Tr(\rho \langle\rho, \Phi(\sigma)\rangle )\\
        =& \langle\rho, \Phi(\sigma)\rangle  -  \innerprod{\rho}{\Phi(\sigma)} \Tr (\rho )
        = 0.
    \end{align*}

    Furthermore, $\forall \, u \in \ker \rho$, we have that $u \in \ker \powh{\rho}$ and so $u^\dagger \dot{\rho} u = 0$.

    Thus $\dot{\rho} \in \tcone_{\da}(\rho)$ by Lemma \ref{lem:_TCone}, and similarly we have that $\dot{\sigma} \in \tcone_{\db}(\sigma)$.
\end{proof}

\section{Auxiliary Theorems and Lemmas}\label{appsecs:theorems}

The Theorems \ref{thm:_LimitSetCompactConnected_ContTime} and \ref{thm:_LimitSetCompactConnected_DiscreteTime} proven in this section are direct generalizations of a fundamental convergence theorem by Losert and Akin (Proposition 1 in \cite{losert1983dynamics}, which was written only for the simplex) to general compact sets. Nevertheless, the proof employed by Losert and Akin in \cite{losert1983dynamics} only really required compactness (i.e., it made use of no other properties of the simplex), and thus could actually be taken wholesale to prove Theorems \ref{thm:_LimitSetCompactConnected_ContTime} and \ref{thm:_LimitSetCompactConnected_DiscreteTime}. We rewrite the theorem statements and proofs here for the sake of clarity and completeness. 
% Corollaries \ref{cor:omegalimits} (in Section \ref{sec:contdynamics}) and  \ref{thm:_hofsigUpdatesAlternating_limitpoints_ClosedConnectedFixedPts} (in Section \ref{sec:discdynamics}) are direct applications of Theorems \ref{thm:_LimitSetCompactConnected_ContTime} and \ref{thm:_LimitSetCompactConnected_DiscreteTime} respectively to the compact set $\da \times \db$.

The notation used here is standard for dynamical systems. $x(t)$ denotes the point that $x$ evolves to after time $t$ has elapsed (or $t$ iterations have passed, in the case of discrete-time dynamical systems). The limit set $\Omega(x)$\footnote{The limit set $\Omega(x)$ depends on the initial condition $x$, but to simplify notation we shall drop the dependence on $x$ in the notation and just write $\Omega$ when the choice of initial condition is unambiguous.} of an orbit $\{x(t)\}_{t \geq 0}$ (or $\{x(t)\}_{t \in \mathbb{N}}$, in the case of discrete-time systems) refers to the set of $\omega$-limits of the orbit, i.e. the set of points $\omega$ for which there exists an increasing sequence $\{t_k\}_{k \in \mathbb{N}}$ that converges to infinity and satisfies $\lim_{k \rightarrow \infty} x(t_k) = \omega$.

\begin{theorem}
\label{thm:_LimitSetCompactConnected_ContTime}
    Consider a continuous-time dynamical system on a compact set $\mathcal{X}$ 
    %$\subseteq \mathbb{C}^n$,
    in a metric space,
    obtained through the differential equation $\dot{x}(t) = F(x(t))$ where $F$ is a continuous function on $\mathcal{X}$. Suppose also that the dynamical system admits a Lyapunov function $u$ (i.e., a  function $u: \mathcal{X} \rightarrow \mathbb{R}$ such that $\fa x \in \mathcal{X}$, $\dot{u}(x) \geq 0$ with equality iff $F(x) = 0$).

    Then the limit set $\Omega$ of an orbit $\{x(t)\}_{t \geq 0}$
    \begin{itemize}
        \item is a compact connected set,
        \item consists entirely of fixed points (i.e., points $x$ for which $F(x) = 0$), and
        \item has the property that the Lyapunov function $u$ is constant over it.\footnote{i.e., $u(\omega) = u(\omega')$ for any $\omega, \omega' \in \Omega$.}
    \end{itemize}
\end{theorem}

\begin{proof}
	\textbf{$u$ is constant over $\Omega$.} First note that $u$ is continuous and $\mathcal{X}$ is compact, so $u(\mathcal{X})$ is bounded. Thus, since $u$ is non-decreasing along orbits, $\{u(x(t))\}_{t\geq 0}$ converges to $u^* := \sup\{u(x(t))\}_{t \geq 0} < \infty.$ \\
    Now consider any $\omega \in \Omega$. There exists a sequence $\{x(t_k)\}_{k \in \mathbb{N}}$ (with $t_k \rightarrow \infty$ as $k \rightarrow \infty$) that converges to $\omega$. $\{u(x(t_k))\}_k$ also converges to $\sup \{u(x(t_k))\}_k = \sup \{u(x(t))\}_{t \geq 0} = u^*$. Thus, since $u$ is continuous, we have that $u^* = \lim_{k \rightarrow \infty} u(x(t_k)) = u (\lim_{k \rightarrow \infty} x(t_k)) = u(\omega)$. \\
    Thus $\fa \omega \in \Omega$, $u(\omega) =  u^*$.

    \textbf{$\Omega$ consists entirely of fixed points.} %\\
    Consider any $\omega = \lim_{k \rightarrow \infty} x(t_k) \in \Omega$, where $\{t_k\}_{k \in \mathbb{N}}$ is an increasing sequence that converges to infinity.
    $\fa s \geq 0$,
    \[
    	\omega (s) = \left(\lim_{k \rightarrow \infty} x(t_k) \right)(s) = \lim_{k \rightarrow \infty} x(t_k + s) \in \Omega.
    \]
    Thus from the already-proven fact that $u$ obtains the same value over $\Omega$, we have that $u(\omega(s)) = u(\omega) \; \fa s \geq 0$. Thus $\omega$ is a fixed point of the dynamics (by the assumption that $u$ is a Lyapunov function, i.e., that $u$ is strictly increasing except at fixed points).

     \textbf{$\Omega$ is compact and connected.} %\\
     $\Omega$ can be written as
     \[
     \Omega = \bigcap_{t\in \mathbb{R}_{\geq 0}} \overline{\{x(s): s \geq t\}}
     \]
     where $\overline{\{x(s): s \geq t\}}$, which denotes the closure of the set $\{x(s): s \geq t\}$, is compact (since it is a closed subset of the compact set $\mathcal{X}$) and connected (since it is the closure of the image of the connected set $[t, \infty)$ under a continuous mapping) for all $t \in \mathbb{R}_{\geq 0}$.
     Thus $\Omega$ is the decreasing intersection of compact, connected sets, and is hence itself compact and connected.
\end{proof}

\begin{theorem}
\label{thm:_LimitSetCompactConnected_DiscreteTime}
    Consider a discrete-time dynamical system on a compact
    % \footnote{Must we say w.r.t to the norm topology?}  
    set $\mathcal{X}$ 
    %$\subseteq \mathbb{C}^n$,
    in a metric space, obtained through the update $x(t+1) = F(x(t))$ where $F: \mathcal{X} \rightarrow \mathcal{X}$ is a continuous function. Suppose also that the dynamical system admits a Lyapunov function $u$ (i.e., a continuous function $u: \mathcal{X} \rightarrow \mathbb{R}$ such that $\fa x \in \mathcal{X}$, $u(F(x)) \geq u(x)$ with equality iff $F(x) = x$).

    Then the limit set $\Omega$ of an orbit $\{x(t)\}_{t \geq 0}$
    \begin{itemize}
        \item is a compact connected set,
        \item consists entirely of fixed points (i.e., points $x$ for which $F(x) = x$), and
        \item has the property that the Lyapunov function $u$ is constant over it.\footnote{i.e., $u(\omega) = u(\omega')$ for any $\omega, \omega' \in \Omega$.}
    \end{itemize}
\end{theorem}

\begin{proof}
    \textbf{$u$ is constant over $\Omega$.} %\\
    First note that since $u$ is continuous and $\mathcal{X}$ is compact, so $u(\mathcal{X})$ is bounded. Thus the non-decreasing sequence $\{u(x(t))\}_{t \in \mathbb{N}}$ converges to $u^* := \sup \{u(x(t))\}_{t \in \mathbb{N}} < \infty.$ \\
    Now consider any $\omega \in \Omega$. There exists a subsequence $\{x(t_k)\}_{k \in \mathbb{N}}$ of the sequence of iterates $\{x(t)\}_{t\in \mathbb{N}}$ that converges to $\omega$. $\{u(x(t_k))\}_k$ also converges to $\sup \{u(x(t_k))\}_k = \sup \{u(x(t))\}_t = u^*$. Thus, since $u$ is continuous, we have that $u^* = \lim_{k \rightarrow \infty} u(x(t_k)) = u (
    \lim_{k \rightarrow \infty} x(t_k)) = u(\omega)$. \\
    Thus $\fa \omega \in \Omega$, $u(\omega) =  u^*$.

    \textbf{$\Omega$ consists entirely of fixed points.} %\\
    Consider any $\omega \in \Omega$ and let $\omega = \lim_{k \rightarrow \infty} x(t_k)$, where $\{t_k\}_{k \in \mathbb{N}}$ is an increasing sequence that converges to infinity. Where $u^* := \sup \{u(x_{t})\}_t$ as previously defined, we have, by the continuity of $F$ and $u$, that
    \begin{align*}
       u(F(\omega))
    =& u(F(\lim_{k \rightarrow \infty} x(t_k)))\\
    =& \lim_{k \rightarrow \infty} u(F(x(t_k)))\\
    =& \lim_{k \rightarrow \infty} u(x(t_k +1))\\
    =& u^* = u(\omega), 
    \end{align*}

    so we must have $F(\omega) = \omega$ by the assumption that $u$ is a Lyapunov function.

    \textbf{$\Omega$ is compact. } %\\
    % If $\omega \in \overline{\Omega}$, then $\fa \step > 0$ $\exists \, z \in \Omega \cap B(\omega, \frac{\step}{2})$ for which $\exists$ infinity many members of $\{x(t)\}_t$ in $B(z, \frac{\step}{2}) \subseteq B(\omega, \step)$.
    % Thus $\omega$ is a limit point of $\{x(t)\}_t$, i.e. $\omega \in \Omega$. \\
    % Thus $\Omega$ is a closed subset of the compact set $\mathcal{X}$, and hence compact.
    $\Omega$ can be written as
     \[
     \Omega = \bigcap_{t\in \mathbb{N}} \overline{\{x(s): s\in \mathbb{N},s \geq t\}}
     \]
     where $\overline{\{x(s): s \in \mathbb{N}, s \geq t\}}$, which denotes the closure of the set $\{x(s): s\in \mathbb{N}, s \geq t\}$, is compact (since it is a closed subset of the compact set $\mathcal{X}$) for all $t \in \mathbb{N}$.
     Thus $\Omega$ is the decreasing intersection of compact sets, and is hence itself compact.

    \textbf{$\Omega$ is connected. } % \\
    Suppose to the contrary that $\Omega$ is the disjoint union of nonempty closed sets $\Omega_1$, $\Omega_2$. $\Omega_1$, $\Omega_2$ are closed subsets of the compact set $\mathcal{X}$ and hence compact, so they are a finite distance $\step > 0$ apart. \\
    For $i=1,2$, let $V_i := \{B(\Omega_i, \frac{\step}{4})\} \cap \{z \in \mathcal{X}: d(F(z), z) < \frac{\epsilon}{2}\}$, where $d(\cdot,\cdot)$ is the distance function and $B(\Omega_i, \frac{\epsilon}{4}) = \{z \in \mathcal{X}: d(z, \Omega_i) < \frac{\epsilon}{4}\}$ is the open set of all points which are at a distance of $<\frac{\epsilon}{4}$ to the set ${\Omega_i}$.\\
    Note that $\Omega_i \subseteq V_i$ for $i=1,2$ since $F(z) = z$ on $\Omega$. Note also that $V_1$, $V_2$ are open in $\mathcal{X}$, with distance $d(V_1, V_2) \geq d(B(\Omega_1, \frac{\epsilon}{4}),B(\Omega_2, \frac{\epsilon}{4})) = \frac{\epsilon}{2}$. In particular, this means that $\fa z \in V_1$, $F(z) \not \in V_2$ (and vice versa), since $d(F(z),z) < \frac{\epsilon}{2} \; \fa z \in V_1 \cup V_2$ . \\
    Now there exists $T$ such that $x(t) \in V_1 \cup V_2$ $\fa t \geq T$, since if not then $\exists$ subsequence $\{x(t_k)\}_k$ that lies within the compact set $\mathcal{X} - (V_1 \cup V_2)$ and hence has a limit point $x^* \in \mathcal{X} - (V_1 \cup V_2)$, which leads to a contradiction since $x^*$ is then a limit point of $\{x(t)\}_t$ that  $\not\in \Omega$. \\
    Suppose then, without loss of generality, that $x(T) \in V_1$. Then since we have established that if $x(t) \in V_1$ then $x(t+1) = F(x(t)) \not\in V_2$, so we must have that $x(t) \in V_1 \; \fa t \geq T$. But this means that no point of $\Omega_2$ is a limit point of $\{x(t)\}_t$, which is a contradiction. Hence $\Omega$ is connected.
\end{proof}

\section{Additional Experiments}\label{appsecs:additionalexperiments}
\begin{figure}[!htb]
    \centering
    \begin{minipage}{.4\linewidth}
      \centering
      \includegraphics[width=.95\linewidth]{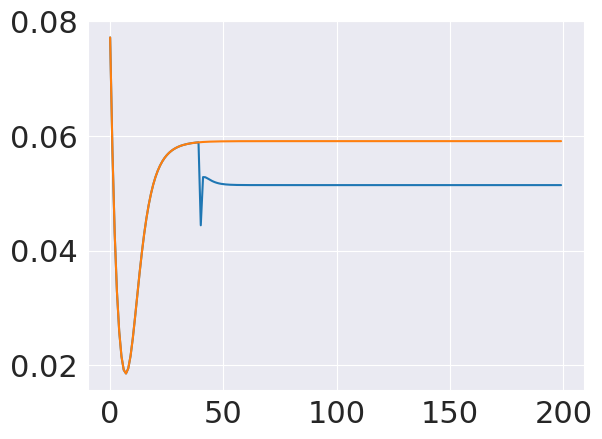}
      Single random perturbation
    %   \label{fig:sub2}
    \end{minipage}
    \begin{minipage}{.4\linewidth}
      \centering
      \includegraphics[width=.95\linewidth]{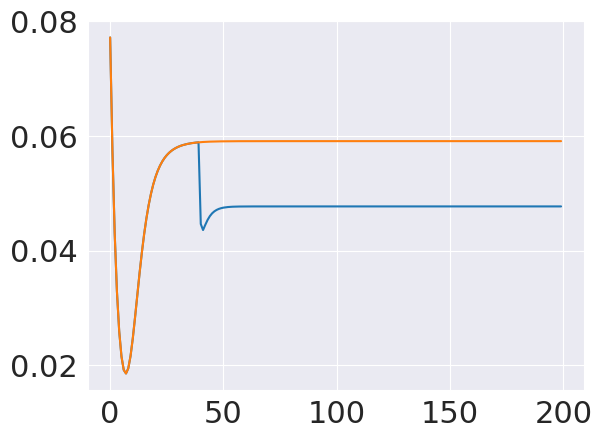}
      Single best response perturbation
    %   \label{fig:sub1}
    \end{minipage}
    \caption{Exploitabilities of \ref{lin-MMWU} with single perturbation.}
    \label{fig:singleperturb}
\end{figure}

In Figure \ref{fig:perturb}, we showed the resultant exploitabilities of \ref{lin-MMWU} after introducing perturbations to the algorithm. In order to generate the plots, we require 3 additional parameters -- maximum allowable exploitability, denoted by $\mathrm{exp}_{max}$ and disturbance/perturbation amount, denoted by $\delta$. $\mathrm{exp}_{max}$ is an upper bound on the maximum allowable exploitability of the dynamics, and $\delta$ is a scalar in $[0,1]$ that denotes how much each player perturbs their current strategy $\rho$ or $\sigma$ by. If randomized, we generate a random Hermitian matrix of suitable dimension, $\mathcal{A}$, and the perturbed strategy is given by $\rho^* = ([1-\delta)\rho + \delta \mathcal{A}]/\tr((1-\delta)\rho + \delta \mathcal{A}))$. Otherwise, the perturbed strategy is given by $\rho^* = [(1-\delta)\rho + \delta \Phi(\rho)]/\tr((1-\delta)\rho + \delta \mathcal{A}))$, which gives a perturbation in the direction of the best response. In Figure \ref{fig:singleperturb}, we also show exploitability plots for \ref{lin-MMWU} with only one perturbation and $\delta = 0.1$, showing that the dynamics go to a fixed point of lower exploitability. We did not perform extensive hyperparameter tuning for this experiment, but from the experiments it is clear that even a single perturbation would improve the performance of \ref{lin-MMWU}.

\begin{figure}[!htb]
    \centering
    \includegraphics[width=0.65\linewidth]{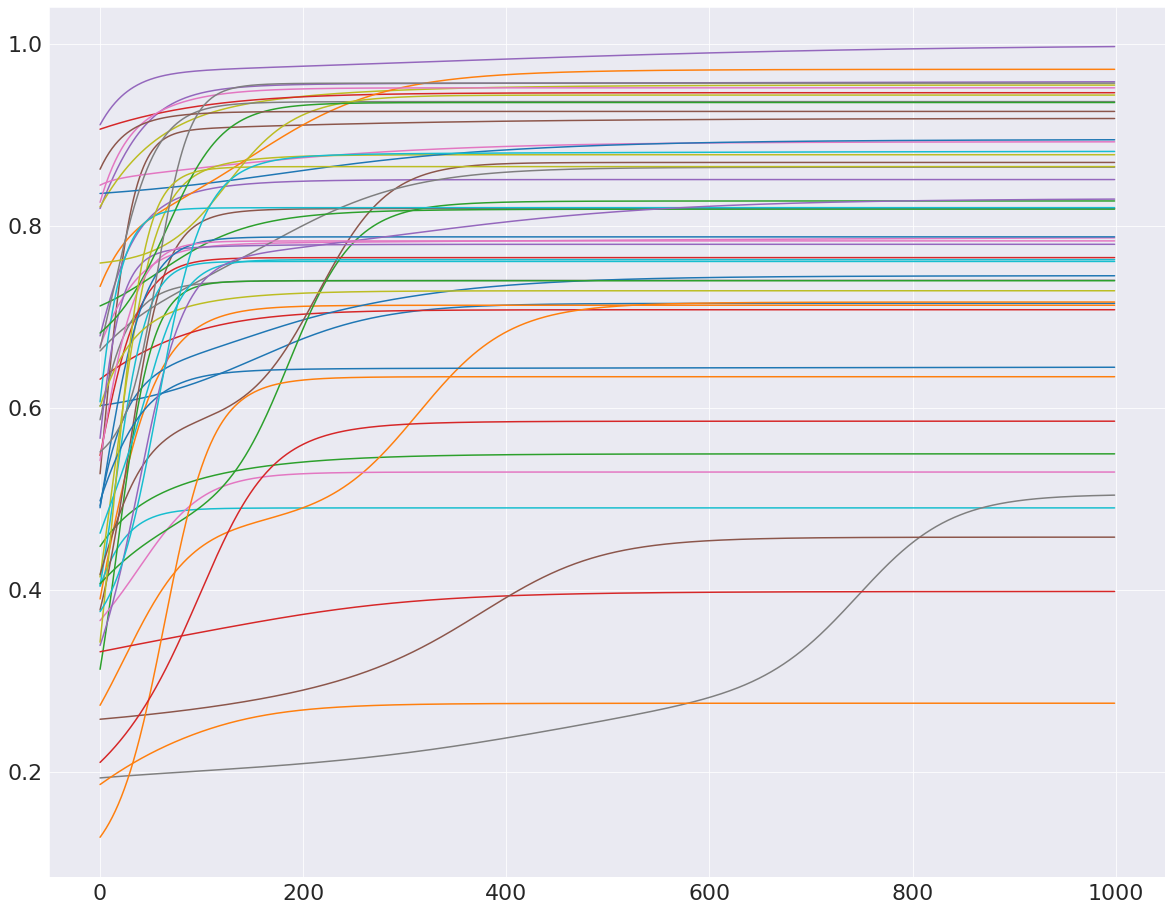}
    \caption{Utility function values for continuous \ref{lin-QREP} dynamics applied to 50 randomly generated quantum common-interest games. In all trajectories, we see that utility increases over time.}
    \label{fig:commoninterest}
\end{figure}

As already determined in the main text, the \ref{lin-QREP} dynamics are a gradient flow, and thus, the common utility function is non-decreasing along its trajectories.  In Figure \ref{fig:commoninterest}, we experimentally verify this fact for the case $q=1$, where we  run the continuous dynamics for $50$ randomly generated quantum common-interest games with randomly generated initial conditions. Our experiments corroborate our theoretical findings, namely that  the utility for both players (plotted is the first player's utility) is strictly increasing unless they are at a fixed point.

\begin{figure}[!htb]
    \centering
    \begin{minipage}{.45\linewidth}
      \centering
      \includegraphics[width=.95\linewidth]{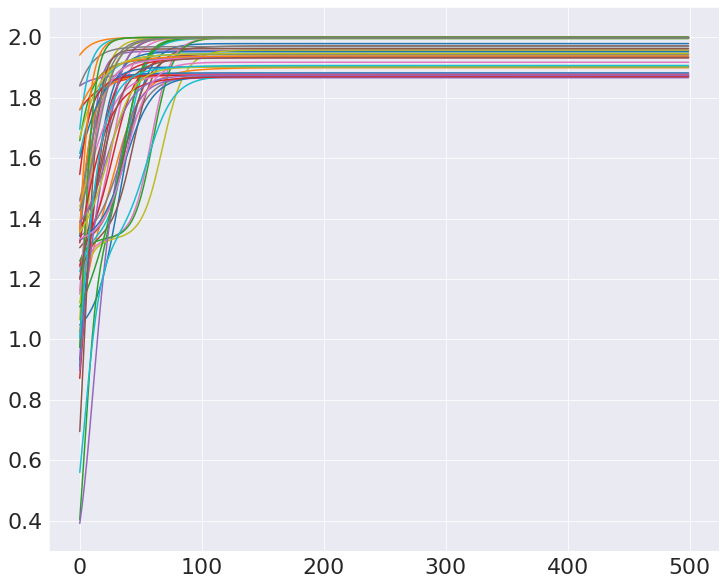}
      Continuous Dynamics (\ref{lin-QREP}).
    %   \label{fig:sub2}
    \end{minipage}
    \begin{minipage}{.45\linewidth}
      \centering
      \includegraphics[width=.95\linewidth]{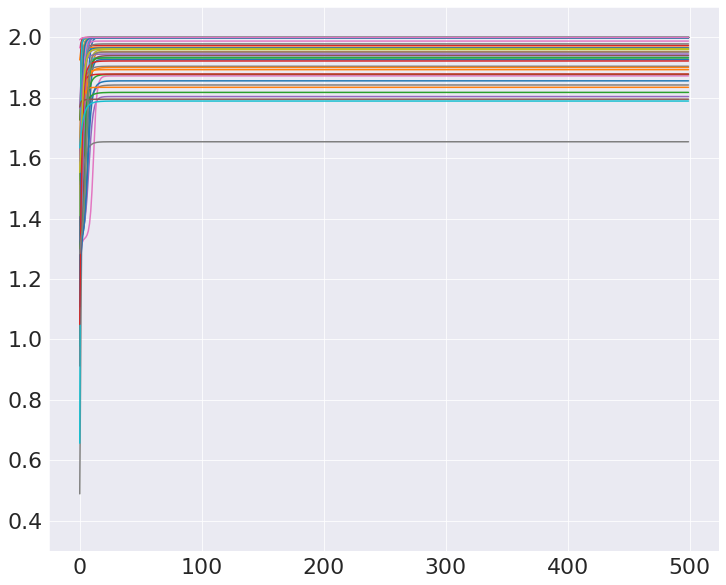}
      Discrete Dynamics (\ref{lin-MMWU}).
    %   \label{fig:sub1}
    \end{minipage}
    \caption{Utility value of continuous vs discrete dynamics (50 random initializations) when applied to a quantum common-interest game.}
    \label{fig:bos}
\end{figure}

Next, we compare the convergence properties of the continuous and discrete dynamics. Our theoretical guarantees for both \ref{lin-QREP} and \ref{lin-MMWU} are that their limit points are a compact connected set of fixed points (Theorem \ref{thm:_a_quantShah}) 
% and Corollary \ref{thm:_hofsigUpdatesAlternating_limitpoints_ClosedConnectedFixedPts}). 
Moreover, we know that the set of fixed points of both dynamics are equivalent (Theorem \ref{thm:_FixedPointsSame_ContTime_DiscreteTime}). We showcase this set of results by using a representative game example (with Nash equilibrium utility of $2$, see Figure \ref{fig:bos}). Indeed, when applied to this game, both \ref{lin-QREP} and \ref{lin-MMWU} converge to a range of utilities which correspond to local optima of the corresponding \ref{BSS} instance, and furthermore this range is similar between the continuous and discrete~dynamic. 

% For the continuous dynamics with $q=1$ \eqref{lin-QREP} and the discretized dynamics (\ref{eqn:_DQREP}) we have results for convergence to a compact connected set of fixed points. We see from a representative game example (with a Nash equilibrium utility value of $2$) that the continuous and discrete dynamics perform similarly (Figure \ref{fig:bos}), converging to a range of utilities which correspond to local optima of the corresponding \ref{BSS} problem instance.

% the discrete dynamics (\ref{update_rule:_hofsig_densitymatrix_potentialgame}) converge to a wider range of maximum utility values compared to the continuous dynamics (\ref{eqn:_aQREP_family}), all tests of which converge to the maximum utility of $2$ (Figure \ref{fig:bos}).

\begin{figure}[!htb]
    \centering
    \begin{minipage}{.45\linewidth}
      \centering
      \includegraphics[width=.95\linewidth]{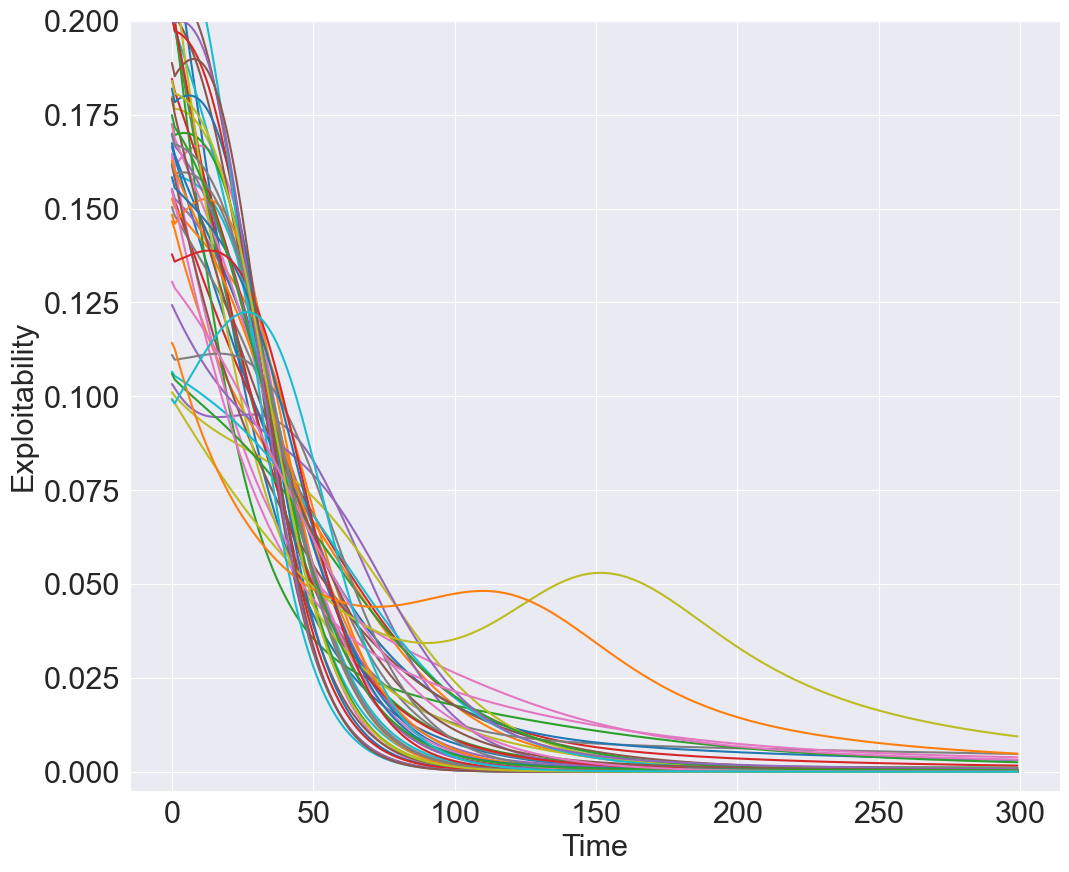}
      Exploitability
    %   \label{fig:sub2}
    \end{minipage}
    \begin{minipage}{.45\linewidth}
      \centering
      \includegraphics[width=.95\linewidth]{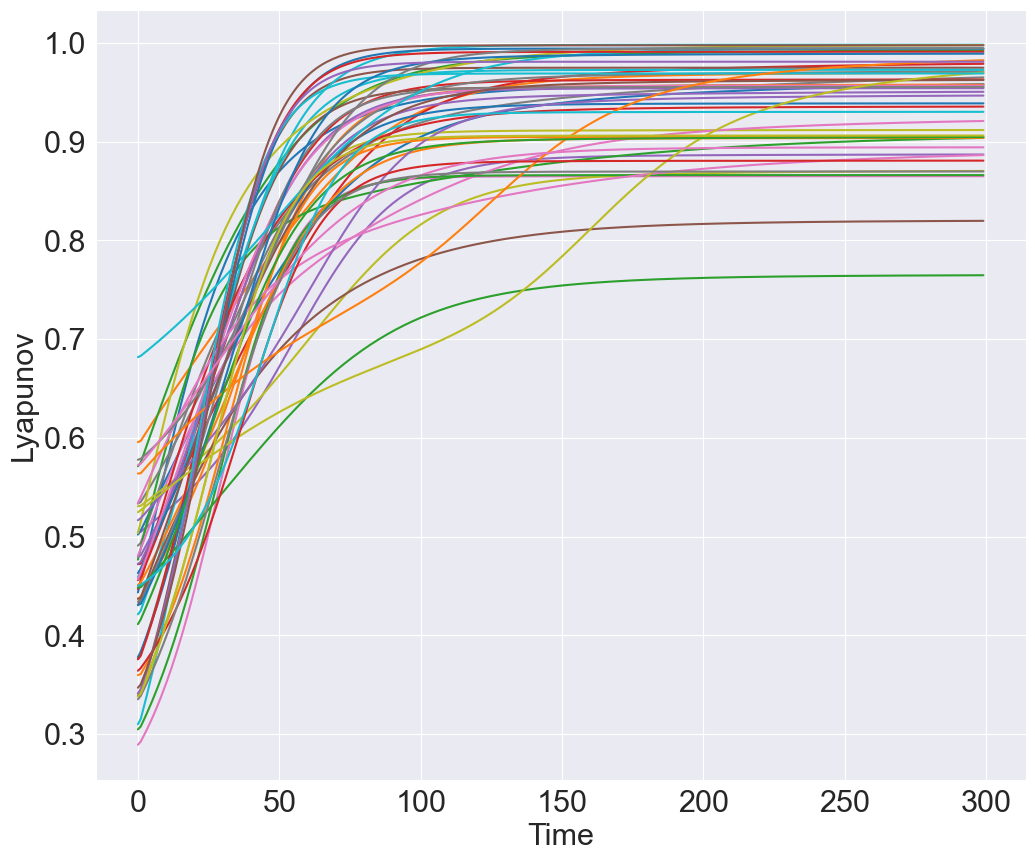}
      Utility function
    %   \label{fig:sub1}
    \end{minipage}
    \caption{Exploitability and utility function values for \ref{MMWU} dynamics applied to 50 randomly generated quantum CIGs.}
    \label{fig:expMMWU}
\end{figure}

Additionally, we perform experiments using \ref{MMWU} (the exponential variant of \ref{lin-MMWU}) with stepsize $\epsilon = 0.1$ on the same set of randomized games as in Figure \ref{fig:exploitlinmmwu}. We show in Figure \ref{fig:expMMWU} that empirically, \ref{MMWU} exhibits similar properties to \ref{lin-MMWU}: utility is increasing over time, and exploitability goes close to zero in a large fraction of the test cases. Moreover, we see that the two ``bad'' cases (orange and brown trajectories) are actually able to escape the peaks of high exploitability and reach lower exploitability over time. We leave this interesting observation to future work.
% However, in the classical setting, \ref{exp} has been shown to exhibit chaotic behavior when applied to CIGs \cite{palaiopanos2017multiplicative}. While showing such a result is outside of the scope of this paper, we nevertheless cannot rule out the possibility of chaotic behavior and thus focus our efforts on \ref{lin-MMWU} instead.

\begin{figure}[!htb]
    \centering
    \begin{minipage}{.45\linewidth}
      \centering
      \includegraphics[width=.75\linewidth]{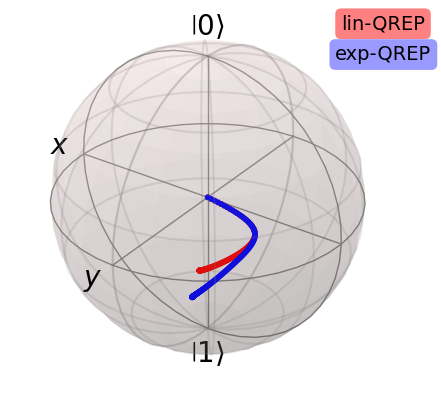}
      % Continuous Dynamics (\ref{lin-QREP}).
    %   \label{fig:sub2}
    \end{minipage}
    \begin{minipage}{.45\linewidth}
      \centering
      \includegraphics[width=.75\linewidth]{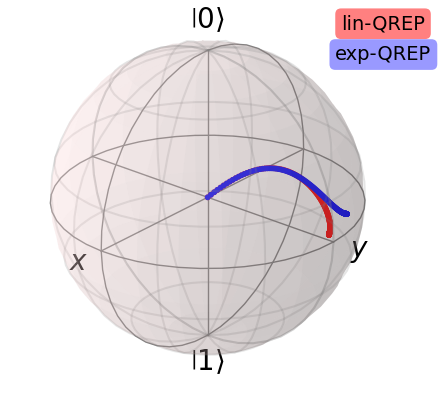}
      % Discrete Dynamics (\ref{eqn:_DQREP}).
    %   \label{fig:sub1}
    \end{minipage}
    \caption{Example simulations where trajectories of \ref{lin-QREP} and quantum replicator dynamics from \cite{jain2022matrix} (\ref{exp-QREP}) diverge.}
    \label{fig:comparing}
\end{figure}

Finally we compare our formulation of continuous time dynamics with the exp-QREP dynamics (which they call \emph{quantum replicator dynamics}) derived in \cite{jain2022matrix}. Their formulation is derived by taking the continuous-time limit of MMWU, which makes it distinct from our \ref{lin-QREP} dynamics. In Figure \ref{fig:comparing}, we run both formulations with the same randomized $\mathcal{H}_2 \otimes \mathcal{H}_2$ game and uniform initial conditions, showing diverging trajectories.

\end{document}

%% file: macrosquantum.tex
\usepackage[utf8]{inputenc}
\usepackage[T1]{fontenc} 
\usepackage{geometry}
\usepackage{caption}
\usepackage{amsmath}
\usepackage{amssymb}
\usepackage[ruled,vlined]{algorithm2e}
\usepackage{bbold}
\usepackage{mathtools}
\usepackage{etoolbox}
\usepackage[dvipsnames]{xcolor}
\usepackage[english]{babel}
\usepackage{amsthm}
\usepackage{hyperref}
\usepackage{subcaption}
\usepackage{float}
\usepackage{paralist}
\usepackage{bbold}
\usepackage{multirow,array}
\usepackage{url}            % simple URL typesetting
\usepackage{booktabs}       % professional-quality tables
\usepackage{amsfonts}       % blackboard math symbols
\usepackage{nicefrac}       % compact symbols for 1/2, etc.
\usepackage{microtype}      % microtypography
\usepackage{bm}
\usepackage{amsmath,amsfonts,amssymb, amsthm}
\usepackage[ruled,vlined]{algorithm2e}
\usepackage{algorithmic}

\usepackage[parfill]{parskip}
\usepackage{physics}
\usepackage{braket}
\usepackage{xfrac}
\usepackage[singlelinecheck = false]{caption}
\usepackage{float}
\usepackage{graphicx}
\usepackage[justification=centering]{caption}
\usepackage{subcaption}
\usepackage{soul}

\usepackage[overload]{empheq} 

\usepackage{mathtools}
\DeclarePairedDelimiterX{\infdivx}[2]{(}{)}{%
  #1\;\delimsize\|\;#2%
}

\DeclareMathOperator*{\argmax}{arg\,max}

\def\bal{\begin{aligned}}
\def\eal{\end{aligned}}

\newcommand{\la}{\langle}
\newcommand{\ra}{\rangle}

\newcommand{\lam}{\lambda}

\DeclareMathOperator{\id}{\mathbb{1}}

\newcommand{\R}{\mathbb{R}}

\def\be{\begin{equation}}
\def\ee{\end{equation}}

\newcommand*\linesmall[0]{\vspace{12pt}}

\newcommand*\innerprod[2]{\left\langle #1, #2 \right\rangle}

\newcommand*\new[1]{{#1}^{new}}

\newcommand*\fa[0]{ \forall \;}

\newcommand*\diag[1]{\mathrm{diag}\left(#1\right)}
\newcommand*\powh[1]{#1^{\sfrac{1}{2}}}

\newcommand*\grads[1]{\textbf{grad}_{#1} \,}
\newcommand*\M[0]{\mathcal{M}} %manifold

\usepackage{scalerel}

\newtheorem{lemma}{Lemma}[section]
\newtheorem{theorem}{Theorem}[section]

\newtheorem{corollary}{Corollary}[section]

\newtheorem*{theorem*}{Theorem}
\newtheorem{remark}{Remark}[section]

\usepackage{graphicx}
\usepackage{caption}

\usepackage[font={small,it}]{caption}
\usepackage[justification=centering]{caption}

\newcommand\da{D(\mathcal{A})}
\newcommand\db{D(\mathcal{B})}

\newcommand*\adj[1]{{#1}^\dagger}

\newcommand{\A}{\mathcal{A}}
\newcommand{\B}{\mathcal{B}}

\newcommand{\BR}{\mathrm{BR}} % best response
\newcommand{\BRa}{\BR_{\A}}
\newcommand{\BRb}{\BR_{\B}}
\newcommand{\ida}{\id_{\A}}
\newcommand{\idb}{\id_{\B}}

\newcommand{\tcone}{\mathrm{T}}
\newcommand{\feasdir}{\mathrm{dir}}
\newcommand{\psdcone}{\mathrm{PSD}}

\makeatletter
\newcommand{\leqnomode}{\tagsleft@true}
\newcommand{\reqnomode}{\tagsleft@false}
\makeatother

\newcommand{\replin}{\ref{replicator}}

\renewcommand{\be}{{\rm BE}}

\DeclareMathOperator{\range}{range}

\newcommand*\ps{\Phi(\sigma)}
\newcommand*\rps{\innerprod{\rho}{\ps}}
\newcommand*\rh{\powh{\rho}}
\newcommand*\rpr{\rh \ps \rh}

\newcommand*\upu{U^\dagger \ps U}
\newcommand*\supu{[\upu]}
\newcommand*\uu[1]{U^\dagger #1 U}
\newcommand*\vrr[1]{[#1]_{[r] \times [r]}}

\newcommand*\h{\frac{1}{2}}

\DeclareMathOperator{\fc}{face}

\DeclareMathOperator{\Diag}{Diag}

%% file: introquantumv6.tex
% \textcolor{blue}{Instead of starting with gamification, start with learning in games, then common-interest/potential games, then learning in common-interest/potential games? Then say how this has been unstudied in the quantum literature. Then bring up the BSS and the gamification angle. (We want to shift focus to studying properties of learning dynamics in common-interest games as opposed to studying algorithms for BSS>)}
The study of game theory has long revolved around the characterization and computation of Nash equilibria in various classes of games. A popular perspective is that of learning in games, where agents play repeated rounds of a game and update their strategies using information obtained from past interactions.
A natural question which arises is when these learning processes give rise to Nash equilibria, and it turns out that in classical two-player zero-sum games and common-interest games (or the more general class of potential games), simple learning dynamics suffice to give rise to Nash equilibria. This connection between learning in games and game-theoretic equilibria has been leveraged in machine learning applications, such as solving large-scale zero-sum games like Poker~\cite{moravvcik2017deepstack} and Go~\cite{silver2016mastering} and understanding the behavior of Generative Adversarial Networks~\cite{goodfellow2014generative, mertikopoulos2018cycles, mertikopoulos2018optimistic}. Many of these success stories have been in zero-sum games, but the focus has recently begun shifting toward understanding learning in games where agents have aligned interests~\cite{dafoe2021cooperative,dafoe2020open}. For instance, many recent works in multi-agent reinforcement learning have explored the challenging frontier of multi-agent coordination~\cite{bard2020hanabi,hu2020other,strouse2021collaborating,leonardos2021global}. As such, studying the convergence of learning dynamics to equilibria in identical/common-interest games, where players share a common utility function, is an important endeavor with many potential applications.

More recently, the study of learning in quantum games where players have access to quantum resources has also begun to gain traction. In classical (normal-form)  games, mixed strategies correspond to probability simplex vectors that capture classical randomness over a finite set of pure strategies; quantum games are a natural extension of the classical game theory framework and have many formulations, see e.g. \cite{eisert1999quantum, gutoski2007toward, bostanci2022quantum, 2012QSGT}. The study of learning in quantum games has so far primarily taken place in the two-player \emph{zero-sum} setting where players select density matrices as strategies and one player obtains a payoff from a bilinear utility function which corresponds to a measurement on the joint state, while the other player gets the negative payoff \cite{jain2009parallel, vasconcelos2023quadratic, jain2022matrix}. Beyond the two-player zero-sum setting, \cite{lotidis2023learning,lin2023no} have studied continuous- and discrete-time learning in general-sum quantum games. However, learning in quantum games where agents share a common utility function has yet to be explored in the literature.

%\cite{lotidis2023learning} showed that a broad class of continuous-time Follow-The-Quantum-Regularized-Leader dynamics (a generalization of the classical Follow-The-Regularized-Leader framework) exhibits constant regret in general quantum games.

In this work, we introduce quantum common-interest games (quantum CIGs) and study learning dynamics---i.e., dynamics that the players use to update their strategies over repeated rounds of play---in this class of games. In a (two-player) quantum CIG, there are two agents---which for concreteness we name Alice and Bob---who control quantum registers \(\mathcal{A}\) and \(\mathcal{B}\) and have strategies given by density matrices in \(D(\mathcal{A})\) and \(D(\mathcal{B})\) respectively. Upon playing the strategy profile \((\rho, \sigma) \in D(\mathcal{A}) \times D(\mathcal{B})\), both players receive a common utility \(u(\rho, \sigma) = \la R, \rho \otimes \sigma \ra\), where \(R\) is a Hermitian matrix assumed to be positive definite that we refer to as the \textit{game operator}. 
% Using the Choi-Jamiołkowski isomorphism defined in \eqref{CJ}, it is useful to express the utility function as \(u(\rho, \sigma) = \la \rho, \Phi(\sigma^\top) \ra\),~as
% \begin{align*}
% \la \rho, \Phi(\sigma^\top) \ra &= \langle \rho, \mathrm{Tr}_\mathcal{B} (R(\mathbb{1}_\mathcal{A} \otimes \sigma)) \rangle = \langle \rho \otimes \id_\mathcal{B}, R(\mathbb{1}_\mathcal{A} \otimes \sigma) \rangle = \langle R, \rho \otimes \sigma \rangle,
% \end{align*}
% where \(R\) is the Choi matrix of \(\Phi\). 
The canonical solution concept is the \textit{Nash equilibria} (NE) of the game, which are the strategy profiles \((\rho, \sigma) \in D(\mathcal{A}) \times D(\mathcal{B})\) such that Alice's and Bob's strategies are best responses to each other,~i.e.,
\[ u(\rho, \sigma) \geq u(\rho', \sigma) \ \forall \; \rho' \in D(\mathcal{A}) \quad \text{ and } \quad u(\rho, \sigma) \geq u(\rho, \sigma') \ \forall \; \sigma' \in D(\mathcal{B}). \]

Beyond multi-agent interaction, common-interest games also describe decentralized optimization of a common objective function (the utility). Quantum CIGs thus have a natural link to the 
Best Separable State problem (BSS), which corresponds to linear optimization over the convex hull of bipartite product states \(\rho \otimes \sigma\), i.e., 
 \begin{equation}\label{BSS}\tag{BSS}
        \max \{ \Tr(R(\rho \otimes \sigma)) : \ \rho \in D(\mathcal{A}), \, \sigma \in D(\mathcal{B}) \}
\end{equation} 
where \(R\) is a fixed Hermitian matrix. The BSS problem is a crucial challenge in quantum information theory, closely tied to entanglement detection \cite{grotschel2012geometric, ioannou2006computational, gurvits2003classical, gharibian2008strong}. Furthermore, we prove in this paper that KKT points of a BSS problem instance correspond to Nash equilibria of its corresponding quantum CIG. Hence, learning dynamics for quantum CIGs can be used as decentralized algorithms for the BSS problem. 
% where the quantum states \(\rho\) and \(\sigma\) are updated in a decentralized manner using first-order feedback \(\nabla_\rho \Tr(R(\rho \otimes \sigma))\) and \(\nabla_\sigma \Tr(R(\rho \otimes \sigma))\).

Motivated by the above, in this paper we introduce and study natural extensions of classical game dynamics to the quantum CIG setting.
One of the most important approaches for learning in games relies on the notion of regret, which 
quantifies the difference between the actual payoff an agent receives and the optimal payoff they could have secured in hindsight by playing a single fixed action  \cite{cesa2006prediction}. No-regret algorithms are algorithms for which agents accrue sublinear regret over time, ensuring that the decisions made are not significantly worse than the best fixed strategy in hindsight. A key result in the special case of two-player zero-sum games is that if players use no-regret algorithms to update their strategies, then the limit points of their time-averaged strategy profiles
% , where \(u_\mathcal{A}(x, y) + u_\mathcal{B}(x, y) = 0\), 
are approximate Nash equilibria. 
% This result has produced several important results in machine learning and beyond \cite{dafoe2021cooperative, dafoe2020open, moravvcik2017deepstack, silver2016mastering}. 
However, beyond the setting of zero-sum games, the limit points of the time-averaged strategy profile of players using no-regret algorithms are approximate coarse correlated equilibrium \cite{roughgarden2010algorithmic}, which may still include 
% highly inefficient (e.g., 
% non-rationalizable)
% non-rationalizable strategies that do not make sense for rational players to play 
strategies that no rational players will play
\cite{viossat2013no,heliou2017learning}. 
% Furthermore, no-regret learning has also been shown to exhibit cyclical and even chaotic behavior in common-interest games \cite{mertikopoulos2018cycles,palaiopanos2017multiplicative}.

% The key result for no-regret learning
% % that underlies this direction of research 
% is the folk theorem that time-averaged trajectories of no-regret play in any game, so in particular also for a common-interest game, converge to a coarse correlated equilibrium \cite{roughgarden2010algorithmic}. Moreover, in the important special case of two-player zero-sum games,
% % , where \(u_\mathcal{A}(x, y) + u_\mathcal{B}(x, y) = 0\), 
% no-regret learning converges to a Nash equilibrium, a result with important applications in machine learning \cite{dafoe2021cooperative, dafoe2020open, moravvcik2017deepstack, silver2016mastering}. However, these results for no-regret learning are not satisfactory for our setting of common-interest games for two main reasons: firstly, the set of coarse correlated equilibria might still include highly inefficient (e.g., non-rationalizable) strategies, which are not necessarily reasonable for both players to play \cite{viossat2013no,heliou2017learning}, and secondly, no-regret learning can exhibit cyclical and even chaotic behavior in common-interest games \cite{mertikopoulos2018cycles,palaiopanos2017multiplicative}.
% Towards exploring pointwise convergence of learning dynamics to Nash equilibria in common-interest games, a line of research explores \emph{Lyapunov}-type analysis 
Beyond no-regret, a line of research based on \emph{Lyapunov}-type analysis \cite{monderer1996potential,swenson2018best,krichene2015online,palaiopanos2017multiplicative} has been able to establish pointwise convergence of learning dynamics to Nash equilibria in the specific setting of common-interest games.
% relies on \emph{Lyapunov}-type analysis \cite{monderer1996potential,swenson2018best,krichene2015online,palaiopanos2017multiplicative}.
% In this setup, several natural no-regret dynamics can exhibit cyclic behavior \cite{add}. 
The common utility that players receive serves as a natural choice of Lyapunov function---i.e., a function whose value increases along trajectories of the dynamics---in common-interest games because it inherently aligns all players' incentives towards a shared objective. An example of dynamics in common-interest games for which the common utility is a Lyapunov function is the \emph{best response dynamics}, where the players make sequential unilateral updates to the pure action that maximizes the common utility \cite{roughgarden2010algorithmic,tardos2007network,swenson2018best}. 
% In classical normal-form games, a pure action always lies in the best response set, and so the best response update can be restricted to pure actions, which means that at each timestep the players are playing a pure strategy profile. 
Since at every step this common utility increases and there is only a finite number of pure strategy profiles, this process must necessarily terminate in finite time at a strategy profile from which no player can improve the common utility via unilateral deviation, which is thus a Nash equilibrium.
% Along this process will, due to the finite number of pure strategy profiles, eventually converge to a strategy profile from which no player can improve the common utility via unilateral deviation, which is a Nash equilibrium.
A smoother update that maintains a distribution over actions (instead of jumping between pure actions) is the \emph{linear multiplicative weights update}, which plays each of the pure actions with a probability proportional to a weight that is updated according to how it performed. 
% For this the fact that {\em the utility is non-decreasing along trajectories}, while not guaranteeing convergence on its own, implies based on basic dynamical systems theory that the set of limit points of dynamics satisfying this property forms a compact, connected set of fixed points which attain the same utility \cite{sandholm2010population}. It can further be proven that 
In common-interest games, it has been shown that under the assumption that fixed points are isolated, any trajectory of the linear multiplicative weights update that is initialized in the interior converges to Nash equilibria~\cite{palaiopanos2017multiplicative}. 

{\bf Contributions.}  In this paper, we introduce and study the performance of quantum analogues of both the best response update and the linear multiplicative weights update (as well as an analogue of the continuous-time replicator dynamics) in quantum common-interest games, proving analogues of classical convergence theorems and explaining points of difference in the quantum setting and why they arise.

In Section \ref{sec:related}, we provide the necessary background on learning in classical and quantum games, which this work builds upon and extends, as well as quantum preliminaries. In Section \ref{sec:QCIG}, we  introduce quantum common-interest games (quantum CIGs). We demonstrate that any instance of the Best Separable State  problem  can be interpreted as a quantum CIG, where the Karush-Kuhn-Tucker  points of the BSS instance correspond to Nash equilibria of the game. 

Our first section on dynamics is Section \ref{sec:brdynamics}, where we show that best response dynamics converge to the set of Nash equilibria in~two-player quantum CIGs.
Subsequently, in Section \ref{sec:contdynamics} we introduce our proposed (continuous-time) linear quantum replicator dynamics and (discrete-time) linear matrix multiplicative weights update and study their convergence properties in quantum CIGs. We show that Nash equilibria are fixed points, and limit points of the dynamics are also fixed points (and hence a superset of the Nash equilibria). Our continuous-time dynamic is a noncommutative generalization of the celebrated replicator dynamics \cite{sandholm2010population,taylor1978evolutionary}, while our discrete-time dynamic is a noncommutative extension of the linear multiplicative weights update \cite{palaiopanos2017multiplicative}. Crucially, we find that the classical result of \cite{palaiopanos2017multiplicative} for convergence to Nash equilibria in classical CIGs does \emph{not} extend to the quantum setting.

% In Sections \ref{sec:contdynamics} and \ref{sec:brdynamics} respectively, we study continuous and discrete-time dynamics for learning in a QCIG. We show that if players update their states according to any of our proposed dynamics, the common utility strictly increases, limit points are fixed points, and interior fixed points are Nash equilibria. Our continuous-time dynamics are noncommutative generalizations of the celebrated replicator dynamics \cite{sandholm2010population,taylor1978evolutionary}, while our discrete-time dynamics are noncommutative extensions of the Baum-Eagon dynamics \cite{baum1967inequality} and the linear multiplicative weights updates \cite{palaiopanos2017multiplicative}. Furthermore, we show that best response dynamics converge to the set of Nash equilibria in~QCIGs.

Throughout the paper, we also assess the performance of our dynamics through extensive simulations. 
% We show that our continuous-time dynamics empirically converge to Nash equilibria. 
In Section \ref{sec:bss}, we also evaluate our discrete-time algorithms on the BSS problem, demonstrating that they closely approximate the optimal value.

\section{Background and Related Work on Learning in Games
% Related Work on Learning in Classical and Quantum Games
}\label{sec:related}

\subsection{Classical common-interest and potential games}
In a classical normal-form game, each player selects an action or, more generally, a distribution over actions. Each player then receives a payoff based on their utility function, which maps the set of pure action profiles (tuples of actions played by each player) to the real numbers. A common-interest game (CIG) is one where every player has the same utility function.

For concreteness, we focus on the two-player case, where Alice and Bob each have a finite set of pure actions (\(\mathcal{A}\) and \(\mathcal{B}\), respectively) and choose as their strategies probability distributions \(x \in \Delta(\mathcal{A})\) and \(y \in \Delta(\mathcal{B})\) to sample their pure actions from. If they select strategies  \((x, y)\), their expected utility~is
\[ 
u(x, y) = x^\top Ay = \sum_{i,j} x_i y_j A_{ij}, 
\]
where \(A_{ij}\) is the payoff if Alice selects pure (i.e., deterministic) strategy \(i\) and Bob selects pure strategy \(j\). common-interest games  model situations where players have aligned incentives and work towards a common goal, making them applicable in various cooperative scenarios.

The canonical solution concept in games is a Nash equilibrium, i.e., a strategy profile \((x^*, y^*)\) that is stable under unilateral player deviations. This means
\[ u(x^*, y^*) \ge u(x, y^*) \quad \forall \ x \in \Delta(\mathcal{A}) \quad \text{ and } \quad u(x^*, y^*) \ge u(x^*, y) \quad \forall \ y \in \Delta(\mathcal{B}). \]
In classical CIGs, where both agents are maximizing the same utility function, there is a natural connection between the game (with a common payoff matrix \(A\)) and the bilinear optimization problem 
$$ \max \{ x^\top A y : \ x \in \Delta(\mathcal{A}), \, y \in \Delta(\mathcal{B}) \},$$
over the product of simplices. Specifically, the optimization problem's KKT points correspond to Nash equilibria of the game \cite{sandholm2010population}. 
% It is well known that computing equilibria even in two-player games is a hard task \cite{chen2009settling, daskalakis2009complexity}. 

A closely related and very expressive class of games are potential games \cite{monderer1996potential}, characterized by the existence of a potential function $V$ that tracks each player's change in utility due to unilateral deviations, i.e., 
$$ 
u_i(s, s_{-i}) - u_i(s', s_{-i})
        =
        V(s, s_{-i}) - V(s', s_{-i}) \quad \forall \ s, s', s_{-i},
$$  
where $u_i$ is the utility function of the $i$-th player. Potential games are one of the most studied game models and have extensive applications including Cournot competition \cite{monderer1996potential} and congestion games \cite{rosenthal1973class} (see., e.g., \cite{marden2009cooperative, zeng2018potential, he2019game, della2016potential} for various theoretical and engineering applications). Although the definition of CIGs is more narrow, CIGs are  ``equivalent'' to potential games in that every potential game has a corresponding CIG with the same Nash equilibria. Additionally, for any first-order dynamic—i.e., dynamics that use only first-order information—the trajectories in the potential game and its corresponding  CIG  are identical. This is because  each player's utility can be separated into a coordination term (which is the same for all players and equal to the potential $V$) and a dummy term (that only depends on the other players), i.e.,
\begin{equation*}
        u_i(s)=V(s)+D_i(s_{-i}),
\end{equation*}
e.g., see \cite{la2016potential}.
Due to this decomposition, for each player \(i\), the gradients of \(u_i(s)\) and \(V(s)\) with respect to their own strategy are equal. Consequently, the trajectories of players' strategies under first-order learning dynamics will be identical, whether they are playing the potential game or the CIG where each player receives utility~\(V\).

\subsection{Learning dynamics in classical games}\label{sec:classicaldynamics} 
% For concreteness, we restrict ourselves to the setting of a normal-form (two-player) repeated game in our exposition of classical learning dynamics. In this game, Alice and Bob each have a finite set of pure actions (\(\mathcal{A}\) and \(\mathcal{B}\), respectively) and receive a common expected utility of \(x^\top  A y\), where $x\in \Delta(\A), y \in \Delta(\B)$. We shall state the dynamics for Alice only as they can easily be extended to dynamics for Bob or dynamics in games with more players by replacing the $Ay$ with the gradient of the player's utility function with respect to the players strategy.

Learning in games takes a dynamic perspective in which agents    play a game  repeatedly, in either discrete or continuous time, and ``learn'' over time how to adjust their strategies as the system equilibrates. In this context, the players' algorithms are called the learning dynamics, and  determine the next iterates \( x^{t+1} \) and \( y^{t+1} \) in discrete time, or the rate of change of each strategy, \( \dot{x} \) and \( \dot{y} \), in continuous time. 
% Therefore, computing a good Nash equilibrium—i.e., one with high utility—is equivalent to finding a good KKT point. However, the setup of ``offline''  optimization differs from the  setting  of games in that  agents need to achieve this in a decentralized manner. Specifically, this work goes beyond the static setup where players engage in a strategic interaction modeled as a game and, perhaps even with knowledge of the payoff matrices and strategy sets, try to compute an equilibrium.  This work takes a dynamic perspective in which agents    play a CIG  repeatedly, in either discrete or continuous time, and ``learn'' over time how to adjust their strategies as the system equilibrates. In this context, the players' algorithms are called the learning dynamics, and  determine the next iterates \( x^{t+1} \) and \( y^{t+1} \) in discrete time, or the rate of change of each strategy, \( \dot{x} \) and \( \dot{y} \), in continuous time.

The field of evolutionary game theory (see, e.g., \cite{hofbauer2003evolutionary}), which studies how the makeup of interacting ecological populations evolve over time, lends itself naturally to this field of learning in games due to the presence of explicit evolutionary dynamics.  One property that learning dynamics coming from evolutionary game theory try to capture is that if a strategy attains more than the average utility then the probability that it is played should increase, mirroring natural selection (populations that are fitter than average should proliferate). In continuous time, which is
the setting of much of evolutionary game theory (studying the relative growth of populations sizes over time), this notion is perfectly captured by the \emph{replicator dynamics}, which is perhaps the most well-studied first-order continuous-time learning dynamic.  In our two-player setting, the replicator dynamics for the first player are given by the differential equation
\begin{equation}\label{replicator}\tag{\rm{REP}}
\dot{x}_\ell=x_\ell((Ay)_\ell - x^\top Ay), 
\end{equation}%\quad \text{ and }  \quad \dot{y}_i=y_i((A^\top x)_i-x^\top Ay),$$
where $x_{\ell}$ is the probability assigned to the $\ell$-th pure action, or via the equivalent closed-form formulation
\begin{equation*}
%\label{rep-exp}
% \tag{\rm{REP}}
x_\ell={\exp(s_\ell(t))\over \sum_i\exp(s_\ell(t))}, \quad s_\ell(t)=\int_0^t(Ay(\tau))_\ell d\tau,
\end{equation*}
see, e.g. \cite{taylor1978evolutionary,hofbauer1998evolutionary,bomze1983lotka,weibull1997evolutionary, cressman2014replicator}. The dynamics for the second player can be analogously written.

There are several approaches for discrete-time learning (i.e., through rounds of repeated play) in normal-form games, and we review the ones most relevant to us. The {\em linear multiplicative weights update} is given by
\begin{equation}
    x_\ell \leftarrow x_\ell {1 + \step (Ay)_\ell \over \sum_\ell x_\ell (1 + \step (Ay)_\ell)} \label{linear}\tag{\rm{lin-MWU}},
\end{equation}
where $\step$ is an adjustable stepsize in $(0, +\infty]$. The limiting case where the stepsize $\step \to +\infty$ gives the update
\begin{equation}
 x_\ell \leftarrow x_\ell{(Ay)_\ell \over x^\top Ay},
\label{BE}
\tag{\rm{BE}}
\end{equation}
which is a special case of the Baum-Eagon (BE) update \cite{baum1967inequality, hofbauer2003evolutionary} for polynomial optimization over the product of simplices.  \ref{linear} and \ref{BE} both adjust the  probability of playing a strategy \(\ell\) based on the strategy's performance \((Ay)_\ell\) relative to the average performance \(x^\top A y\), and can be seen as discretizations of the \ref{replicator} dynamics. Similar to the \ref{replicator} dynamics, both  \ref{BE} and \ref{linear}  have the property that they increase the weight of a strategy when it performs better than average and decrease it when it performs worse. Since \ref{linear} and \ref{BE} have similar properties that are relevant to us, we shall regard \ref{BE} as a special case of \ref{linear} with stepsize $\step = +\infty$ and henceforth only mention \ref{linear} in our discussions.

%Thus while \ref{BE} and \ref{linear} are usually seen as distinct (though related) dynamics in the literature, we shall discuss mainly \ref{linear} and make inclusions or exclusions for the limiting case \ref{BE} where~needed.

A discretization of \ref{replicator} that is based on its closed-form exponential formulation is the {\em exponential multiplicative weights update}, defined as
\begin{equation}
    x_\ell \leftarrow x_\ell { \exp(\step(Ay)_\ell)\over \sum_\ell x_\ell \exp(\step(Ay)_\ell)}
    \label{exp} \tag{\rm{exp-MWU}}.
\end{equation}
The expression of \ref{exp} can be obtained by writing the exponential closed-form formulation of \ref{replicator} recursively for the case where the payoffs are observed at discrete times. Both \ref{linear} and \ref{exp} fall within the {\em multiplicative weights} family of learning algorithms which adjust strategies' weights according to their performance, see e.g. \cite{arora2012multiplicative,palaiopanos2017multiplicative, freund1997decision}.

Finally, another natural algorithmic approach to learning in games is the alternating best response dynamics
\begin{equation}
\tag{BR} \label{BRc}
\begin{split}
x(t+1) &\in \argmax_{x \in \Delta(\mathcal{A})} u(x, y(t)), \\
y(t+1) &\in \argmax_{y \in \Delta(\mathcal{B})} u(x(t+1), y).
% \new{x} &\in \argmax_{x' \in \Delta(\mathcal{A})} u(x', y), \\
% \new{y} &\in \argmax_{y' \in \Delta(\mathcal{B})} u(\new{x}, y').
% \rho_{t+1} &\in \argmax_{\rho'}\langle \rho', \Phi(\sigma_t)\rangle, \\
% \sigma_{t+1} &\in \argmax_{\sigma'}\langle \rho_{t+1}, \Phi(\sigma')\rangle.
\end{split}
\end{equation} 
see e.g. \cite{roughgarden2010algorithmic}. In this setting, players update their strategies in an alternating fashion: at time \(t+1\), the updating player selects a strategy that is a best response to the other player's strategy at time \(t\). In a normal-form game, the updates can always be chosen to be pure strategies as the set of maximizers of the linear utility function over the simplex will always contain a vertex of the simplex.

As a first-order check that these dynamics can be used for learning Nash equilibria of a (common-interest) game, it can be easily verified that Nash equilibria are fixed points of all of the \ref{replicator}, \ref{linear}, \ref{exp}, and \ref{BRc} dynamics \cite{hofbauer1998evolutionary, roughgarden2010algorithmic}. However, convergence to Nash equilibria is not as simple to achieve, and often requires additional conditions. For \ref{replicator}, it is known (as part of the ``folk theorem'' of evolutionary game theory) that the limit points of trajectories initialized in the interior of the strategy space---i.e., from mixed strategies with full support---are Nash equilibria, which has the implication that any convergent interior trajectory goes to a Nash equilibrium (see e.g., \cite{hofbauer1998evolutionary, cressman2014replicator}). In addition, it has been shown that, in all except a measure-zero set of common-interest games and initializations, convergence to (pure) Nash equilibria is achieved \cite{kleinberg2009multiplicative, panageas2016average}. 

For the discrete-time dynamics on the other hand, it is known that under the assumption that fixed points are isolated, trajectories of \ref{linear}  that are initialized in the interior converge to Nash equilibria \cite{palaiopanos2017multiplicative}. Alternating \ref{BRc} does converge in finite time to pure Nash equilibria in common-interest games (see e.g., \cite{roughgarden2010algorithmic, tardos2007network}), but it is worth noting that the canonical proof of this---which we have written in our introduction---relies heavily on the finiteness of the pure strategy profile space that \ref{BRc} stays within.

% s known for the replicator dynamics (\ref{replicator}) that 
% % that replicator dynamics (\ref{replicator}) converge to Nash equilibria \cite{hofbauer1998evolutionary, kleinberg2009multiplicative, panageas2016average}. 
% % This is based on results establishing an equivalence between the set of fixed points of \ref{replicator} and the set of Nash equilibria of the CIG. 
% In discrete time, it is known that the alternating \ref{BRc} converges to pure Nash equilibria in CIGs (see e.g., \cite{roughgarden2010algorithmic, tardos2007network}), while for \ref{linear} it is known that under the assumption that fixed points are isolated, trajectories initialized in the interior converge to Nash equilibria \cite{palaiopanos2017multiplicative}. 

Both \ref{linear} and \ref{exp} are no-regret algorithms, and thus is it known that if the players use them then the limit points of the time average of the strategy profiles are approximate coarse correlated equilibria \cite{roughgarden2010algorithmic}. However, as stated in the introduction, the set of coarse correlated equilibria can include non-rationalizable strategies, which are strategies that are eliminated during iterative elimination of dominated strategies and thus unsatisfactory~\cite{viossat2013no,heliou2017learning}. Moreover, \ref{exp} (specifically in the fixed stepsize regime which we consider) has been shown to exhibit chaotic, non-convergent behavior in common-interest games \cite{palaiopanos2017multiplicative}. It is for this reason that we focus on \ref{replicator}, \ref{linear}, and \ref{BRc} as the learning dynamics whose quantum analogues we want to study as learning dynamics in quantum common-interest games.

\subsection{Learning dynamics in quantum games}%\label{sec:quantumrelatedwork}
The majority of the literature on quantum games investigates  the potential advantages of using quantum strategies over classical ones. To this end, researchers have developed quantum versions of well-known games such as the Prisoner's Dilemma \cite{eisert1999quantum} and Matching Pennies~\cite{meyer1999quantum}. 
In addition, an increasing amount of research has focused on  studying quantum notions of equilibria, i.e.,   states that remain stable against unilateral player deviations~\cite{2012QSGT}, determining  their tractability \cite{bostanci2022quantum}, and obtaining structural characterizations  of equilibrium sets~\cite{ickstadt2023semidefinite, ickstadt2023semidefinite}. 
Beyond the analysis of specific games, various attempts have been made to establish more general theories of quantum games that aim to unify the existing works, see e.g.,  \cite{gutoski2007toward, chiribella2009theoretical}.

In contrast, there are relatively few works that investigate learning in quantum games. Most existing results focus on a specific zero-sum setting where players select density matrices $\rho$ and $\sigma$ and receive bilinear utilities $u_i(\rho, \sigma) = \text{Tr}(R_i(\rho \otimes \sigma^\top))$
subject to the constraint 
% that 
$u_1(\rho, \sigma) + u_2(\rho, \sigma) = 0$. (The payoffs can also be expressed explicitly as bilinear functions $u_i(\rho, \sigma) = \langle \rho, \Phi_i(\sigma) \rangle$,  where $R_i$ is the Choi matrix \eqref{CJ} of the super-operator $\Phi_i$.) The study of learning in quantum games has drawn much inspiration from learning in classical games. As an initial foray, \cite{jain2009parallel} proved that a noncommutative version of multiplicative weights updates called the Matrix Multiplicative Weights Update (MMWU) can be used to compute approximate Nash equilibria in two-player quantum zero-sum games, and subsequently, \cite{vasconcelos2023quadratic} introduced algorithms for solving quantum zero-sum games that use an extra-gradient mechanism to obtain a quadratic speedup on the rate found in \cite{jain2009parallel}. Most recently, \cite{lin2023no} showed that the noncommutative variant of discrete-time no-regret learning exhibits time-average convergence to quantum coarse correlated equilibria in general quantum games. From a continuous-time perspective, \cite{jain2022matrix,lotidis2023learning} showed that the continuous-time limit of MMWU and generalizations thereof exhibit cyclical (i.e., non-convergent) behavior in two-player zero-sum quantum games. \cite{lotidis2023learning} also showed that a broad class of continuous-time Follow-The-Quantum-Regularized-Leader dynamics (a generalization of the classical Follow-The-Regularized-Leader framework) exhibits constant regret in general quantum games, and proved a modified analogue of the classical evolutionary folk theorem for replicator dynamics which draws connections between Nash equilibria of the game and ``stable points'' of the dynamics.
% Meanwhile, \cite{ickstadt2022semidefinite,ickstadt2023semidefinite} study structural properties of Nash equilibria in quantum zero-sum and general games.

We now state two of the learning dynamics that have been used for learning in quantum games, as they are analogues of some of the classical learning dynamics shown in Section \ref{sec:classicaldynamics} and are closely related to the quantum learning dynamics we shall introduce and study in this paper. The first is the Matrix Multiplicative Weights Update, which (for the $\rho$ player) is given~by:
  \begin{equation}\label{MMWU}\tag{MMWU}
 \rho(t+1) \leftarrow  
 \frac{
    \exp\left(\step \sum_{\tau=1}^t\Phi(\sigma(\tau))\right)
    }{
    \Tr(\exp\left(\step \sum_{\tau=1}^t\Phi(\sigma(\tau))\right))
    },
    \end{equation}
 and converges (in the time-average sense) to Nash equilibria in quantum zero-sum games \cite{jain2009parallel}.  \ref{MMWU} is a generalization of \ref{exp} to the quantum setting and and was first  introduced for  online optimization over the set of density matrices \cite{arora2012multiplicative,kale2007efficient, tsuda2005matrix}.
MMWU has found many applications: important examples include solving SDPs \cite{arora2007combinatorial}, proving that QIP=PSPACE  \cite{jain2011qip}, finding balanced separators \cite{orecchia2012approximating}, and spectral sparsification~\cite{allen2015spectral}.

% \caption{Matrix Multiplicative Weights Update}
   %\STATE {\bfseries Initialize}:  $A_0 = \mathbb{1}_\mathcal{A}$, $\rho_0 = A_0/\mathrm{Tr}(A_0)$, $B_0 = \mathbb{1}_\mathcal{B}$, and $\sigma_0 = B_0/\mathrm{Tr}(B_0)$.

 Recently,  \cite{jain2022matrix} also introduced the exponential quantum replicator dynamics (exp-QREP),  a (continuous-time) quantum generalization of the exponential formulation of the replicator dynamics (\replin) given by:
 % \paragraph{exp-QREP}
\begin{equation}
\rho = \frac{\exp(S(t))}{\mathrm{Tr}(\exp(S(t)))},\quad S(t) = \int_0^t \Phi(\sigma(\tau))d\tau.
 \label{exp-QREP}\tag{exp-QREP}
\end{equation}
% \begin{equation}
% { d\rho \over dt} = \frac{d}{dt}\left(\frac{\exp(A)}{\mathrm{Tr}(\exp(A))}\right),\quad A(t) = \int_0^t \Phi(\sigma(\tau))d\tau.
%  \label{exp-QREP}\tag{exp-QREP}
% \end{equation}
Note that  these dynamics are simply called the quantum replicator dynamics in \cite{jain2022matrix}. 
The main result in \cite{jain2022matrix} is that the \ref{exp-QREP}  dynamics exhibit a type of periodic behavior called Poincar\'e recurrence when applied to quantum zero-sum games. \ref{MMWU} can be obtained as a discretization of the \ref{exp-QREP} dynamics, 
%can be discretized
in the same manner that the discrete-time exponential MWU is obtained from  the exponential variant of the continuous-time replicator dynamics (\replin) in the classical regime. 
Thus, while noncommutative generalizations of the exponential variants of the classical learning dynamics are known and been studied for learning in quantum games, the linear variants have yet to be studied. Moreover, while works such as \cite{lin2023no,lotidis2023learning,ickstadt2023semidefinite,ickstadt2023semidefinite} have studied learning in general quantum games and quantum zero-sum games, quantum common-interest games remain an important class of games which have yet to be studied in the literature.

% \edits{I removed the comparisons between the learning dynamics here because the notation was extremely confusing. I think it's best to put it after the classical dynamics section.}

In this paper, we complete the picture of analogues (see Table~\ref{table:quantumdy_incomplete}) of the classical replicator and multiplicative update learning dynamics we showed in Section \ref{sec:classicaldynamics}  
by introducing the Linear Quantum Replicator Dynamics (\ref{lin-QREP}) and the Linear Matrix Multiplicative Weights Update (\ref{lin-MMWU}) and studying their convergence properties, along with those of the best response dynamics (\ref{BR}), in the class of quantum common-interest games. The analysis of \ref{MMWU} remains out of the scope of this paper, since its classical counterpart (\ref{exp}) has been shown to exhibit chaotic behavior in classical CIGs, and its regret-minimizing property does not suffice to guarantee convergence to Nash.

\begin{table*}[ht]
\centering
\begin{tabular}{|c|c|c|}
% \toprule
\hline
& Continuous-time & Discrete-time \\
\hline
Linear variant & \ref{lin-QREP} (this work)  &  \ref{lin-MMWU} (this work)
% \begin{tabular}{c} \ref{BE} \\ \ref{linear} \end{tabular}
% ref{BE}, \ref{linear} 
\\
\hline
Exponential variant & \ref{exp-QREP} \cite{jain2022matrix} & \ref{MMWU} \cite{kale2007efficient,arora2005fast,tsuda2005matrix}\\
\hline
% \multirow{2}{*}{\textbf{Problem Dimensions}}        
% \bottomrule
\end{tabular}
\caption{Replicator variants and discretizations thereof for learning in quantum games, in analogy to the classical dynamics discussed in Section \ref{sec:classicaldynamics}.}
% Table \ref{table:classicaldy}.}
\label{table:quantumdy_incomplete}
\end{table*}

\subsection{Quantum preliminaries}

%A  {\em finite dimensional complex Euclidean space} refers   to the vector space~$\mathbb{C}^n$  (for some $n\ge 1$) equipped with the canonical inner product on $\mathbb{C}^n$.
%We denote by $\{e_i\}_{i=1}^n$ the standard orthonormal basis of~$\C^n$, {and by $e$ the vector of {all $1$'s} of appropriate dimension}.
%Given two   complex Euclidean spaces  $\A, \B$ we  denote by $\mathcal{L}(\A,\B)$ the space of linear operators from $\A$ to $\B$ which  we endow  with the Hilbert-Schmidt inner product
%$\la A,B\ra:=\tr(A^\dagger B)$ for $A, B\in $\mathcal{L}(\A,\B)$. For an operator $A\in \mathcal{L}(\A,\B)$ we denote its {\em adjoint} operator by $X^\dagger\in \mathcal{L}(\B,\A)$ and its {\em transpose} by $X^\sfT\in \mathcal{L}(\B,\A)$.
%$\mathcal{L}(\A,\B)$.
Finally, we review some quantum preliminaries as well as geometrical properties regarding the minimal face of a point in the set of density matrices, which is the strategy space of our quantum games.

A $d$-dimensional quantum register is mathematically described as the set of unit vectors  in a   $d$-dimensional Hilbert space $\mathcal{H}.$
The \emph{state} of a qudit quantum  register $ \mathcal{H}$ is represented by a \emph{density matrix}, i.e.,  a $d\times d$ Hermitian positive semidefinite matrix with trace equal to one. The state space of a quantum register $\mathcal{H}$ is denoted  by  $D(\mathcal{H})$.
%to denote the set of all density matrices associated with a register that is described by $\mathcal{H}$. One can naturally view such density matrices as linear operators acting on $\mathcal{H}$.
When two quantum registers with associated spaces $\mathcal{A}$ and $\mathcal{B} $ of dimension $n$ and $m$ respectively are considered as a joint quantum register, the associated state  space is given by the density operators  on the tensor product space, i.e., $D(\mathcal{A}\otimes \mathcal{B})$.  If the two registers are independently prepared in states described by $\rho$ and $\sigma$, then the joint state is described by the  density matrix $\rho \otimes \sigma\in \mathbb{C}^{nm\times nm}$.

To interact with a quantum register,  we need to measure it. One mathematical formalism of the process of measuring a  quantum system is the POVM,  defined as a set of positive semidefinite operators $\{P_i\}_{i=1}^m$ such that $\sum_{i=1}^mP_i=\mathbb{1}_\mathcal{H}$, where $\mathbb{1}_\mathcal{H}$ is the identity matrix on $\mathcal{H}$. If the quantum  register  $\mathcal{H}$ is in a state described by density matrix $\rho\in D(\mathcal{H})$, upon performing the measurement $\{P_i\}_{i=1}^m$ we get the outcome $i$ with probability $\langle P_i, \rho \rangle$,
where
$\langle A, B\rangle = \Tr(A^\dag B)$ is the \emph{Hilbert-Schmidt inner product} defined on  the linear space of Hermitian matrices.  Note that $\langle A, B\rangle$ is a real number for any Hermitian matrices $A$ and $B$, and is   non-negative if $A$ and $B$ are positive~semidefinite.

 Given a finite-dimensional Hilbert   space $\mathcal{H}=\mathbb{C}^n$, we denote by $\text{L}(\mathcal{H})$ the set of linear operators acting on $\mathcal{H}$, i.e.,   the set of all $n\times n$ complex matrices over $\mathcal{H}$.
%  , and by   $\text{Pos}(\mathcal{H})$ the cone of positive semidefinite (PSD) $n\times n$ matrices.
 A linear operator that maps matrices to matrices, i.e.,  a mapping  $\Phi:\mathrm{L}(\mathcal{B}) \to \mathrm{L}(\mathcal{A})$, is called a {\em super-operator}. The adjoint  super-operator $\Phi^\dagger:\mathrm{L}(\mathcal{A}) \to \mathrm{L}(\mathcal{B})$  is uniquely determined by the equation
$    \langle A, \Phi(B)\rangle = \langle \Phi^\dagger(A), B\rangle
$.  A super-operator $\Phi:\mathrm{L}(\mathcal{B})\to\mathrm{L}(\mathcal{A})$ is    {\em positive} if it maps PSD matrices  to PSD matrices. % In addition, $\Phi^*$ is positive if and only if $\Phi$ is positive.
There exists a  linear bijection between  matrices $R\in \mathrm{L}(\mathcal{A}\otimes\mathcal{B})$ and super-operators $\Phi:\mathrm{L}(\mathcal{B})\to\mathrm{L}(\mathcal{A})$ known as the {\em Choi-Jamio\l{}kowski isomorphism}. Specifically, for a super-operator $\Phi$  its {\em Choi matrix}~is:
\begin{equation}\label{CJ}
    C_\Phi= \sum_{1\leq i,j\leq m} \Phi(E_{i,j}) \otimes E_{i,j}\in \mathrm{L}(\mathcal{A}\otimes\mathcal{B}),
\end{equation}
where $\{E_{i,j}\}_{i,j=1}^m$ is the standard orthonormal basis of $\mathrm{L}(\mathcal{B}) = \mathbb{C}^{m\times m}$. Conversely, given an operator $R=\sum_{1\le i,j\le m}A_{i,j}\otimes E_{i,j}\in \mathrm{L}(\mathcal{A}\otimes\mathcal{B})$, we can define $\Phi_R:\mathrm{L}(\mathcal{B})\to\mathrm{L}(\mathcal{A})$ by setting $\Phi_R(E_{i,j})=A_{i,j}$ from which it easily follows that $C_{\Phi_R}=R$. Explicitly, we~have
\begin{equation}\label{eqn:superoperator}
    \Phi_R(B) = \mathrm{Tr}_\mathcal{B} (R(\mathbb{1}_\mathcal{A}\otimes B^\top)),
    \end{equation}
    where the partial trace
    $ \mathrm{Tr}_\mathcal{B}:\mathcal{L}(\A \otimes \B)\to \mathcal{L}(\A)$
    is the {\em unique} function
 that satisfies:
\begin{equation*}\label{basic:ptrace}
\mathrm{Tr}_\mathcal{B}(A\otimes B)=A\Tr(B) \;  \forall \ A, B.
\end{equation*}
Moreover, the  adjoint map is $\mathrm{Tr}_\mathcal{B}^\dagger(A)=A\otimes \mathbb{1}_\B$.
Lastly,  a superoperator $\Phi$ is completely positive (i.e., $\mathbb{1}_m\otimes \Phi$ is positive for all $m\in \mathbb{N}$) iff the Choi matrix of $\Phi$ is positive semidefinite. In particular, if the Choi matrix of the super-operator $\Phi$ is PSD, it follows that $\Phi$ is positive.

\paragraph{Geometry of the set of density matrices.} We round off this section with some properties regarding the faces of the set of density matrices. This shall be important to us as the space of strategies available to each player in a quantum game is the set of density matrices of a given dimension.

For a given quantum register $\A$, the \emph{minimal face} (see, e.g., \cite{barvinok2002course}) of a density matrix $\rho \in \da$, which is the smallest face of $\da$ that contains $\rho$, is
\begin{equation}
\label{eq:facerr}
\begin{split}
	\fc_{\da}(\rho) &\equiv \{ X \succeq 0: \; \tr(X) = 1, \; \range(X) \subseteq \range(\rho)\} \\
	&= \{X \succeq 0: \tr(X) = 1, \; U^\dagger X U \text{ is supported on the upper } r \times r \text{ submatrix}\},
\end{split}
\end{equation}
where the second equality is due to the fact that, writing the spectral decomposition $\rho = \sum_{i=1}^n \lambda_i u_i u_i^\dagger$ where $S := \{i: \lambda_i > 0\} = [r]$ and 
% Let $U_r := \begin{pmatrix} u_1 & \ldots & u_r \end{pmatrix}$ and $U_{n-r} := \begin{pmatrix} u_{r+1} & \ldots & u_n \end{pmatrix}$ so that
% $U := \begin{pmatrix} U_r & U_{n-r} \end{pmatrix}$ is unitary. 
$U := \begin{pmatrix} u_1 & \ldots & u_n \end{pmatrix}$ is unitary, we have for any $ X \succeq 0$ that
\begin{equation*}
\begin{split}
	\range(X) \subseteq \range(\rho) = \ker(\rho)^\perp
	&\Longleftrightarrow X u_i = 0 \; \forall \ i > r \\
	&\Longleftrightarrow \text{If } i > r \text{ or } j > r \text{ then } [U^\dagger X U]_{ij} = u_i^\dagger X u_j = 0 \\
	&\Longleftrightarrow U^\dagger X U \text{ is supported on the upper } r \times r \text{ submatrix}.
\end{split}
\end{equation*}

The relative interior of the minimal face of $\rho$ on $\da$ is given by
\begin{equation}
\label{eq:minface}
{\rm relint} \, \fc_{\da}(\rho)=\{X\succeq 0: \; \tr(X)=1, \; \range(X)= \range(\rho)\}.
\end{equation}

Note that these are in direct analogy to the classical setting, where the minimal face of a probability distribution $x \in \Delta(\mathcal{A})$ is the set $\{x': x'_\ell \geq 0 \ \forall \; \ell, \; \sum_{\ell} x'_\ell = 1, \; \textrm{supp}(x') \subseteq \textrm{supp}(x)\}$ and its relative interior is the set $\{x': x'_\ell \geq 0 \ \forall \; \ell, \; \sum_{\ell} x'_\ell = 1, \; \textrm{supp}(x') = \textrm{supp}(x)\}$.